\documentclass[acmsmall,screen]{acmart}

\AtBeginDocument{%
  \providecommand\BibTeX{{%
    \normalfont B\kern-0.5em{\scshape i\kern-0.25em b}\kern-0.8em\TeX}}}

\setcopyright{acmcopyright}
\copyrightyear{2022}
\acmYear{2022}
\acmDOI{10.1145/3510422}

\usepackage{lipsum}
\usepackage[normalem]{ulem}
\usepackage{booktabs}
\usepackage{enumitem}
\usepackage{comment}
\usepackage{graphicx}
\usepackage{multirow}
\usepackage{float}
\usepackage{listings}
\usepackage{soul}
\usepackage{xcolor}
\usepackage{amsmath}
\usepackage{tikz}
\usepackage{romannum}
\usepackage{algorithm,algpseudocode,amsmath}
\usepackage[font=scriptsize,labelfont={scriptsize}]{subfig}
\usepackage[export]{adjustbox}
\usepackage{multicol}

\begin{document}

\title{ERASE: Energy Efficient Task Mapping and Resource Management for Work Stealing Runtimes}





\author{Jing Chen}
\email{chjing@chalmers.se}
\author{Madhavan Manivannan}
\email{madhavan@chalmers.se}
\author{Mustafa Abduljabbar}
\email{musabdu@chalmers.se}
\author{Miquel Peric{\`a}s}
\email{miquelp@chalmers.se}
\affiliation{%
  \institution{Chalmers University of Technology}
  \city{Gothenburg}
  \country{Sweden}}

\begin{abstract}
Parallel applications often rely on work stealing schedulers in combination with fine-grained tasking to achieve high performance and scalability.  
However, 
reducing the total energy consumption
in the context of work stealing runtimes is still challenging, particularly when using asymmetric architectures with different types of CPU cores.
A common approach
for energy savings
involves dynamic voltage and frequency scaling (DVFS) wherein throttling is carried out based on factors like task parallelism, stealing relations and task criticality. This paper makes the following observations: 
(i) leveraging DVFS on a per-task basis is impractical when using fine-grained tasking and in environments with cluster/chip-level DVFS;
(ii) task moldability, wherein a single task can execute on multiple threads/cores via work-sharing, can help to reduce energy consumption; and (iii) mismatch between tasks and assigned resources (i.e.~core type and number of cores) can detrimentally impact energy consumption. 
In this paper, we propose ERASE (EneRgy Aware SchedulEr), an intra-application task scheduler on top of work stealing runtimes that aims to reduce the total energy consumption of parallel applications. 
It achieves energy savings by guiding scheduling decisions based on per-task energy consumption predictions of different resource configurations.
In addition, ERASE is capable of adapting to both given static frequency settings and externally controlled DVFS. 
Overall, ERASE achieves up to 31\% energy savings and improves performance by 44\% on average, compared to the state-of-the-art DVFS-based schedulers.
\end{abstract}


\acmJournal{TACO}
\acmYear{2022} \acmVolume{1} \acmNumber{1} \acmArticle{1} \acmMonth{1} \acmPrice{}\acmDOI{10.1145/3510422}



\maketitle

\section{Introduction}

Reducing energy consumption of multiprocessor systems is a major requirement to enhance battery life, reduce the cost of cooling and overall improve system's scalability and reliability.
Asymmetric multiprocessor systems featuring different types of cores with different performance and power characteristics have been introduced to reduce the total energy consumption while still enabling high performance.
Examples of such systems include~ARM's big.LITTLE architecture~\cite{Jeff2012AdvancesIB}, Intel’s hybrid Lakefield CPU~\cite{Lakefield} and Apple's A12-A14 bionic processors~\cite{AppleA12}. However, effectively scheduling parallel applications on asymmetric multiprocessors remains an open challenge. 
Random work stealing is a well-known approach for scheduling task-parallel applications targeting load balancing and scalability~\cite{blumofe-jacm99, ChenQuan-IPDPS2012-WorkStealing/Asymmetry/Performance} that has been implemented in several production runtimes, such as Cilk~\cite{frigo-cilk5}, TBB~\cite{tbb_scheduler} and OpenMP's explicit tasks~\cite{openmp50-api}. 
In this paper, we show that random work stealing does not produce energy efficient schedules, particularly on asymmetric architectures.
The two principles on which random work stealing is based, namely the work-first principle and random victim selection, are neither aware of task characteristics (such as the task's arithmetic intensity) nor of the cores' performance and energy profiles. 
As a result, random work stealing could detrimentally impact energy consumption and performance of parallel applications on asymmetric platforms.
 
A few recent proposals have targeted energy efficient task scheduling by introducing extensions on top of work stealing~\cite{HarisRibic-WorkStealing/PercoreDVFS/Tempo, Christopher-ISCA2016-Asymmetric/PercoreDVFS/Parallelism,EmilioCastillo-IPDPS2016-CATA/Criticality/EDP,HarisRibic-AEQUITAS}. 
The main idea is to throttle DVFS depending on factors like task parallelism, stealing relations and/or task criticality.
The reliance on DVFS limits their applicability due to several reasons.
Firstly, DVFS switching overheads limit the potential of schemes that leverage per-task throttling.
Studies have shown that DVFS transition delay is around 100 microseconds
~\cite{DVFS_Overhead, EmilioCastillo-IPDPS2016-CATA/Criticality/EDP,Christopher-ISCA2016-Asymmetric/PercoreDVFS/Parallelism,AMD_ZEN2_Energy}. 
Exposing fine-grained task parallelism,
with task execution times being in the order of microseconds, 
enables scalability and load balancing in multicore and manycore architectures~\cite{Qthreads,Carbon}. 
However, applying DVFS on a per-task basis 
for energy savings would
incur unacceptably large overheads for fine-grained tasking.

Secondly, per-core DVFS control entails significant hardware cost and leads to low conversion efficiency due to the overheads introduced by the on-chip voltage regulators~\cite{PerCorevsClustered,TejaswiniKolpe-DATE2011-PowerManagement/ClusterDVFS}.
Many systems turn to cluster-level DVFS where voltage settings can only be controlled for a subset of cores (referred to as core-cluster). 
This applies both to embedded platforms such as the ODROID XU-3~\cite{xu3} and NVIDIA Jetson products~\cite{nvidia}, and to specific server processors from Intel and AMD~\cite{intel-skylake,amd-zen}. 
With cluster-level DVFS, however, multiple tasks attempting frequency changes within the same cluster will result in destructive interference, since the decision taken to reduce energy consumption of a task mapped to a specific core can affect concurrently running tasks on the same cluster. 
A similar problem occurs in processors with multi-threaded cores, where one thread's DVFS actions impact the performance of other threads running on the same core. 

A third concern with application-level DVFS techniques is that DVFS is most often not under the control of the application. In a multiuser OS, many processes share a subset of cores, and DVFS control is commonly restricted to the kernel~\cite{eas-linux}, the system administrator~\cite{linux-userspace-dvfs}, or power management frameworks such as GEOPM~\cite{eastep-hpc17.geopm}. 
Consequently, an energy efficient runtime designed to be reactive to both given static frequency settings and externally controlled DVFS, instead of relying on actively changing DVFS settings, has the potential to be a more general solution to the problem of energy-aware scheduling.

In this paper, we address the problem of reducing the total energy consumption of task-based applications running on symmetric and asymmetric multicore platforms for given core frequencies. 
We focus on 
designing an intra-application task scheduler based on
the most popular choices for task-based runtimes and multicore organizations, namely work-stealing~\cite{blumofe-jacm99} and core-clusters with common frequency~\cite{intel-skylake,amd-zen}.
Our proposal \textbf{ERASE} (EneRgy Aware SchedulEr) leverages the following novel insights:

\begin{itemize}[topsep=2pt,leftmargin=5.5mm]
    \item[(1)]  \textit{Task moldability}~\cite{MoldableResourceAllocation,  ChingChiLin-ICPADS2018-Energy/virtualmachine/moldable, Wuwei-IPDPS2015-CholeskyDAG/Distributed,Miquel-ElasticPlaces}, i.e.~executing a single task on multiple threads/cores via work-sharing can 
    help to reduce the energy consumption by decreasing resource oversubscription or making use of otherwise idle resources. 
    Moldability helps to exploit the internal parallelism within a single task by supporting 1:M mapping (i.e.~a single task to multiple threads/cores) in addition to the traditional 1:1 mapping (i.e.~a single task to a single thread/core). To enable moldable execution, the runtime partitions the task's workload and dynamically maps them to M ($\geq$1) resources. 
    \item[(2)] \textit{Task-type aware execution}   exploits information about different task characteristics (such as arithmetic intensity) to reduce energy consumption. We explore a combination of task moldability and task-type awareness to figure out the optimal resource assignment for minimizing the total energy consumption. Specifically, we identify the optimal execution place (core type, number of cores) for each task, since mismatch between task properties and assigned resources on asymmetric architectures can negatively impact the total energy consumption and performance.
\end{itemize}

The design goals of ERASE are (i) simplicity, to enable low overhead implementation, and (ii) adaptability, to enable its effective application across 
given static frequency settings and externally controlled DVFS on multicore platforms. 
In a nutshell, ERASE reduces energy consumption of an application by attempting to execute each task in the application with the lowest possible energy consumption. 
ERASE comprises four essential components: (i) \textit{online performance modeling}, (ii) \textit{power profiling}, (iii) \textit{core activity tracing} and (iv) \textit{a task mapping algorithm}. 
Online performance modeling monitors task execution and continuously updates a performance model that provides performance predictions for incoming tasks.
Power profiling provides the runtime with estimates of CPU power consumption with respect to different resource configurations (i.e.~number/type of cores) for given core frequencies.
Core activity tracing continuously tracks the activities (i.e.~work stealing attempts) and status (i.e.~active or sleep) of each core and infers the instantaneous task parallelism, which gives the task mapping algorithm a hint for attributing power consumption to concurrently running tasks accurately.
Finally, the task mapping algorithm integrates the aforementioned information and guides scheduling decision for each task based on the energy estimates of running the task on different resource configurations.

The
results obtained on two different platforms (i.e. NVIDIA Jetson TX2 and Tetralith) show 
that our proposal is a general approach for 
reducing the total energy consumption of executing task-based applications
and the proposed models in ERASE achieve reasonable accuracy.
In summary, the main contributions of this paper are as follows:
\begin{itemize}[leftmargin=*]
    \item We propose ERASE: an energy efficient task scheduler that combines power profiling, performance modeling and core activity tracing for energy efficient mapping (i.e.~choosing the cluster) and resource management (i.e.~selecting the number of cores per task).
    The proposal exploits the insights of task moldability, task-type awareness and instantaneous task parallelism detection for guiding scheduling decisions to reduce the total energy consumption of parallel applications. 
    \item We describe how to integrate ERASE on top of work stealing runtimes, using the XiTAO~\cite{xitao} runtime for the prototype implementation. 
    \item We compare ERASE to state-of-the-art scheduling techniques on top of the runtime and the evaluation shows that ERASE achieves up to 31\% energy savings and outperforms the state-of-the-art by 44\% on average.
\end{itemize}

The remainder of this paper is organized as follows. 
Section~\ref{background} presents the background and motivation for the proposed approach.
Section~\ref{approach} formulates the problem of energy efficient task scheduling 
and then discusses the design of ERASE in detail.
Section~\ref{implementation} describes how ERASE is integrated on top of a work stealing runtime.
Section~\ref{setup} introduces the experimental setups, which are used for evaluation in Section~\ref{Perf_Eva}.
Section~\ref{RelatedWork} presents related work, Section~\ref{conclusion} concludes the work. 
\section{Background and Motivation}  \label{background}
We firstly provide the necessary background in Section~\ref{definitions}. 
In Section~\ref{Type_Moldability}, we study the impact of task moldability and task-type aware scheduling on energy consumption to motivate the approach adopted by ERASE.
Next, we show that schedulers that leverage per-core DVFS to reduce energy consumption have limited effectiveness on architectures with cluster-level DVFS, in Section~\ref{percore_effectiveness}.

\subsection{Background} \label{definitions}
\begin{figure}[t]
\centering
\vspace{-3mm}
\includegraphics[width=0.75\columnwidth]{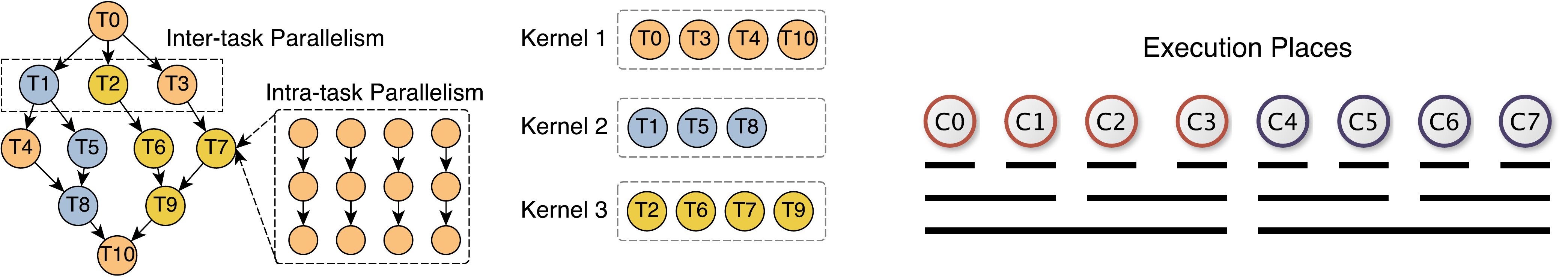}
\vspace{-3mm}
\caption{The illustration of a task DAG, kernels and possible execution places.} 
\vspace{-5mm}
\label{fig_dag_example}
\end{figure}

In this paper, we focus on parallel applications that are expressed as task directed acyclic graphs (DAGs), where the nodes denote tasks and the edges denote the dependencies between tasks.
Figure~\ref{fig_dag_example} shows an example task DAG.
A DAG implicitly expresses the inter-task parallelism (i.e.~the independent tasks that can be executed in parallel). 
We allow each task to furthermore contain internal parallelism, in such case moldability can be used to execute a single task over multiple resources. 
Given our focus on HPC applications, tasks are commonly used to represent numerical kernels. 
Each kernel consists of one or multiple tasks, and the routines executed by different tasks of the same kernel are identical.
In Figure~\ref{fig_dag_example}, the task DAG contains three different kernels.
Given a platform, an \textit{execution place} refers to execution resources that are assigned to each task.
They commonly share tightly coupled resources, such as cache levels, memory channels, NUMA nodes, etc.~in order to reduce interference.
Figure~\ref{fig_dag_example} shows an example of supported execution places on an asymmetric platform with two clusters and four cores in each cluster.

\subsection{Impact of Task-type Awareness and Task Moldability on Asymmetric Platforms} \label{Type_Moldability}

\begin{figure*}[t]
\vspace{-4mm}
\centering
\subfloat[\scriptsize{MatMul - Energy}]{\includegraphics[width=0.25\textwidth]{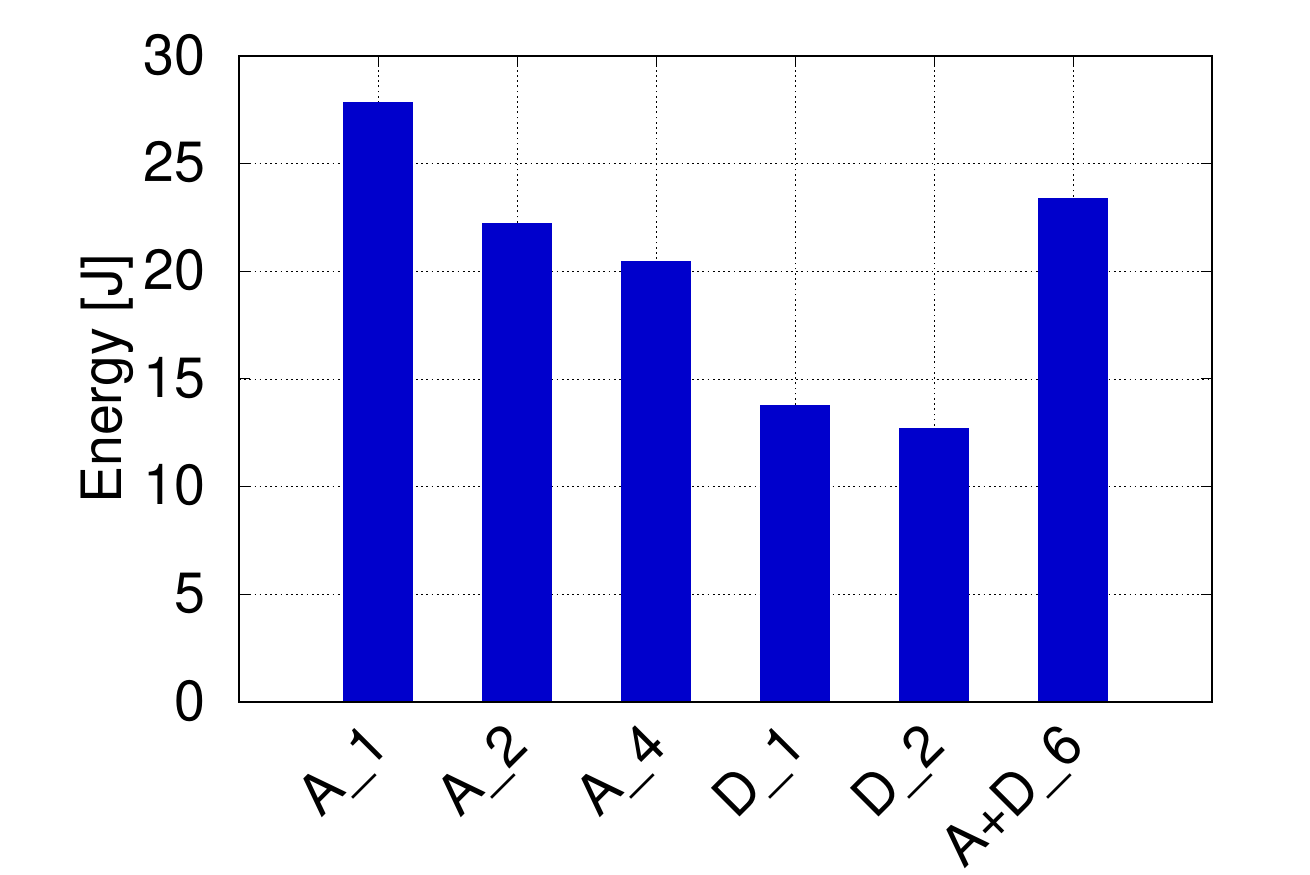}
\label{fig_Matmul_E}}
\subfloat[\scriptsize{MatMul - Execution Time}]{\includegraphics[width=0.25\textwidth]{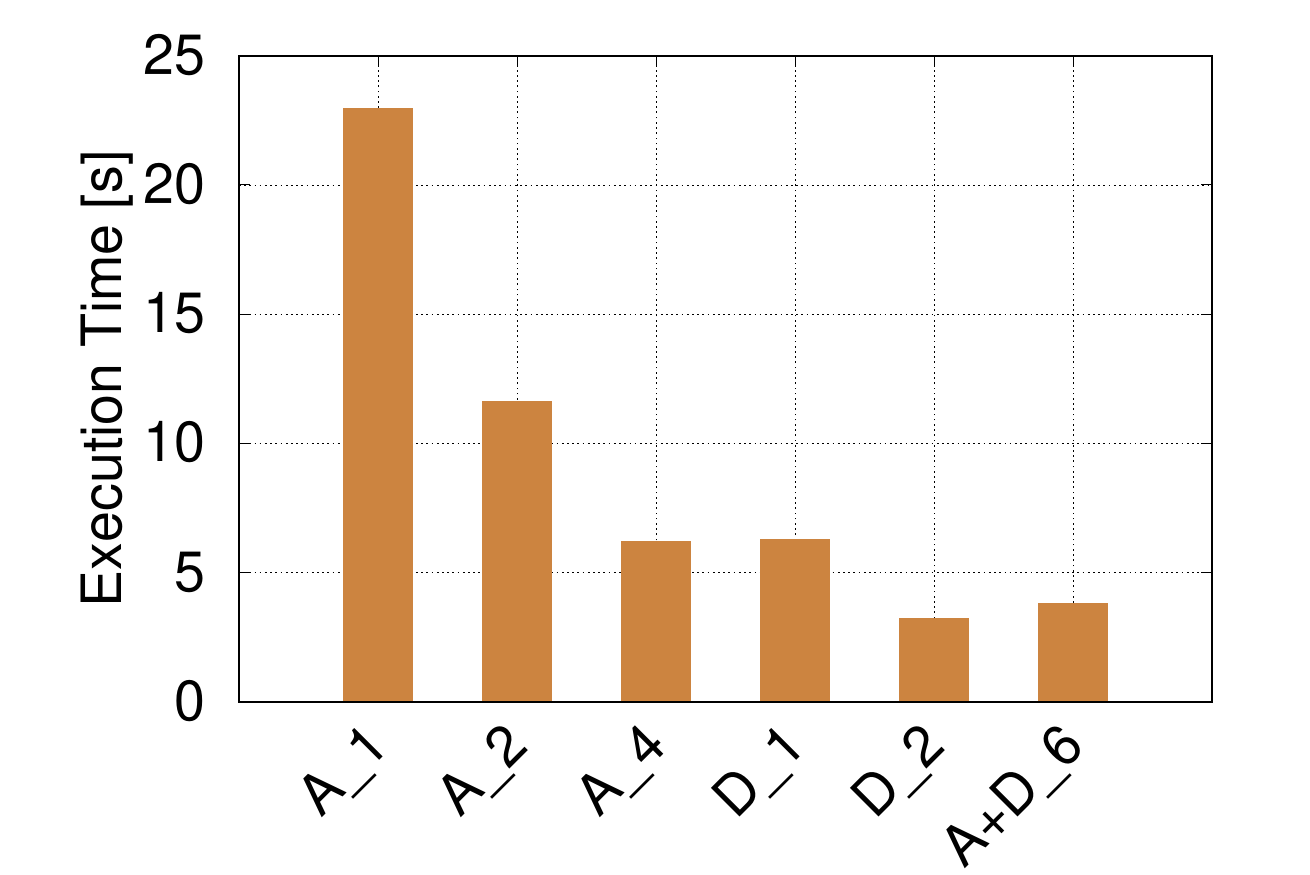}
\label{fig_Matmul_P}}
\subfloat[\scriptsize{Copy - Energy}]{\includegraphics[width=0.25\textwidth]{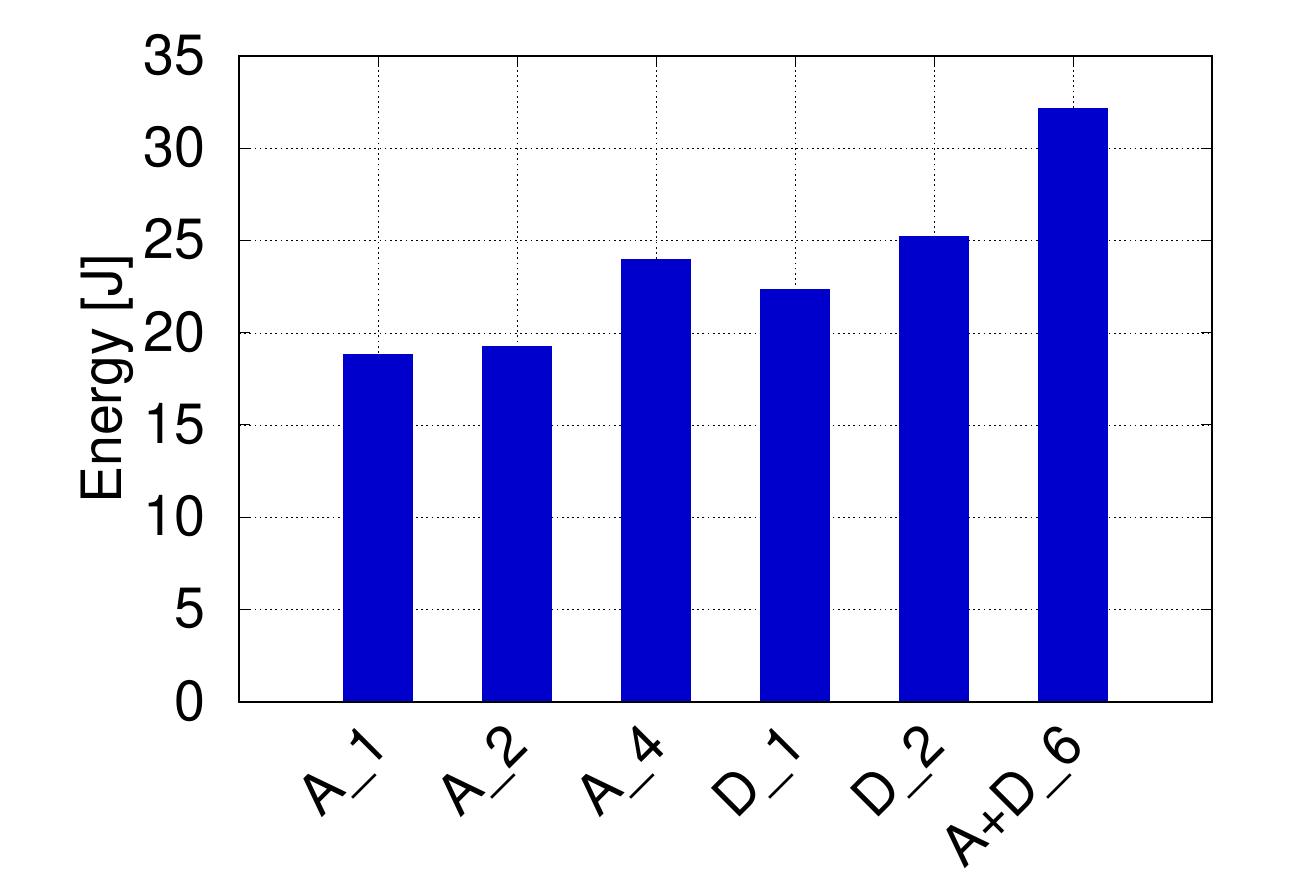}
\label{fig_Copy_E}}
\subfloat[\scriptsize{Copy - Execution Time}]{\includegraphics[width=0.25\textwidth]{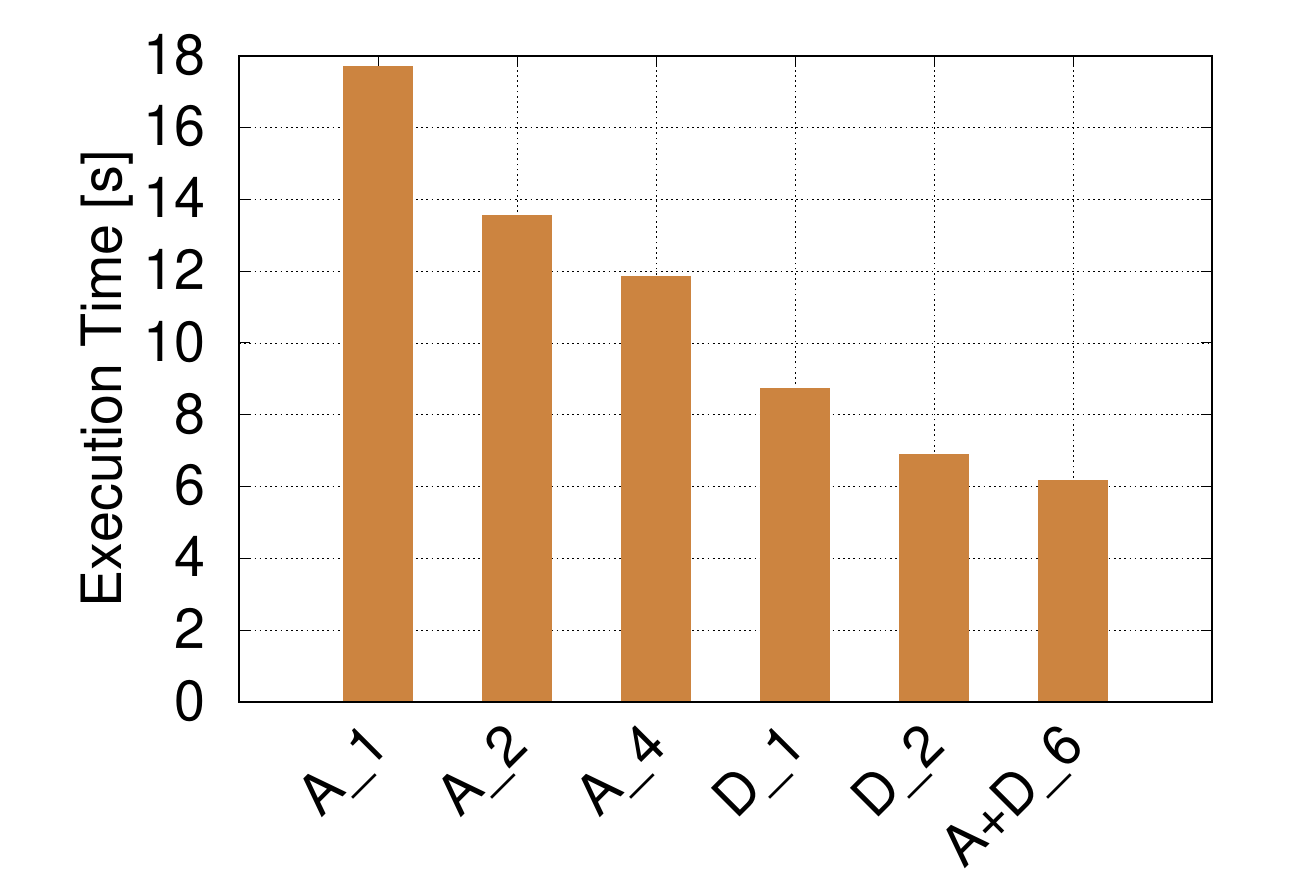}
\label{fig_Copy_P}}
\vspace{-3mm}
\caption{Energy consumption and execution time of MatMul and Copy when running with different configurations on Jetson TX2. }
\vspace{-6mm}
\label{fig:MO}
\end{figure*}

To showcase the impact of task-type aware execution and task moldability on energy consumption and performance, we select two representative benchmarks: \textit{Matrix Multiplication} (abbr.~MatMul) - compute-bound and \textit{Memory Copy} (abbr.~Copy) - memory-bound and run on the NVIDIA Jetson TX2 board. 
It consists of two CPU clusters: Denver and A57. 
The former comprises a high performance dual-core NVIDIA Denver CPU, while the latter comprises a comparatively lower performance quad-core ARM CPU. 
Additional details about the benchmarks and the platform are provided in Section~\ref{setup}.
Figure~\ref{fig:MO} presents energy consumption and execution time when running the benchmarks
with different configurations. 
The labels in the x-axis follow the format \texttt{x\_y}, where \texttt{x} denotes the cluster name and \texttt{y} denotes the number of cores in the cluster for executing each task in the benchmark.
When running with a specific configuration, the rest of the cores are disabled in order to obtain an accurate measure of the energy consumption.
For instance, 
A\_1 represents the case when tasks execute only on a single A57 core and all the other cores are disabled, while
A\_4 represents the case when a benchmark executes only on four A57 cores (with both the Denver cores disabled) and each task executes on all four cores to exploit task moldability.
The label A+D\_6 represents the configuration where each task executes on all six cores.

Figure~\ref{fig_Matmul_E} shows that executing a matrix multiplication task on high performance Denver core(s) consumes less energy compared to that on lower performance A57 core(s). 
This can be attributed to the 3.5$\times$ increase in Instruction Per Cycle (IPC) in a Denver core compared to an A57 core. 
It can be seen that energy consumption is lowest when running on two Denver cores, i.e.~D\_2, which achieves 8\% energy savings in comparison to using a single Denver core (D\_1). 
This is because running with D\_2 achieves linear speedup of 2, while the power consumption of D\_2 only increases by 87\%  comparing with D\_1. 
Additionally, 
it is clear in Figure~\ref{fig_Matmul_P} that the most energy efficient configuration (D\_2) is also the fastest, which illustrates that task moldability can help to reduce energy consumption and improve performance simultaneously. 
The results also show that D\_2 achieves 46\% energy savings and improves performance by 15\% in comparison to A+D\_6, which indicates the importance of selecting the appropriate core type and number of resources over simply attempting to use all available resources.  

Figure~\ref{fig_Copy_E} shows that, for memory copy, running on a single A57 core consumes less energy than running on a Denver core. 
This can be attributed to the fact that running on a fast out-of-order core, i.e.~Denver, does not provide sufficient performance benefit to compensate the additional power costs.
However, unlike matrix multiplication, the energy consumption of A\_2 and A\_4 is higher than A\_1. 
This is because the power cost of using additional A57 cores cannot be offset by performance benefits. 
Hence, the configuration that consumes the least energy is the slowest.
Additionally, the results in Figure~\ref{fig_Copy_P} show that attempting to use all available resources achieves the lowest execution time but leads to the highest energy consumption.


In summary, these results indicate that the 
best configuration for minimizing energy consumption
is task-type dependent, and that exploiting task moldability can help
to further reduce energy consumption
(e.g.~for matrix multiplication in this case), which motivate us to consider task-type awareness and task moldability together when designing energy efficient work stealing runtimes.

\subsection{Impact of Per-core DVFS Scheduling on Cluster-level DVFS} \label{percore_effectiveness}


In this section, we study (1) the impact of task granularity on energy when using DVFS; and (2) how much energy efficiency is lost when a runtime assuming per-core DVFS scheduler is deployed on a platform that features only cluster-level DVFS.
The results motivate us to consider the negative impact of DVFS on fine-grained task DAGs and to devise scheduling strategies that take the platform DVFS into account.
To showcase the effects, we evaluate the Criticality Aware Task Acceleration scheduler (\textbf{CATA})~\cite{EmilioCastillo-IPDPS2016-CATA/Criticality/EDP} on Jetson TX2.
CATA dynamically tunes the frequency of each core based on incoming task criticality and the available power budget at the moment to improve performance and energy efficiency, i.e.~Energy-Delay Product (EDP). 
We reproduce CATA on the Jetson TX2 with two frequency levels (2.04GHz, 0.35GHz) and set a power cap whereby only one of the clusters on TX2 can operate in maximum frequency.

Table~\ref{tab:cata_erase} compares the EDP and execution time of CATA with RWS, a random work stealing scheduler that we use as the baseline,
and the scheduler proposed in this work (ERASE). 
The latter is given to provide an idea of the potential energy reductions that can be achieved when a scheduler is aware of the platform DVFS characteristics. All three schedulers are evaluated with three synthetic benchmarks featuring a degree configurable parallelism ($dop$) that ranges from 2 to 8. Additional details about the schedulers, the benchmarks and the platform are provided in Section~\ref{setup}.

\begin{table*}[t]
\scriptsize
\vspace{-3mm}
\caption{EDP and execution time comparisons between CATA, RWS and ERASE on Jetson TX2}
\vspace{-3mm}
\begin{center}
\begin{tabular}{p{1.2cm}p{.6cm}p{.59cm}p{.59cm}p{.59cm}p{.59cm}p{.59cm}p{.59cm}p{.59cm}p{.59cm}p{.59cm}p{.59cm}p{.59cm}p{.6cm}} \toprule
\textbf{Application}& & \multicolumn{4}{c}{\textbf{MatMul}} & \multicolumn{4}{c}{\textbf{Copy}} & \multicolumn{4}{c}{\textbf{Stencil}} \\ \midrule
\textbf{Parallelism}& & dop=2 & dop=4 & dop=6 & dop=8 & dop=2 & dop=4 & dop=6 & dop=8 & dop=2 & dop=4 & dop=6 & dop=8 \\ \midrule
\multirow{3}{*}{\textbf{EDP}} & CATA & 27138.5 &	16832.6 &	11270.4 &	7693.9 &	5362.2 	& 2486.8 &	2158.7 	& 2089.2 &	2275.0 &	1990.7 &	1062.0 &	736.5  \\ \cmidrule(r){2-14} 
& RWS & 15723.0 &	6898.1 &	4267.4 	& 3315.4 &	7266.2 & 4230.4 &	3932.3 & 	2937.9 &  	5648.8 	& 2475.2 &	1828.5 	& 1377.1  \\ \cmidrule(r){2-14} 
& ERASE & 1979.9 &	1463.4 & 	1433.4 & 	1338.6 & 	2140.5 & 	1886.2 & 2096.6 & 2020.7 & 310.0 & 260.8 & 280.5 & 262.2  \\ \midrule
\multirow{3}{1.2cm}{\textbf{Execution Time[s]}} & CATA & 178.59 &	114.50 &	85.57	 & 66.15 & 69.83 &	36.83 &	31.10	& 29.71 & 34.19 &	28.52 &	20.43 &	17.56\\ \cmidrule(r){2-14}
& RWS & 102.23 & 56.84 & 37.97 & 28.93 & 108.16 & 61.16 & 53.06 & 40.94 & 57.96 &	30.74 &	24.54 &	19.57 \\ \cmidrule(r){2-14} 
& ERASE & 25.85	& 19.76 &	19.41 & 19.31 & 36.74	& 31.83	& 32.37	& 31.48 & 9.87 &	8.41 &	9.00 &	8.46\\ \bottomrule
\end{tabular}
\label{tab:cata_erase}
\end{center}
\vspace{-4mm}
\end{table*}

The results show that when CATA is applied for fine-grained task DAGs on a platform that features only cluster-level DVFS, it does not consistently achieve the best performance or lowest EDP, even when compared to a simple RWS strategy. 
The comparison with 
ERASE furthermore shows that large improvements are possible.
The limited benefit from CATA can be attributed to the following reasons:
(1) DVFS control, triggered from user space, leads to reconfiguration serialization and becomes a performance bottleneck; 
(2) When per-core DVFS is employed on platforms that only support cluster-level DVFS, it leads to interference in DVFS settings among cores in the same cluster; and
(3) Fine-grained tasks further exacerbate the DVFS overheads due to the reconfiguration serialization issue.
Overall, these results indicate that DVFS-aware task schedulers need to be carefully designed to avoid DVFS reconfiguration overheads and the interference that results from multiple cores requesting different frequencies.
\section{Energy Efficient Task Scheduler} \label{approach}
We first discuss the problem of minimizing the total energy consumption when executing a task DAG.
Next, we propose our energy efficient task scheduler (ERASE) and provide an overview of the essential modules that constitute ERASE, followed by the description of all modules in detail and an explanation of how they interact with each other to enable energy efficient task scheduling.
Finally, we present the complexity and overheads analysis for ERASE.

\subsection{Problem Formulation} \label{problem_statement}

We assume a platform that includes $m$ core-clusters $\{\beta_{1}, ..., \beta_{m}\}$, where each cluster comprises $p$ cores of the same type.
With task moldable execution, each task can be executed with variable number of cores of the same type $\in$ \{$\gamma_{1}, \gamma_{2}, ..., \gamma_{p}$\}.
Given a task DAG that includes $\tau$ tasks $\{T_{0}, ..., T_{\tau-1}\}$, 
the problem is to minimize the total energy consumed for executing the entire DAG on this platform.

\subsection{ERASE Overview}
ERASE assumes that the total energy consumption (abbr.~TE) of running the DAG on such a platform consists of two components: 
the energy consumed for running each task and the energy consumed in the idle periods due to the dependencies between tasks.
Thus, the problem of minimizing the total energy consumption of executing a task DAG can be partitioned into two parts: (1) minimizing the energy consumed by each task; (2) minimizing the energy consumed in idle periods, as shown in equation~\ref{E_minimization}.
\begin{equation} \label{E_minimization} \footnotesize
    min\ TE = \sum\limits_{i=0}^{\tau-1} min\ E_{T_{i}} + \sum\limits_{j=1}^{m} \sum\limits_{k=1}^{p} min\ E_{idle}
\end{equation}

\textbf{Minimizing the energy consumed by each task.}
We show in Section~\ref{Type_Moldability} that with different execution places,
the energy consumption varies greatly.
Thus, to minimize the energy consumption when running a task, it is necessary to figure out the best execution place for the task.
There are two knobs at runtime for the configuration selection: (i) the choice of cluster and (ii) the number of cores.
Therefore, for a specific task $T_{i}$, energy minimization can be expressed as:
\begin{equation}\label{energypertask}  \footnotesize
    min\ E_{T_{i}} = min(\underbrace{E_{T_{i}}(\beta_{1},\gamma_{1}), E_{T_{i}}(\beta_{1},\gamma_{2}), ..., E_{T_{i}}(\beta_{1},\gamma_{p})}_{\beta_{1}}, \underbrace{E_{T_{i}}(\beta_{2},\gamma_{1}), 
    ..., E_{T_{i}}(\beta_{2},\gamma_{p})}_{\beta_{2}}, \underbrace{E_{T_{i}}(\beta_{3},\gamma_{1}), ..., E_{T_{i}}(\beta_{m},\gamma_{p})}_{\beta_{3}\ to\ \beta_{m}}).
\end{equation}
According to equation~\ref{energypertask}, the runtime needs to evaluate all different execution places to obtain the configuration that consumes the least energy for each task.
For each possible configuration, energy consumed by a task is the product of execution time and power consumed during the task execution.
It is crucial that the runtime can predict the execution time and the power consumption before execution. 
Thus, designing a scheduling system that achieves energy minimization goal requires a performance model for execution time prediction and a power model for power prediction, respectively.



\textbf{Minimizing the energy consumed in idle periods.} 
In a work stealing runtime, cores that become idle (i.e.~because there are no tasks in their work queues) 
attempt to steal tasks from other queues for load balancing. 
However, when not enough ready tasks are available, the idle cores continuously attempt work stealing without success, which can lead to energy waste.
The problem of minimizing  energy consumption during idle periods requires the runtime to be able to detect the cores' instantaneous utilization and dynamically put idle cores to sleep.
Meanwhile, during the sleeping periods, it is possible that ready tasks are assigned to the idle cores or that there are tasks available for work stealing. 
If the idle cores cannot be woken up in time to deal with the tasks, it could cause performance degradation and energy increase.
The challenge is to determine the sleep duration such that energy consumption during the period is minimized with minimal performance impact. This is described in Section~\ref{implementation}.

\begin{figure*}[!t]
\centering
\vspace{-2mm}
\includegraphics[width=0.65\textwidth]{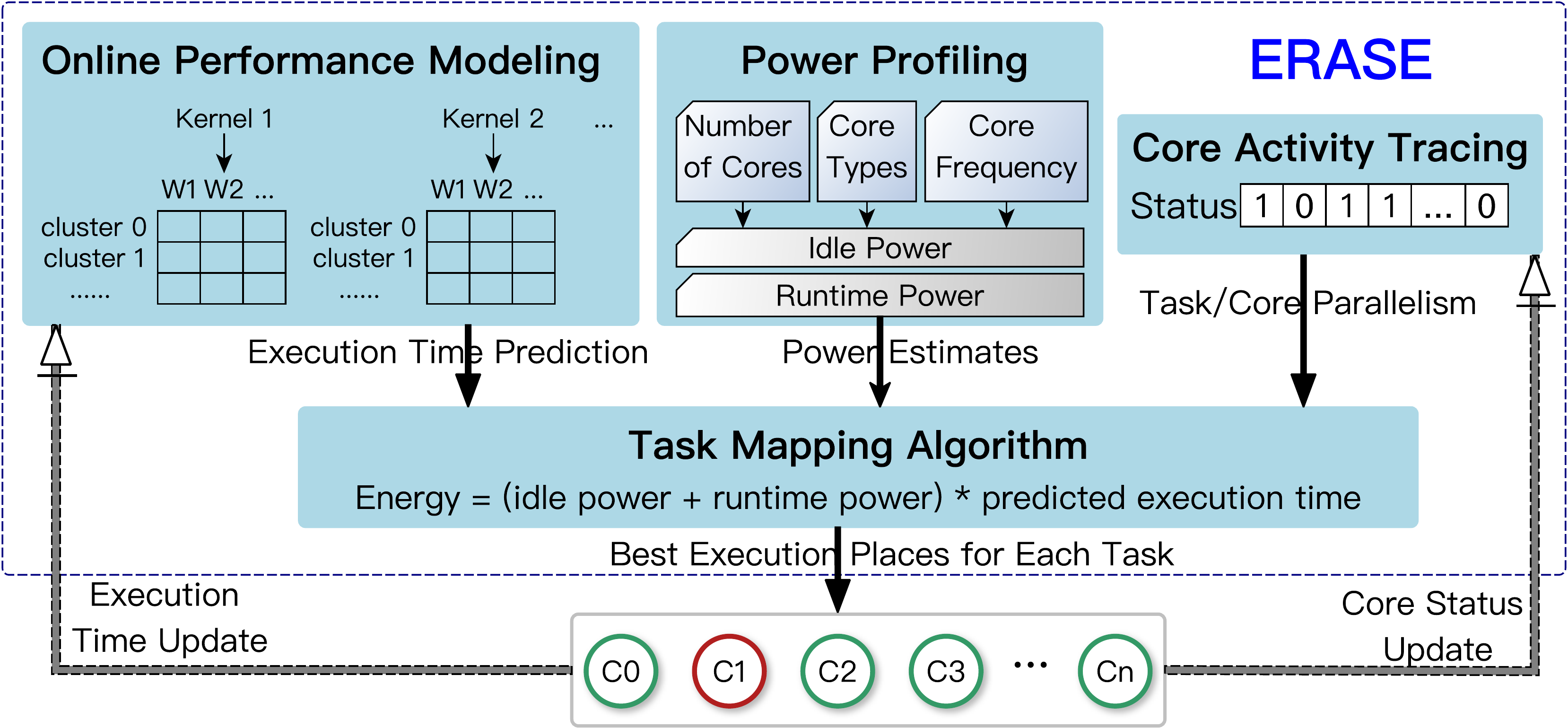}
\vspace{-3mm}
\caption{An overview of ERASE comprising four modules. C$_x$ denotes the core id, W$_y$ denotes the possible resource widths. In status, ``1'' denotes that C$_x$ is in active state while ``0'' denotes the core is in sleep state.}
\vspace{-5mm}
\label{fig:runtimeoverview}
\end{figure*}

To address the aforementioned challenges, we propose an energy efficient task scheduler called ERASE.
Figure~\ref{fig:runtimeoverview} highlights the four essential modules in ERASE. 
To enable per-task energy estimation, the scheduler predicts the performance and power consumption when a task is mapped onto an execution place.
Performance prediction is provided by the online performance modeling module,
as discussed in Section \ref{performance}. 
The information about power consumption estimation is obtained through power profiling as discussed in Section~\ref{charac}. 
The core activity tracing module, discussed in Section \ref{idlenesstracer}, dynamically tracks how many cores are effectively utilized and how many are idle at any given moment, which provides the instantaneous task parallelism.
The task mapping algorithm, presented in Section \ref{sec:par_tracer}, leverages information provided by the aforementioned modules to estimate the energy consumption on different execution places and
map each task on the execution place that consumes the least energy.
These modules cooperate with each other and enable the runtime to determine the best execution place for each task for the total energy savings.
In ERASE, each possible execution place is expressed as a tuple (leader core, resource width).
The \textit{leader core} denotes the starting core identifier (id) of the allocated resources.
The \textit{resource width} denotes the number of cores assigned to a task to enable moldable execution.

\subsection{Online Performance Modeling} \label{performance}

The online performance modeling module enables ERASE to predict the execution time of a task when mapped on an execution place. 
In ERASE we adopt a history-based model since it is a simple yet effective method for estimating performance and detecting performance heterogeneity and system interference for fine-grained tasking applications~\cite{Interference_SRMPDS2020, hetero_predictive_runtime_Hipeac2009, estimate_time_Supercomputing2013, MDR_ICS11}.
The online performance module is implemented as a look-up table for each kernel.
The size of each look-up table equals the product of the number of clusters and the number of available resource widths on the platform. 
For the example shown in Figure~\ref{fig_dag_example}, ERASE utilizes a 2$\times$3 performance look-up table for each of the kernels (3 in total) to predict the performance on the Jetson TX2 platform.
This module only requires information about the number of clusters and their organization into core-clusters with shared caches, which can readily be obtained using tools like $hwloc$.
\begin{figure}[t]
\vspace{-2em}
\begin{algorithm}[H] \footnotesize
\caption{\small \textbf{Online Performance Modeling Algorithm}}\label{perf_prediction}
\begin{algorithmic}[1]
\State Initialize all entries to zeros
\For{i in cluster indexes} 
\For{j in resource widths of cluster i}
\State temp\_time = task->access\_performance\_table(i, j) 
\If{temp\_time == 0}  
    \State task->leader = starting core id of cluster i + (rand() \% (number of cores / j)) $\times$ j
    \rlap{\smash{$\left.\begin{array}{@{}c@{}c@{}c@{}}\\{}\\{}\\{}\\{}\\{}\\{}\\{}\\{}\\{}\\{}\end{array}\color{black}\right\}%
          \color{black}\begin{tabular}{l}Phase 1: Training\end{tabular}$}}
    \State task->width = j
    \State return
\EndIf 
\EndFor
\EndFor 
\State Time\_prediction = task->access\_performance\_table(i, j) \Comment{Phase 2: Prediction given cluster i, resource width j}
\If{lc is leader core \&\& lc $\in$ cluster i}
    \State task->update\_performance\_table(i, j, execution\_time) \rlap{\smash{$\left.\begin{array}{@{}c@{}c@{}}\\{}\\{}\end{array}\color{black}\right\}%
          \color{black}\begin{tabular}{l}Phase 3: Update corresponding entry after execution\end{tabular}$}}
\EndIf
\end{algorithmic}
\end{algorithm}
\vspace{-3em}
\end{figure}

Algorithm~\ref{perf_prediction} shows the three phases in the module: training, prediction and update.
In training phase, all the entries of a look-up table are initialized to zeros to guarantee that all execution places are visited at least once (lines 1-11). 
The execution time prediction for a task 
is obtained by accessing the corresponding value of the specific execution place in the table as shown in line 12.
After a task completes the execution, the corresponding entry in the table is updated by the leader core of a task to reduce cache migrations (lines 13-15).
Each table is organized to fit into cache lines where each cluster only accesses one cache line indexed with the cluster id, which helps avoid false sharing.
To avoid large fluctuations in the values in the table, which could result from short isolated events, entries are updated by computing a weighted average of the previous and the new value.

The online performance module is capable of reacting to the externally controlled DVFS changes.
The performance look-up tables record the CPU cycles and the execution time consumed during task execution, then speculate the frequency from cycles and execution time.
By comparing to the previous frequency record,
the online performance module is able to detect the frequency change instantly. 
Once the frequency change is detected, the module will reset the look-up table entries to zeros and retrain the model.
We evaluate the overhead of retraining the performance modeling tables and the result shows that with fine-grained tasking retraining introduces little overhead (see Section~\ref{dyna_reaction} for details).
Nevertheless, for coarse-grained tasking, the scheme would have limited effectiveness due to larger training overhead, since the execution time spent on training phase occupies higher percentage of the total execution time.


In comparison to other complex models based on an offline linear regression or a trained neural network, the history-based model can be implemented without any offline training and with low overhead, but one has to take into account that its performance depends on a fine-grained partitioning of the application into tasks. 
The scheme has the following advantages.
First, the table size is relatively small. For instance, the training phase on the TX2 platform only requires five tasks' execution times to fill in all the entries of a table.
Second, in a parallel application, each kernel generally contains a large number of tasks and each task of the kernel contributes to the update of the table. 
Thus, the number of tasks for training is negligible in comparison to the total amount of tasks in an application.
We show that the module provides reasonably accurate predictions for different applications and across multiple platforms (refer to Section \ref{acc_eval} for details).


\subsection{Power Profiling} \label{charac}

The goal of this module is to estimate power consumption when scheduling a task on an execution place. 
A simple approach to estimate power consumption is to run a single task (one from each kernel) on an execution place and profile the power consumption values obtained from the power sensors as reference. 
Since the number of tasks is usually much higher than the number of possible execution places (as is the case with long-running applications and/or with fine-grained tasking), the online profiling approach would have a relatively low overhead. 
However, there are limitations with power measurement support on existing systems that preclude the use of online profiling approach described above. 

Firstly, the power measurement supports, that are available on the platforms we present in Section~\ref{platform}, provide composite power consumption values for the entire chip/socket instead of values at the per-core level. 
This makes it difficult to isolate the power consumption for a single task that executes on a subset of cores, especially when there are concurrently running tasks. 
Secondly, our analysis of the power measurement overheads on Jetson TX2 shows that it takes tens of milliseconds before the power sensor begins to accurately capture the change in power consumption since the values reported are averaged over a time window. 
With the execution time of fine-grained tasks being in the order of microseconds, which is much lower in comparison, the online power profiling approach becomes ineffective. 
Due to these limitations, we adopt a power characterization approach based on profiling representative microbenchmarks to help estimate the power consumption.
This power profiling step just needs to be done once for a specific platform, for example at install-time or boot-time, and has no impact on execution time.

%

\textbf{Task Type Detection.}
We consider the tasks of an application to be broadly grouped into one of the three representative types in ERASE: compute-bound, memory-bound and cache-intensive. 
Most modern architectures provide Performance Monitoring Counters (PMC) to monitor the activity of cores, such as cpu-cycles, cache-misses, cache-references and number of retired instructions per cycle.
The availability of PMCs differs between different platforms.
In this work, we utilize the arithmetic intensity (\textit{AI}), i.e.~the ratio of floating point operations over memory traffic~\cite{RooflineModel}, to determine online the task type category for a specific kernel.
We record the number of cycles and the last level cache misses during the task execution by using the Linux system call \textit{perf\_event\_open()}~\cite{perf_event_open}.
The two performance counters required to implement this calculation are widely available on the majority of platforms.
AI is calculated as follows:
\begin{equation}\label{AI_equation}  \footnotesize
\vspace{-1mm}
    AI = \frac{number\ of\ cycles \times FLOPs/cycle}{number\ of\ cache\_misses \times 64\ bytes}.
\end{equation}
We profile a set of configurable synthetic microbenchmarks (refer to Section~\ref{syntheticdag} for more details) offline and record the corresponding PMCs and the average power consumption during the execution when running them with different number/type of cores and frequencies.
After calculating the AI values for all microbenchmarks, 
we employ a $k$-NN (k-Nearest Neighbor) algorithm to cluster them into three groups by computing the euclidean distances of the AIs and finally choose an AI threshold belonging to the frontier region between the clusters. 
The averaged power values from each group are used as power estimates once a task is classified as belonging to a certain task type. 
Further analysis about the accuracy of this approach is presented in Section~\ref{acc_eval}.

The inputs of the power profiling step are the number of cores, core types and available frequency levels on the platform, as shown in Figure~\ref{fig:runtimeoverview}.
During power profiling, two different components are measured for different frequencies: \textit{idle power} and \textit{runtime power}.
Idle power of a platform denotes the power consumed even when no useful work is being performed. 
It can be simply measured when all the cores are online but do not perform any useful activity.
Runtime power is defined as the additional power consumed when performing useful work. 
Note that runtime power only represents the additional power consumed to do useful work and does not include idle power.
Thus, the power consumption for a task is estimated by summing up the runtime power and the idle power consumed by this task.
The runtime power estimate for a task executing on an execution place can be simply obtained from the power profiling results. 
However, idle power obtained with power profiling is the total idle power of the entire chip/cluster.
It cannot be attributed to the task directly because this is shared between concurrently running tasks. 
In other words, each concurrently running task shares a portion of idle power at any given moment, and attributing the entire idle power from power profiling to a single task may lead to inaccurate energy estimation. 
This observation motivates the need for core activity tracing module, described in Section~\ref{idlenesstracer}, which addresses the problem by continuously tracking the number of active cores and giving the task mapping algorithm a hint for attributing the idle power consumption to concurrently running tasks accurately.

\subsection{Core Activity Tracing} \label{idlenesstracer}
In this section, we introduce the core activity tracing module, which solves the problem of idle power sharing described in Section~\ref{charac}. 
The module maintains an integer array \textit{status[]} that stores the instantaneous core states, as shown in Figure~\ref{fig:runtimeoverview}.
The module continuously tracks the number of active cores doing useful work and dynamically updates the \textit{status} entries if cores become idle. 
In order to correctly attribute the idle power among concurrently executing tasks, it is essential to detect the instantaneous task parallelism 
during execution.
Since each task can execute on a variable number of cores, the instantaneous task parallelism detection can be transformed into detecting the \textit{resource occupation} (i.e.~the percentage of the task resource width over the number of all active cores in the same cluster) of each parallel task.
In other words, the idle power consumed by a task is computed by splitting the total idle power obtained from profiling proportionally to each task's resource occupation among the running tasks in parallel. 




\subsection{Task Mapping Algorithm} \label{sec:par_tracer}

The task mapping module leverages the other modules described previously to determine the most energy efficient execution place for each task. 
The pseudo-code for the module is shown in Algorithm~\ref{taskmapping}.
The inputs to the algorithm are the ready tasks that are woken up by a core and the outputs are the predicted execution places that will consume the least energy consumption for these tasks.
In Algorithm~\ref{taskmapping}, we show a case that is tailored for a platform with two clusters ($A$ with $a$ cores and $B$ with $b$ cores). 
The algorithm can be easily extended to other platforms.

\begin{figure}[t]
\vspace{-2em}
\begin{algorithm}[H] \footnotesize
\caption{\small \textbf{Task Mapping Algorithm}}\label{taskmapping}
\begin{algorithmic}[1]
\State \textbf{Input:} \textit{ready tasks that are woken up by the core}; \textbf{Output:} \textit{best execution place for each task}
\For{each ready task}
\For{each execution place}  \Comment{Randomly select an execution place from each (cluster, width) configuration}
\State NumActCores\_A = status[0]+ status[1] + ...+ status[a-1] \Comment{Cluster $A$ includes $a$ cores}
\State NumActCores\_B = status[a]+ status[a+1] + ... + status[a+b-1] \Comment{Cluster $B$ includes $b$ cores}
\State idleP\_{temp\_A} = (NumActCores\_B $\textgreater$ 0)? idleP\_A : idleP\_chip 
\For{core i in current evaluated execution place}  \Comment{Check if all cores in the execution place are all active}
    \If{status[i] is 0}
        \State NumActCores\_A++ \Comment{Selected cores are currently sleeping but could later be used for execution}
    \EndIf
\EndFor
\State resource\_occupation = resource width / NumActCores\_A \Comment{Resource occupation}
\State idleP = idleP\_{temp\_A} $\times$ resource\_occupation \Comment{Idle power prediction}
\State runtP = task->power\_profile (task type, core type, frequency, resource width) \Comment{Runtime power prediction}
\State execution time = task->visit\_performance\_table(cluster A, resource width)   \Comment{Execution time prediction}
\State energy = ( idleP + runtP ) $\times$ execution time  \Comment{Energy prediction}
\If{energy \textless minimum}
    \State minimum = energy; update best placement with new execution place 
\EndIf
\EndFor
\EndFor
\end{algorithmic}
\end{algorithm}
\vspace{-3em}
\end{figure}


The algorithm iterates over all possible (cluster, resource width) configurations to find the one that consumes the least energy for each task (line 3). 
We randomly select one execution place from each possible (cluster, resource width) configuration for the energy prediction.
We assume, in this example for the purposes of illustration, that the current evaluated configuration is in cluster A. 
The first step is to obtain information about the number of active cores on each cluster (lines 4-5). 
This information is provided by the core activity tracing module discussed in Section~\ref{idlenesstracer}. 
The idle power value (\textit{idleP\_temp\_A}) attributed to tasks running on cluster A depends on the number of active cores on cluster B (line 6).
If there are active cores on cluster B, the parallel tasks running on cluster A share the idle power of cluster A alone. 
Otherwise, the whole cluster B is in sleep and the tasks running on cluster A share the idle power of the entire chip. 
The information about idle power consumption, i.e.~\textit{idleP\_chip} and \textit{idleP\_A}, is obtained from the power profiling module.
As this is an estimation before task execution, it is possible that the selected cores are in sleep state.
Therefore, if the ready task is going to execute on a configuration that contains cores that are currently in sleep state, their status will be updated after waking up and the number of active cores increases in this case, which is shown in lines 7-11.
Line 12 computes the resource occupation for the task relative to other tasks on cluster A, and line 13 estimates the idle power consumed by the task if it is scheduled in the execution place.
Next, line 14 estimates the runtime power for the task by accessing power profiling results based on the task type, core type, current detected cluster frequency and the resource width.
Execution time prediction is obtained from the online performance modeling module, as shown in line 15.
Finally, the algorithm estimates the total energy consumption if the ready task is scheduled to this execution place (line 16).
The algorithm iterates through all possible configurations to determine the one that consumes least energy for each ready task (lines 17-19).

Algorithm~\ref{taskmapping} illustrates the steps to determine the most energy efficient execution place for a single ready task. 
The algorithm is run every time ready tasks are released. 
As multiple cores may release tasks in parallel, the algorithm may execute concurrently.  
It should be noted that the online performance model does not explicitly model the effect of inter-task concurrency for predicting the execution time. 
However, the impact of concurrency is implicitly captured in the execution time history, which reflects on the performance table and influences future predictions. 
While the power profiling module explicitly models concurrency for accurate idle power prediction, it does not take this into account for runtime power prediction. 
When sufficient bandwidth is available, the runtime power consumed by a task is dominated by the core's and private cache's performance, while the impact of interference from other co-running tasks is negligible.
Our evaluation shows that the models are able to predict performance, power and energy consumption across a range of benchmarks and concurrency settings with reasonable accuracy (see section \ref{acc_eval} for details).

\subsection{ERASE Complexity and Overhead Analysis} \label{complexity}
In this section, we analyze the complexity and the execution overheads of ERASE.

\textbf{Time Complexity.}
For each task, Algorithm~\ref{taskmapping} includes two nested loops. 
Firstly, ERASE traverses all possible execution configurations to find the one that consumes the least energy (line 3).
Considering a platform includes $n$ cores in total, $m$ core-clusters such that each cluster
comprises $\frac{n}{m}$ cores, the number of available resource widths of each cluster is ${\log_{2} \frac{n}{m}}$. 
Therefore, the total number of possible configurations on the platform 
equals ${m \times \log_{2} \frac{n}{m}}$.
The \textit{for} loop in line 7 traverses the selected core(s) in each execution place, which could be $\mathrm{2^{0}, 2^{1}, ..., \frac{n}{m}}$.
The time complexity of the loop is $O(\frac{n}{m})$.
Therefore, the time complexity of Algorithm~\ref{taskmapping} for each task becomes $O(n \times \log_{2} \frac{n}{m})$.
The outermost loop in line 2 handles all the ready tasks in a DAG.
When taking number of tasks ($nr\_task$) into account, the time complexity becomes $O(nr\_task \times n \times \log_{2} \frac{n}{m})$.
Regarding the performance model, the number of tasks utilized for model training equals the product of the number of kernels in an application and the total number of execution places.
Hence, the time complexity of building up the performance model per kernel is $O(m \times \log_{2} \frac{n}{m})$.
The power model is built offline. Accessing power model entries for a given execution place takes constant time.
Consequently, the time complexity of accessing the power model is $O(1)$.

\textbf{Space Complexity.}
In ERASE, the space taken by the performance model equals the product of the number of kernels in an application and the performance look-up table size. 
Since the number of kernels remains unchanged regarding the number of cores $n$, just the look-up table size (i.e.~$m \times \log_{2}\frac{n}{m}$) increases with larger platforms.
Therefore, the space complexity of the performance model is
$O(m \times \log_{2}\frac{n}{m})$.
The power profiling module uses three offline power models and each power model profiles the idle power and the runtime power of all possible execution places. 
Thus, the space complexity of the power model is also $O(m \times \log_{2}\frac{n}{m})$.

\textbf{Overheads.}
We evaluate the task scheduling overheads by calculating the time consumed in running the ERASE framework.
The total execution time of an application can be partitioned into three portions on each core: the accumulation of task execution time on the core, sleeping time and the time portion that runs the ERASE framework.
We compute the average overhead by averaging the total overheads of all cores, which can be expressed by:
\vspace{-2mm}
\begin{equation}  \footnotesize
    \vspace{-1mm}
    Overhead = \frac{\sum\limits_{i=1}^{n} 1 - \frac{\sum TaskExecTime_{i}+SleepTime_{i}}{Total\ Execution\ Time}}{n}.
\end{equation}
The evaluation on both platforms shows that the overheads are rather small. 
For instance, with MatMul on TX2, the overhead is 0.23\% on average across multiple degrees of parallelism.
\section{Integration of ERASE with A Work Stealing Runtime} 
\label{implementation}


\begin{figure}[t]
\centering
\vspace{-3mm}
\includegraphics[width=0.83\columnwidth]{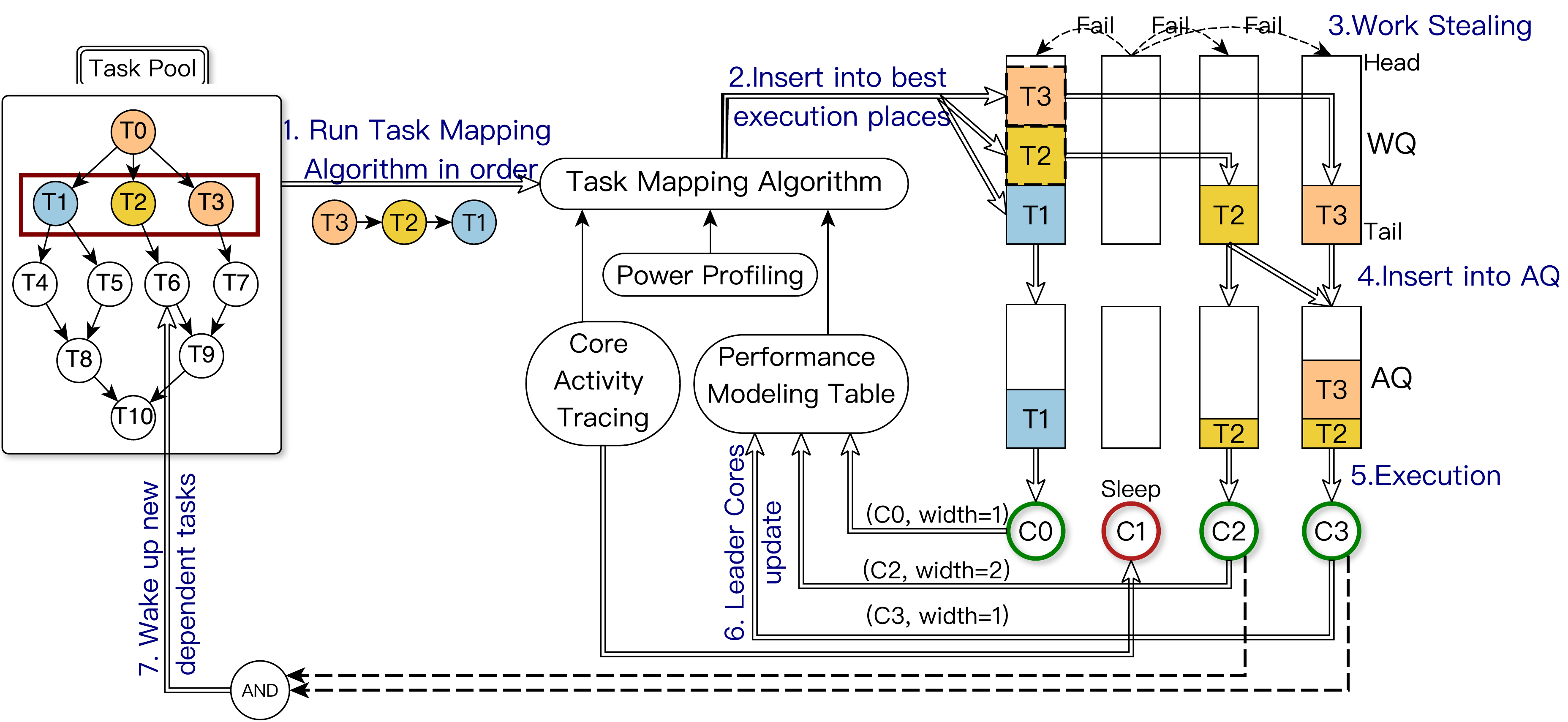}
\vspace{-3mm}
\caption{An illustration of task execution flow with ERASE in XiTAO runtime, when core 0 finishes the execution of T0 and releases three dependent tasks.} 
\vspace{-4mm}
\label{fig:runtimeflow}
\end{figure}

In this section, we describe how ERASE can be integrated into a work stealing environment.
We use the open source XiTAO\footnote{\url{https://github.com/CHART-Team/xitao.git}} runtime to exemplify the integration. The overall architecture of ERASE within XiTAO is shown in Figure~\ref{fig:runtimeflow}.
In XiTAO, each logical core manages two queues: a work queue (WQ), subject to work stealing, and a FIFO assembly queue (AQ), used to implement embedded work-sharing regions.
The WQs hold the tasks for which the best execution place has been computed.
When a task is fetched from the tail of WQ, pointers to the task are then inserted into all AQs representing the execution places for the task. 
The FIFO nature of the AQs guarantees that at some later point each core will fetch these pointers and execute the task.
Note that in an asymmetric platform with multiple clusters, work stealing is only allowed among cores within the same cluster. 
Two cases are to be considered in the event that two ready tasks select the same execution place: (1) if resource width $<$ number of cores in the cluster, other available core(s) in the cluster can steal the second ready task, insert it into the AQs and start execution immediately to avoid 
delay; 
(2) if resource width = number of cores in the cluster, no available resources for the second ready task exist. Hence, the task is inserted into the AQs and it must wait for the completion of the first task to begin execution.

\begin{figure}[t]
\vspace{-2em}
\begin{algorithm}[H]  \footnotesize
\caption{\small \textbf{Exponential Back-off Sleep Approach}}\label{listing:backoff}
\vspace{-5mm}
\begin{multicols}{2}
\begin{algorithmic}[1]
\State idle\_tries = 0, backoff\_param = 0
\While {true}
    \State Check local WQ or try to steal from other WQs
    \If{no available task}
        \State idle\_tries++
        \If{idle\_tries == N}
            \State status[core\_id] = 0
            \State sleep (1 $\ll$ backoff\_param) 
            \State idle\_tries = 0; backoff\_param++ 
        \EndIf
    \Else 
        \State status[core\_id] = 1
        \State idle\_tries = 0;  backoff\_param = 0
        \State execute(task)
    \EndIf
\EndWhile
\end{algorithmic}
\end{multicols}
\vspace{-3mm}
\end{algorithm}
\vspace{-9mm}
\end{figure}

Figure~\ref{fig:runtimeflow} illustrates the execution flow of ERASE implemented in XiTAO using the DAG example.
Here we assume that the four cores are from the same cluster for the purpose of simplicity.
Thus, work stealing between the four cores does not lead to task resource width change due to the identical characteristics.
Task pool collectively represents all the tasks in the sample application.
We capture the state when T0 has finished execution on C0 and the core has released three dependent ready tasks T1, T2 and T3. 
T4 to T10 are pending tasks that will only be released after T1, T2 and T3 finish execution.
When T1, T2 and T3 become ready, the runtime determines the most energy efficient execution place for each task by running Algorithm~\ref{taskmapping} (step \#1). 
When a core releases multiple ready tasks, finding the best task placement for these tasks follows the task creation order.
In this example, the most energy efficient execution places for the tasks T1, T2 and T3 are (C0,1) (C0,2) and (C0,1), respectively. 
The information about the best configuration obtained from the task mapping algorithm is embedded in the task. 
Consequently, all the three tasks are enqueued in the WQ of C0 (step \#2). 
Each core first looks for tasks in its own WQ before attempting to steal from other cores. 
In the case of C0, it finds T1 in the tail of its WQ and therefore inserts the task in its own AQ. 
C2 and C3 do not find tasks in their local WQs and consequently end up stealing tasks T2 and T3 from C0 and inserting them in their respective queues (step \#3). 
After C2 picks up T2 from the local WQ, it inserts T2 in the AQ of C2 and C3 since the best resource width is two (step \#4). 
Similarly, C3 picks up T3 from its local WQ and inserts it in the AQ of C3. Tasks in the AQ are consequently picked up and executed by the respective cores (step \#5).
If the work stealing attempts are continuously unsuccessful and the number of attempts exceeds a threshold, the core status tracked by the runtime is transitioned to ``sleep'' state, as shown for C1.
After the execution, each leader core is responsible for updating the corresponding performance model with its execution time (step \#6).
Note that the execution flow is asynchronous, i.e.~cores that execute a moldable task do not need to wait for the completion of each other.
The last core that finishes the execution is responsible for waking up the dependent tasks.
In this example, we assume that T2 finishes execution of C2 after C3 and therefore wakes up the dependent task T6 (step \#7).
If T1, T2 and T3 complete execution on different cores at the same time, the task mapping algorithm is executed concurrently on each of these cores, which in turn ends up determining the most energy efficient execution place for T4, T5, T6 and T7.

To minimize energy waste from continuous work stealing attempts by idle cores, we adopt an \textit{exponential back-off sleep} strategy~\cite{kwak_exp_backoff}, as shown in Algorithm~\ref{listing:backoff}.
The core activity tracing module keeps track of the number of unsuccessful steal attempts (idle\_tries) for each core and compares it with a runtime threshold parameter $\mathbf{N}$. 
When the idle\_tries matches $\mathbf{N}$, the core is transitioned into sleep status (lines 6-10).
Upon wake up, if the core finds any ready task from its own work queue or makes a successful steal from other work queues, it resets the number of unsuccessful attempts (idle\_tries) and the back-off parameter (backoff\_param) to zeros (lines 12-13). 
Otherwise, it will sleep for an exponentially increasing time (ranging from 1ms to 64ms in this work) if it still can not find any ready task. 
The evaluation in XiTAO shows that in comparison to the case where idle cores continuously attempt work stealing, exploiting the exponential back-off sleep strategy reduces energy by up to 67\% during under-utilization periods with minimal impact on performance.
Sleep optimizations are also used in popular work stealing runtimes such as cilk~\cite{cilk_scheduler} and TBB~\cite{tbb_scheduler}.

\section{Experimental Methodology} \label{setup}
\subsection{Evaluation Platforms} \label{platform}
We evaluate the ERASE scheduler on two different platforms. 

\textbf{Asymmetric platform:} The NVIDIA Jetson TX2 platform features a dual-core NVIDIA Denver 64-bit CPU and a quad-core ARM A57 CPU (each with 2 MB L2 cache).
The board is set to \texttt{MAX-N} nvpmodel mode~\cite{nvpmodel}. 
The Denver and the A57 cores implement the ARMv8 64-bit instruction set and are cache coherent. 
The platform only supports cluster-level DVFS and the two clusters support the same range of frequencies, i.e. between ~MAX (2.04GHz) and MIN (0.35GHz). 

\textbf{Symmetric platform:} We evaluate the scheduler on a dual-socket 16-core Intel Xeon Gold 6130 node (32 cores in total, code-named ``Tetralith'')  that runs Linux version 3.10.0. This machine is part of a larger compute cluster called Tetralith of National Supercomputer Centre at Linköping University~\cite{tetralith}. As common with such infrastructure, frequency settings are controlled by the computing center's administrators and beyond our control. Therefore we do not carry out evaluations that require frequency adaptation on this platform.
Intel’s Running Average Power Limit (\texttt{RAPL})~\cite{Rapl} interface is used for energy measurement and modeling.

\subsection{Power Profiling using INA3221 and RAPL} \label{powerprofiling}

\begin{figure*}[!t]
\vspace{-6mm}
\centering
\subfloat[\scriptsize{Compute-bound:MAX}]{\includegraphics[width=0.25\textwidth]{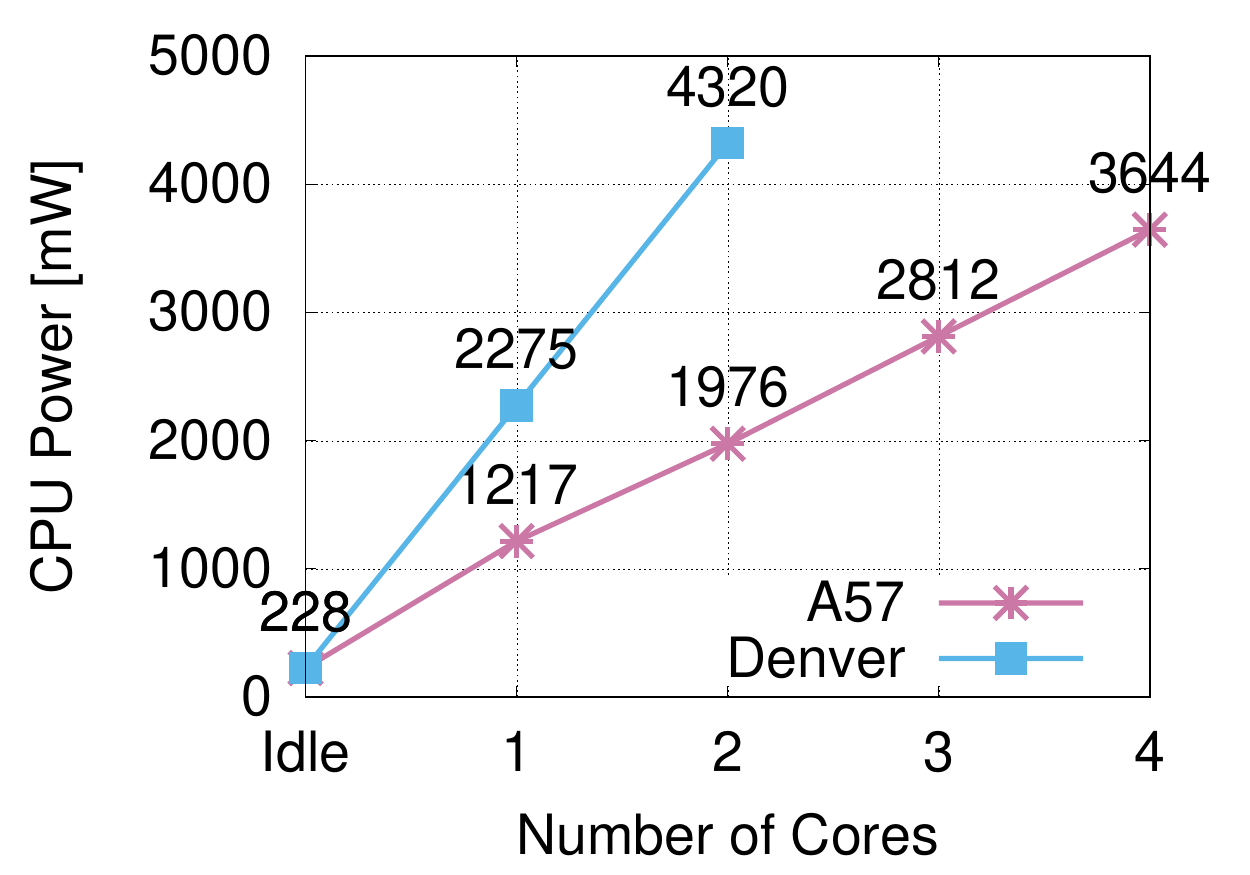}
\label{fig_compute_MAX}}
\subfloat[\scriptsize{Compute-bound:MIN}]{\includegraphics[width=0.25\textwidth]{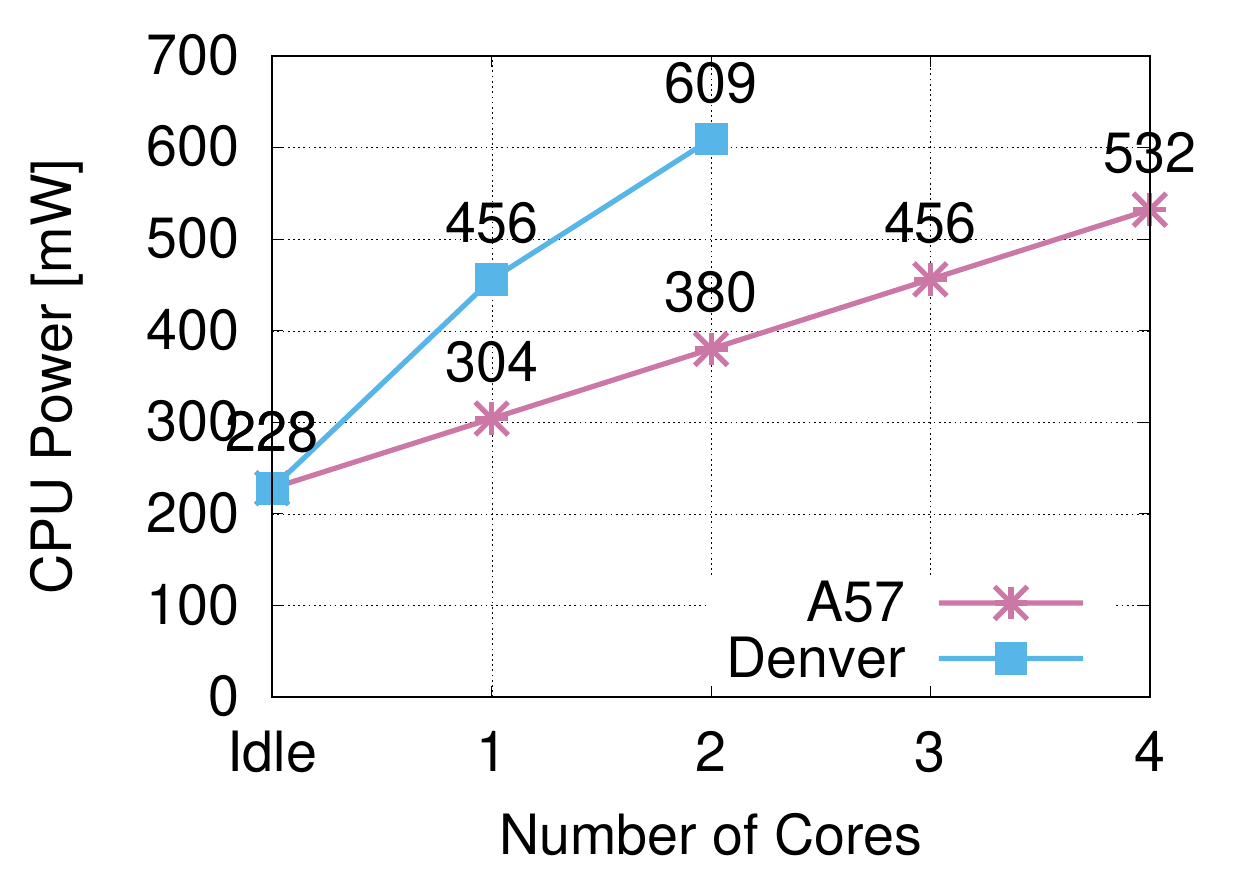}
\label{fig_compute_MIN}}
\subfloat[\scriptsize{Memory-bound:MAX}]{\includegraphics[width=0.25\textwidth]{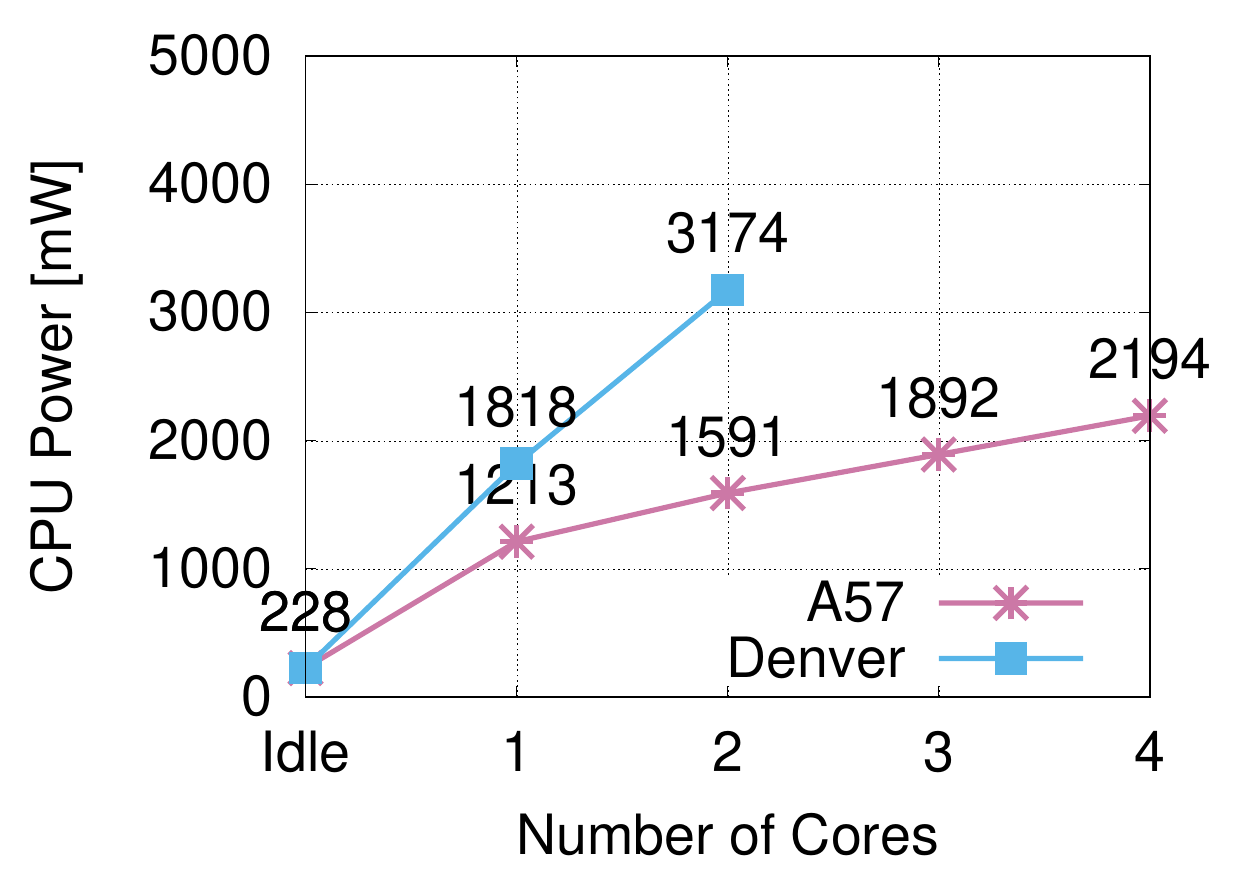}
\label{fig_memory_MAX}}
\subfloat[\scriptsize{Memory-bound:MIN}]{\includegraphics[width=0.25\textwidth]{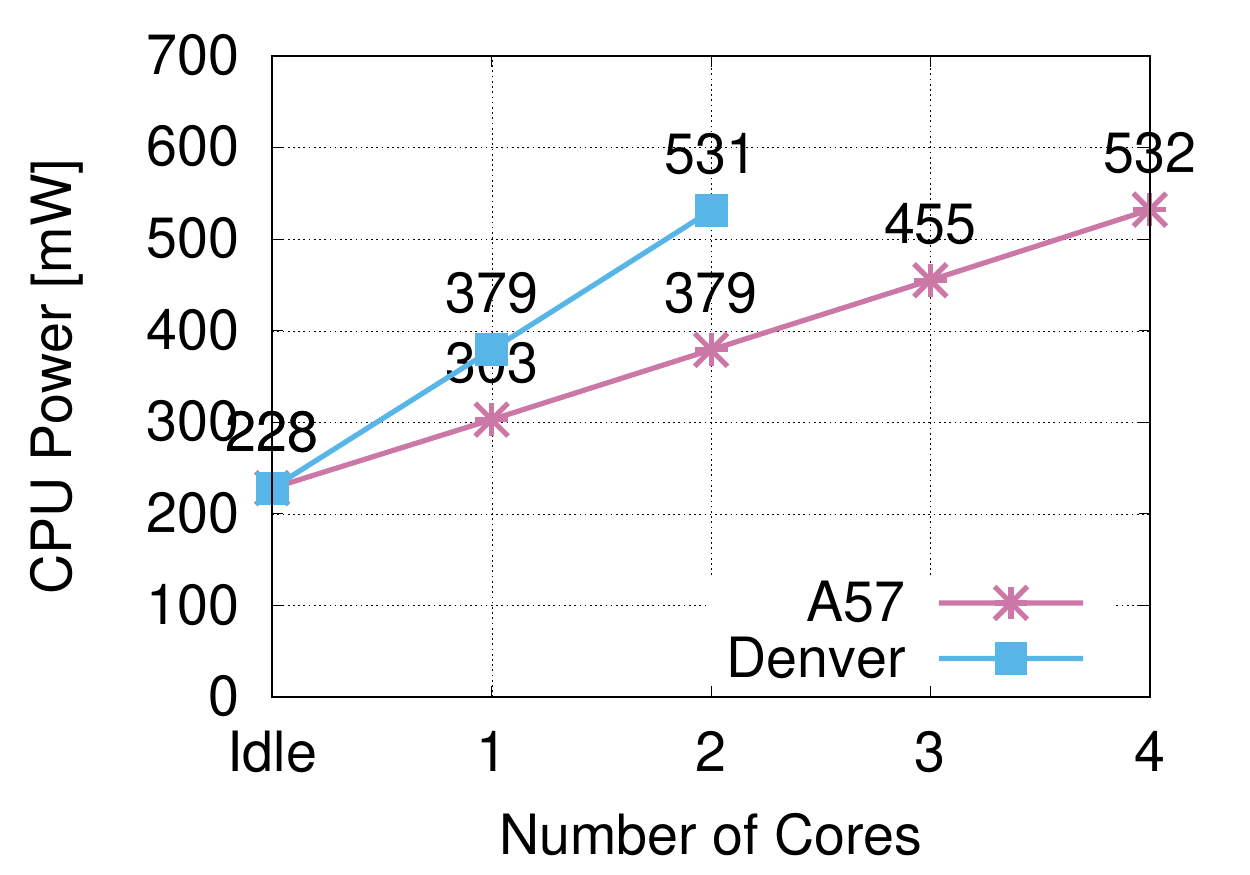}
\label{fig_memory_MIN}}
\vspace{-3mm}
\caption{Power profiling results of a compute-bound microbenchmark and a memory-bound microbenchmark on Jetson TX2 with two frequency levels.}
\vspace{-5mm}
\label{fig:computebound}
\end{figure*}

\textbf{INA3221:} The NVIDIA Jetson TX2 platform, that we use in our evaluation, features an on-board \texttt{INA3221} power sensor. 
We read
the power samples every 5 milliseconds ($\Delta t = 5ms$) and accumulate these samples obtained throughout the duration of application execution to compute the total energy consumption: 
$\textrm{Energy}=\int_{t_{0}}^{t_{s}} P(f,t)dt \approx \sum_{t_{1}}^{t_{s}} [P(f,t_i) - P(f,t_{i-1})] . \Delta t$, where
$f$ denotes frequency.
Figure~\ref{fig:computebound} presents the power profiling results for a compute-bound microbenchmark and a memory-bound microbenchmark with different number/type of cores running on TX2, respectively.
Here, we monitor power consumption for two different frequencies - MAX and MIN as an example.
It can be observed that the total idle power of the entire chip is 228mW, independent of the frequencies and the task type. 
Note that the Denver cluster and the A57 cluster differ in their idle power consumption.
The idle power of the A57 cluster is obtained by powering off all the cores in the Denver cluster, which is evaluated to be $\sim$152mW. 
Since one of A57 cores cannot be powered off on the platform, the idle power of the Denver cluster is obtained by subtracting the A57 cluster idle power from the total idle power of the entire chip, i.e.~228mW - 152mW = 76mW.

Taking the compute-bound microbenchmark as an example, we observe from Figures~\ref{fig_compute_MAX} and~\ref{fig_compute_MIN} that the runtime power consumption increases linearly with the number of cores used within a cluster. 
This linearity is not affected by the underlying frequency configuration in this case. 
With MAX frequency, the runtime power consumption is much higher than idle power consumption.
For example, the runtime power consumption of using one A57 core is 989mW (computed by subtracting 228 from 1217), which accounts for 81.3\% of the total power consumption and dominates the total power consumption in this case.
When using all the four A57 cores, the runtime power occupies an even larger fraction (around 95\%) of the total power consumption.
However, at MIN frequency, the idle power makes up a larger fraction of the total power consumption. 
For example, the idle power constitutes 75\% of the total power consumption when using one A57 core and 55\% of the total when using one Denver core, as shown in Figure~\ref{fig_compute_MIN}. 
This illustrates the importance of accurate power consumption attribution to each concurrently running task.

\textbf{RAPL}: \texttt{RAPL} interface is used to measure the energy and build the power profiles for the scheduler when running on an Intel platform. 
Since \texttt{RAPL} provides the energy readings for each CPU socket/package instead of power, it is necessary to convert the sampled energy values to the average power consumption during a fixed time window (5 milliseconds). 
We obtain the average power consumption using the equation: $Power_{i} = (Energy_{i} - Energy_{i-1})/\Delta t$.

\subsection{Benchmarks} \label{syntheticdag}
We evaluate the proposal using three synthetic benchmarks, two commonly used algebra kernels - dot product and Sparse LU factorization, and four applications from the Edge and HPC domains\footnote{All benchmarks are available in XiTAO github repository, refer to \url{https://github.com/CHART-Team/xitao.git}}.
Table~\ref{tab:benchmarks} lists the configurations of the benchmarks.
The magnitude of task execution time in these benchmarks ranges from 10 microseconds to 100 milliseconds (at 2.04GHz on a Denver core).
We introduce the notion of task criticality,
since several schedulers we compare against (discussed in Section~\ref{schedulers}) rely on task criticality to improve energy efficiency. 
We define \textit{critical path} as the longest path in a DAG. 
The tasks on the critical path are denoted as \textit{critical tasks}.
We define \textit{DAG parallelism} (abbr.~$dop$) as the total amount of tasks divided by the length of the critical path.

\textbf{Synthetic Benchmarks} are constructed with a configurable DAG parallelism and a notion of criticality.
The DAG is constructed in such a manner that the root node releases $dop$ tasks and one of the released nodes further releases $dop$ tasks. This process recursively continues till the total number of tasks spawned reaches the user specified limit. We mark tasks that spawn the maximum number of tasks as critical tasks.
The DAG comprises tasks that are of one of three types: Matrix Multiplication (abbr.~MatMul), Copy or Stencil. 
\textbf{MatMul} represents the \textit{compute-bound} class, where matrices
are pre-allocated and partitioned in N$\times$N tiles. 
The matrix size is set to fit into L1 cache, which aims to minimize the cache-misses as the analysis requires focusing on the CPU power dissipation arising from the arithmetic units. 
\textbf{Copy} represents the \textit{memory-bound} class.
Each task reads and writes a large matrix to memory, effectively creating streaming behavior to access the main memory continuously. 
\textbf{Stencil} represents the \textit{cache-sensitive} class as we set the task size greater than the last level cache but far less than DRAM.
It involves repeated updates of values associated with points on a multi-dimensional grid using the values at a set of neighboring points.

\textbf{Dot Product} computes the sum of the products of two equal-length vectors. 
In XiTAO, we partition the vectors into blocks and mark each block as a single task. 
The DAG is iteratively executed for a predefined number of iterations (100 in this work).

\begin{table*}[!t]
\scriptsize 
\vspace{-3mm}
\caption{The Configurations of Evaluated Benchmarks}
\vspace{-4mm}
\begin{center}
\begin{tabular}{@{}p{1.3cm}p{0.8cm}p{0.8cm}p{0.9cm}p{0.9cm}p{0.8cm}p{0.9cm}p{1.4cm}p{3cm}} \toprule
\textbf{Application} & \multicolumn{2}{c}{\textbf{MatMul}} & \multicolumn{2}{c}{\textbf{Copy}} & \multicolumn{2}{c}{\textbf{Stencil}} & \textbf{VGG-16} & \textbf{Alya}\\ \midrule
\textbf{Platform} &   TX2 &   Tetralith   & TX2   & Tetralith & TX2 & Tetralith  & TX2 & TX2 \\ \midrule
\textbf{Input Size}   &   $2^{6}\times2^{6}$   & $2^{7}\times2^{7}$  & $2^{12}\times2^{12}$ & $2^{13}\times2^{13}$ & $2^{8}\times2^{8}$ & $2^{10}\times2^{10}$ & BlockSize=64 & 50k/100k/ 200k CSR non-zeros \\ \midrule 
\textbf{Nr\_Tasks} &   50000  &    20000 & 20000 & 10000 & 10000 & 10000  & 509 & 7789/ 21360 / 47840 \\ \midrule
\textbf{Criticality} & $\surd$  & $\surd$  & $\surd$  & $\surd$  & $\surd$  & $\surd$  & $\times$ &$\times$   \\ \bottomrule
\end{tabular}
\vspace{2mm} \\
\begin{tabular}{@{}p{1.3cm}p{1.9cm}p{2.6cm}p{2.55cm}p{3.85cm}} \toprule
\textbf{Application} & \textbf{Heat} & \textbf{Sparse LU} & \textbf{Dot Product} & \textbf{Biomarker Infection} \\ \midrule
\textbf{Platform} &  Tetralith & Tetralith & TX2 & TX2  \\ \midrule
\textbf{Input Size}   & Resolution = 40960 & 32 blocks, blocksize=1024 & 200 blocks, blocksize=32k & SampleSize=2, NumBiomarkers 6217 \\ \midrule 
\textbf{Nr\_Tasks} &  32032 & 1512 & 20000 & 6217 \\ \midrule
\textbf{Criticality} & $\times$ & $\surd$ & $\surd$ & $\times$ \\ \bottomrule
\end{tabular}
\label{tab:benchmarks}
\vspace{-3mm}
\end{center}
\end{table*}

\textbf{Sparse LU Factorization} (abbr.~SLU) computes LU matrix factorization over sparse matrices~\cite{BarcelonaBenchmark}
It includes four kernels: LU0, FWD, BDIV and BMOD.
Among them, all LU0, FWD, BDIV are marked as critical tasks and some BMOD tasks are also marked as critical. The matrix is composed of N$\times$N blocks, each of which has a pointer to a sub-matrix of size M$\times$M.

\textbf{Image Classification Darknet-VGG-16 CNN} \label{vgg_setup}
(abbr.~VGG-16) is an inference deep learning algorithm that is typical of mobile and edge devices. 
It is a 16-layered deep neural network implemented as a fork-join DAG that spans all layers. 


\textbf{Alya}
is high performance computational mechanics code developed at Barcelona Supercomputing Center~\cite{Alya}.
The application involves solving partial differential equations, and its parallelization strategy is based on mesh partitioning. 
We test with three different input sizes, i.e.~50K, 100K, 200K non-zeros expressed in compressed sparse row format.

\textbf{Heat Diffusion} (abbr.~Heat) is implemented on a 2D grid by using one of the iterative numerical methods: 2D Jacobi stencil. 
The DAG is iteratively executed for a predefined number of iterations (1000 in this work).
We also experiment with other decompositions, e.g.~pencil (2D), without change in conclusions.  
The Heat DAG is symmetrical, thus no task criticality assignment is performed.


\textbf{Biomarker Infection Research} (abbr.~BioInfection) is a medical use case developed using XiTAO runtime in the LEGaTO project~\cite{BiomarkerAPP}. 
The application serves as an indicator of a biological condition, identify risk factors, examine diseases, predict diagnoses, determine the state of the disease or measure the effectiveness of treatment.
The dataset tested belongs to a pilot study for differentiating between periprosthetic hip infection and aseptic hip prosthesis loosening.

\subsection{Evaluated Schedulers} \label{schedulers}
We evaluate the effectiveness of ERASE by comparing it to different scheduling techniques. The schedulers we evaluate can be broadly categorized into two groups. 
All schedulers are implemented on top of the XiTAO runtime with the exponential back-off sleep strategy described in Section~\ref{implementation}.

The first group comprises schedulers that do not rely on DVFS and are described below: 

(\romannum{1}). Random Work Stealing (\textbf{RWS}) is a widely used baseline scheduler, which tries to keep all cores busy through work stealing and also works well in asymmetric environments.
Note that RWS schedules a task on a single core and does not leverage task moldability.

(\romannum{2}). Criticality-Aware Task Scheduler (\textbf{CATS})
is a performance-oriented scheduler that leverages task criticality initially proposed by Chronaki et.al~\cite{CATS}.
Tasks marked as critical are scheduled on the faster cores, while non-critical tasks are scheduled on the slower cores. 
CATS permits work stealing in specific cases, i.e.~the faster cores can steal among themselves and also from the slower cores.
Note that CATS does not exploit task moldability.

(\romannum{3}). Criticality-Aware Low Cost (\textbf{CALC}) is a scheduler designed to evaluate the benefit of combining task moldable execution with criticality based performance scheduler like CATS. 
For each critical task, the runtime globally compares the parallel execution cost, i.e.~execution time $\times$ resource width,
of all execution places and then selects the best one that minimizes the cost and it cannot be stolen. For non-critical tasks, the runtime only locally determines the best resource width that minimizes the cost while keeping the leader core fixed. 


In the second group, we compare to a scheduler that exploits DVFS on a cluster-level DVFS platform and it is described below:

(\romannum{1}). \textbf{Aequitas}~\cite{HarisRibic-AEQUITAS} is a round-robin time-slicing scheduler that leverages  HERMES~\cite{HarisRibic-WorkStealing/PercoreDVFS/Tempo} to achieve energy efficiency. 
HERMES utilizes workpath sensitive and workload sensitive algorithms to tune the frequency of individual cores based on thief-victim relations and the number of ready tasks available in the WQ. 
Additionally, for architectures with cluster-level DVFS, Aequitas lets each active core within a cluster control DVFS for a short interval in a round-robin manner. 
We reproduce Aequitas on TX2 using five frequency levels (i.e.~2.04GHz, 1.57GHz, 1.11GHz, 0.65GHz and 0.35GHz) as in the original proposal, four thresholds for workload sensitive algorithm 
(~$\lceil\frac{dop+1}{5}\rceil$, $2\lceil\frac{(dop+1)}{5}\rceil$, $3\lceil\frac{(dop+1)}{5}\rceil$, $4\lceil\frac{(dop+1)}{5}\rceil$) and a slicing time interval of 1s. 
We conducted a sensitivity analysis with time slicing intervals of 2s, 1s, 0.5s and 0.25s. 
The evaluation shows that using 1s interval gives the best results. Hence, in this work we present the Aequitas results with 1s interval.
\section{Evaluation} \label{Perf_Eva}
We evaluate the effectiveness and adaptability of ERASE by comparing it to several state-of-the-art scheduling techniques on two different platforms.
We first compare ERASE to a state-of-the-art scheduler that relies on DVFS throttling in Section~\ref{dvfs}.
Then, we compare ERASE to other schedulers that do not rely on DVFS on both TX2 and Tetralith platforms in Sections~\ref{tx2_eva} and~\ref{haswell_eva}.
We also present how fast ERASE can react to a dynamic DVFS environment in Section~\ref{dyna_reaction}.
Lastly, we evaluate the proposed models' accuracy in ERASE, which is shown in Section~\ref{acc_eval}.
We repeat each experiment 100 times and report the average energy and execution time numbers. 
The observed coefficient of variation in the measurements is below 1.3\%.



\subsection{Comparison to the State-of-the-art} 
\label{dvfs}
Table~\ref{tab:aequitas_erase} compares the energy consumed by ERASE and Aequitas~\cite{HarisRibic-AEQUITAS}, when running three synthetic benchmarks ($dop$ ranging from 2 to 8) on the Jetson TX2 platform. 
Since the scheduling strategies have significant impact on performance, we also provide execution time 
comparisons to better understand the trade-off between energy and performance.
A general observation is that ERASE consumes less energy and is faster than resource-agnostic and non-moldable counterparts Aequitas across most of the benchmarks and different degrees of parallelism.

\begin{table*}[t]
\scriptsize
\vspace{-3mm}
\caption{Energy consumption and execution time comparison between Aequitas and ERASE on TX2}
\vspace{-3mm}
\begin{center}
\begin{tabular}{p{1.1cm}p{.65cm}p{.57cm}p{.57cm}p{.57cm}p{.57cm}p{.57cm}p{.57cm}p{.57cm}p{.57cm}p{.57cm}p{.57cm}p{.57cm}p{.6cm}} \toprule
\textbf{Application}& & \multicolumn{4}{c}{\textbf{MatMul}} & \multicolumn{4}{c}{\textbf{Copy}} & \multicolumn{4}{c}{\textbf{Stencil}} \\ \midrule
\textbf{Parallelism}& & dop=2 & dop=4 & dop=6 & dop=8 & dop=2 & dop=4 & dop=6 & dop=8 & dop=2 & dop=4 & dop=6 & dop=8 \\ \midrule
\multirow{2}{*}{\textbf{Energy[J]}} & Aequitas & 92.85	& 87.93 &	84.06 &	80.36 & 71.01 & 53.31 & 52.66 & 57.83 & 45.22	& 41.51 &	39.62 &	35.15 \\ \cmidrule(r){2-14}
& ERASE & 76.59 &	74.06 &	73.85 &	69.32 & 58.26	& 59.26	& 64.77	& 64.19 & 31.41 &	31.01 &	31.17 &	30.99 \\ \midrule
\multirow{2}{1.1cm}{\textbf{Execution Time[s]}} & Aequitas & 80.70 &	43.72 &	34.11 &	32.60 & 39.21 &	28.03 &	28.51 &	25.31 & 25.38 &	20.27 &	16.73 &	12.44 \\ \cmidrule(r){2-14}
& ERASE & 25.85	& 19.76 &	19.41 & 19.31 & 25.85 &	19.76 &	19.41 &	19.31 & 9.87 &	8.41 &	9.00 &	8.46 \\ \bottomrule
\end{tabular}
\label{tab:aequitas_erase}
\vspace{-3mm}
\end{center}
\end{table*}



Overall, ERASE achieves 15\% energy reduction on average and up to 68\% performance improvement for MatMul, compared to Aequitas, which results in 59\% EDP reduction.
For Stencil, ERASE achieves 22\% energy reduction, 49\% performance improvement and therefore 60\% EDP reduction on average than Aequitas. 
In the case of Copy, ERASE is more energy efficient than Aequitas with DAG parallelism of 2 and it reduces energy consumption and EDP by 18\% and 46\%, respectively.


\textbf{Analysis:}
The inefficiency of Aequitas can be attributed to two reasons: (1) cores that acquire domination control (the ability for changing the cluster frequency for a specific timeslice) can decrease the frequency of the whole cluster resulting in delaying execution of critical tasks thereby increasing idleness; (2) Aequitas does not take core asymmetry into account. If critical tasks are scheduled on low performance cores and the cluster’s frequency is decreased, the amount of idleness can increase considerably especially when there are fewer ready tasks in work queues overall.
Aequitas consumes less energy than ERASE 
with Copy of $dop$ = 4 to 8.
Because Copy is memory-bound and frequency reduction helps to save more energy without impacting performance.

\subsection{Evaluation on Asymmetric Platform} \label{tx2_eva}
\begin{figure*}[t]
\vspace{-6mm}
\centering
\subfloat[\scriptsize{Synthetic MatMul - Energy Consumption}]{\includegraphics[width=\textwidth]{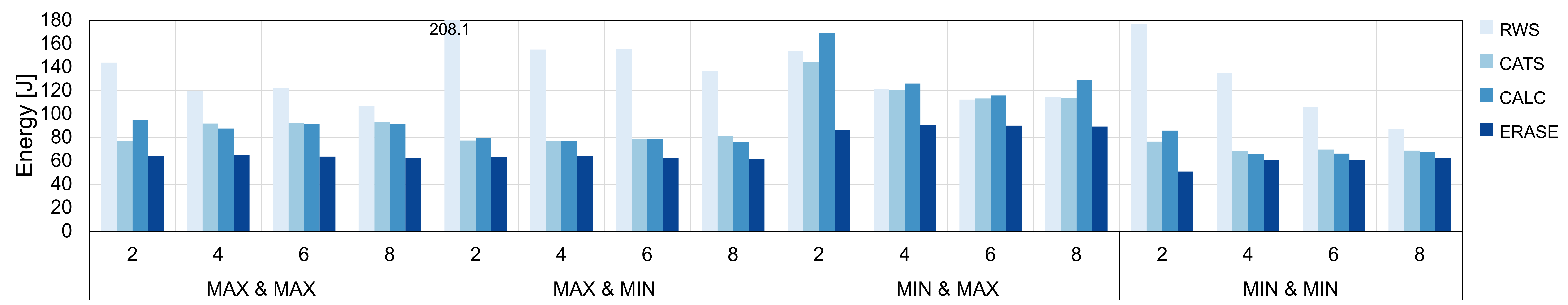}
\label{fig_MM}}
\\
\vspace{-4mm}
\subfloat[\scriptsize{Synthetic MatMul - Execution Time}]{\includegraphics[width=\textwidth]{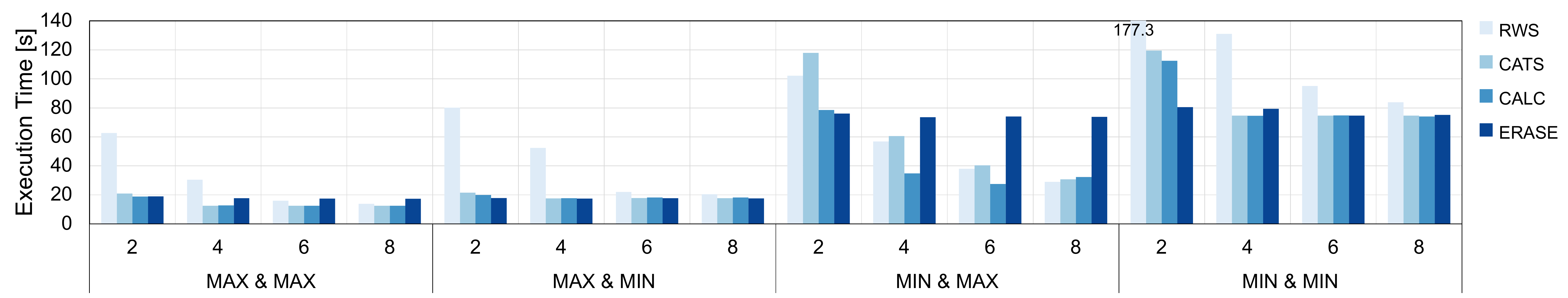}
\label{fig_MM_perf}}
\vspace{-3mm}
\caption{Energy consumption and execution time comparison between RWS, CATS, CALC and ERASE with Synthetic MatMul on Jetson TX2.} 
\vspace{-5mm}
\label{fig_MatMul_tx2}
\end{figure*}

\begin{figure*}[t]
\vspace{-6mm}
\centering
\subfloat[\scriptsize{Synthetic Copy - Energy Consumption}]{\includegraphics[width=\textwidth]{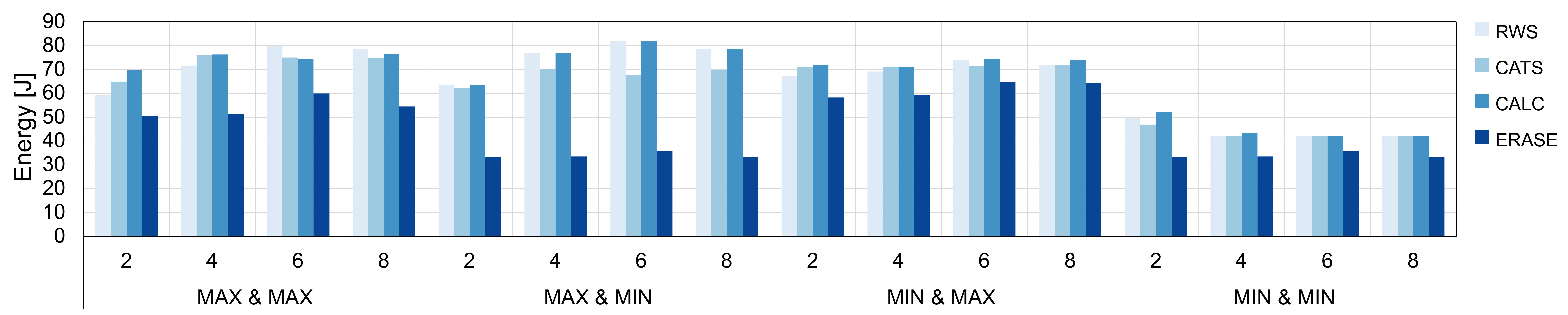}
\label{fig_CP}}
\\
\vspace{-4mm}
\subfloat[\scriptsize{Synthetic Copy - Execution Time}]{\includegraphics[width=\textwidth]{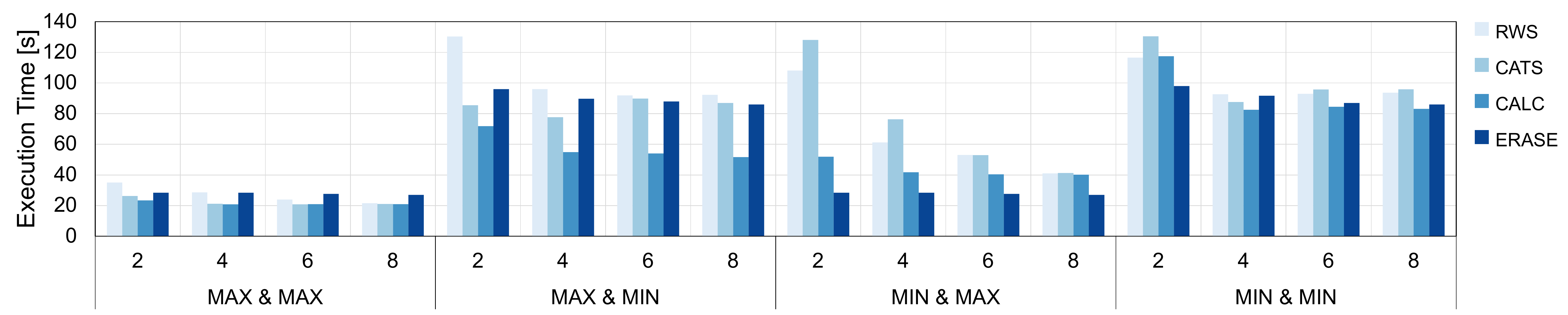}
\label{fig_CP_perf}}
\vspace{-3mm}
\caption{Energy consumption and execution time comparison between RWS, CATS, CALC and ERASE with Synthetic Copy on Jetson TX2.} 
\vspace{-4mm}
\label{fig_Copy_tx2}
\end{figure*}

\begin{figure*}[t]
\vspace{-4mm}
\centering
\subfloat[\scriptsize{Synthetic Stencil - Energy Consumption}]{\includegraphics[width=\textwidth]{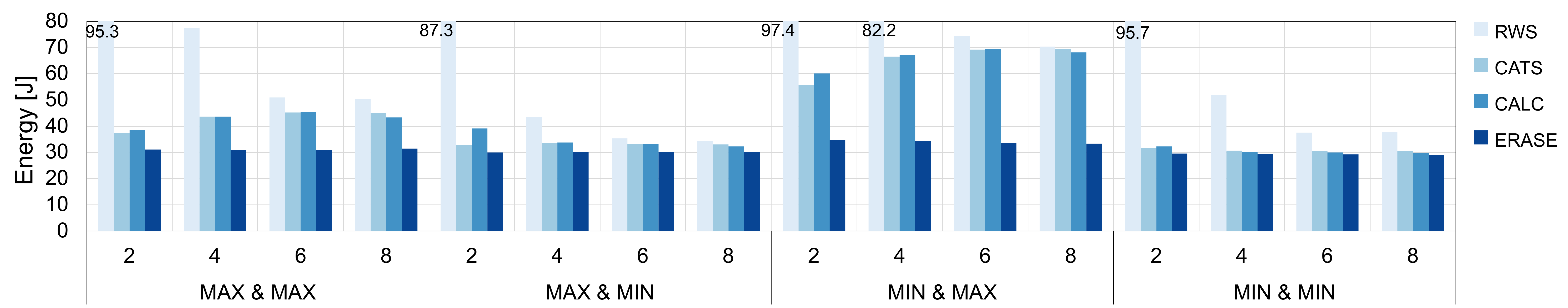}
\label{fig_ST}}
\\
\vspace{-4mm}
\subfloat[\scriptsize{Synthetic Stencil - Execution Time}]{\includegraphics[width=\textwidth]{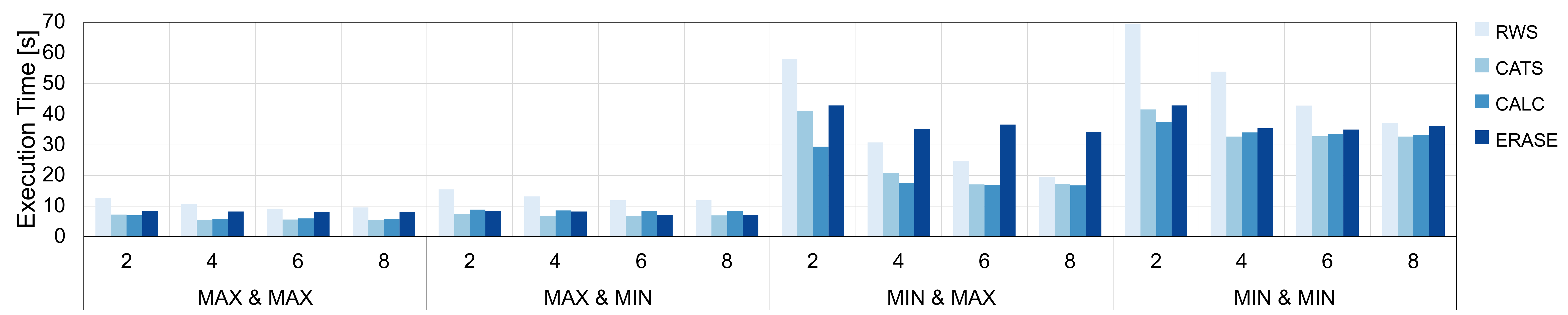}
\label{fig_ST_perf}}
\vspace{-3mm}
\caption{Energy consumption and execution time comparison between RWS, CATS, CALC and ERASE with Synthetic Stencil on Jetson TX2.} 
\vspace{-4mm}
\label{fig_Stencil_tx2}
\end{figure*}

Next, we compare ERASE to RWS, CATS and CALC on TX2. 
These schedulers do not rely on DVFS but use other approaches for improving performance and energy efficiency.
To compare the efficacy of the schedulers at different static frequency settings, we perform the evaluation using the four MIN and MAX frequency permutations on the two clusters, as shown in the x-axis labels of Figures~\ref{fig_MatMul_tx2}, ~\ref{fig_Copy_tx2}, ~\ref{fig_Stencil_tx2} and~\ref{real_applications}. 
For example, MAX \& MIN means that the Denver cluster is set to maximum frequency and the A57 cluster is set to minimum frequency.


\subsubsection{Synthetic Benchmarks}
Figures~\ref{fig_MatMul_tx2}, \ref{fig_Copy_tx2} and \ref{fig_Stencil_tx2} compares the energy and execution time 
of the four schedulers with MatMul, Copy and Stencil over different DAG parallelism, respectively.
The results show that ERASE generally consumes the least amount of energy compared to other three scheduling policies.
Another important observation is that ERASE does consistently well independent of DAG parallelisms and frequency settings. 
While the energy consumption of other three schedulers varies considerably with different frequency settings and is usually much higher with low parallel slackness (i.e.~fewer tasks in all the work queues).
If we consider MatMul in Figure~\ref{fig_MatMul_tx2} as an example, ERASE achieves 47\% energy reduction and improves performance by 8\% on average compared to RWS across different frequency and parallelism settings. 
It is clear that in most cases, the energy reduction with ERASE far exceeds the performance losses, which results in 43\% EDP reduction on average.
In comparison to CATS and CALC, ERASE achieves an average energy reduction of 30\% and an average reduction in EDP of 10\% across different DAG parallelisms.

\begin{figure*}[t!]
\vspace{-8mm}
\centering
\subfloat[\scriptsize{RWS}]{\includegraphics[width=0.2\textwidth]{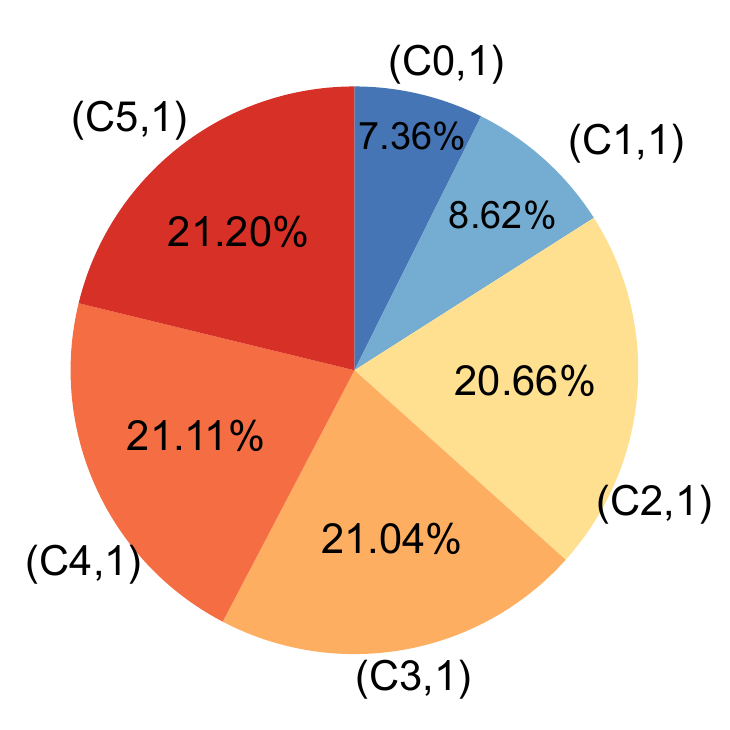}
\label{fig_rws_td}}  \hspace{6mm}%
\subfloat[\scriptsize{CATS}]{\includegraphics[width=0.2\textwidth]{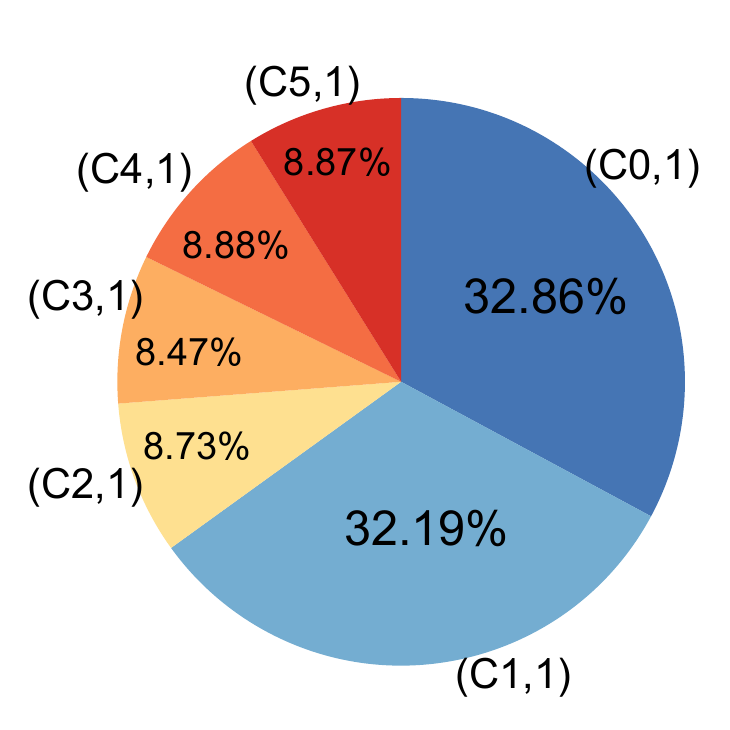}
\label{fig_cats_td}}  \hspace{6mm}%
\subfloat[\scriptsize{CALC}]{\includegraphics[width=0.2\textwidth]{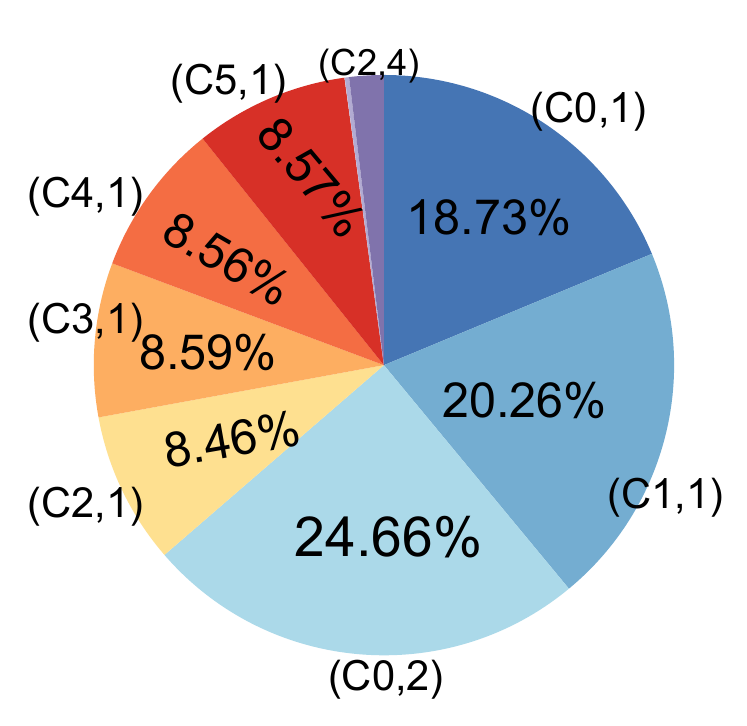}
\label{fig_calc_td}}  \hspace{6mm}%
\subfloat[\scriptsize{ERASE}]{\includegraphics[width=0.2\textwidth]{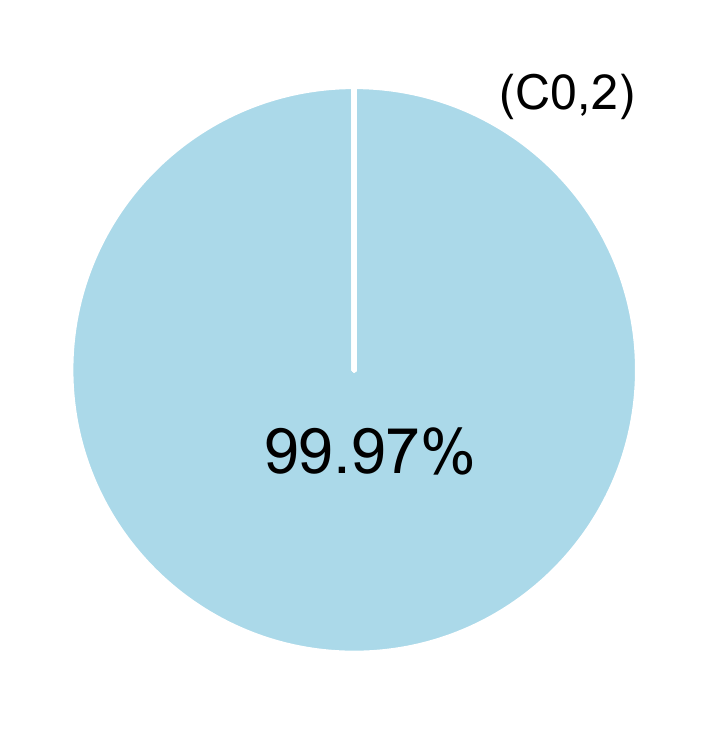}
\label{fig_erase_td}}
\vspace{-3mm}
\caption{Task distribution of RWS, CATS, CALC and ERASE with Synthetic MatMul on Jetson TX2.} 
\vspace{-4mm}
\label{fig_mm_taskdistribution}
\end{figure*}

\begin{figure*}[t!]
\vspace{-4mm}
\centering
\subfloat[\scriptsize{RWS}]{\includegraphics[width=0.25\textwidth]{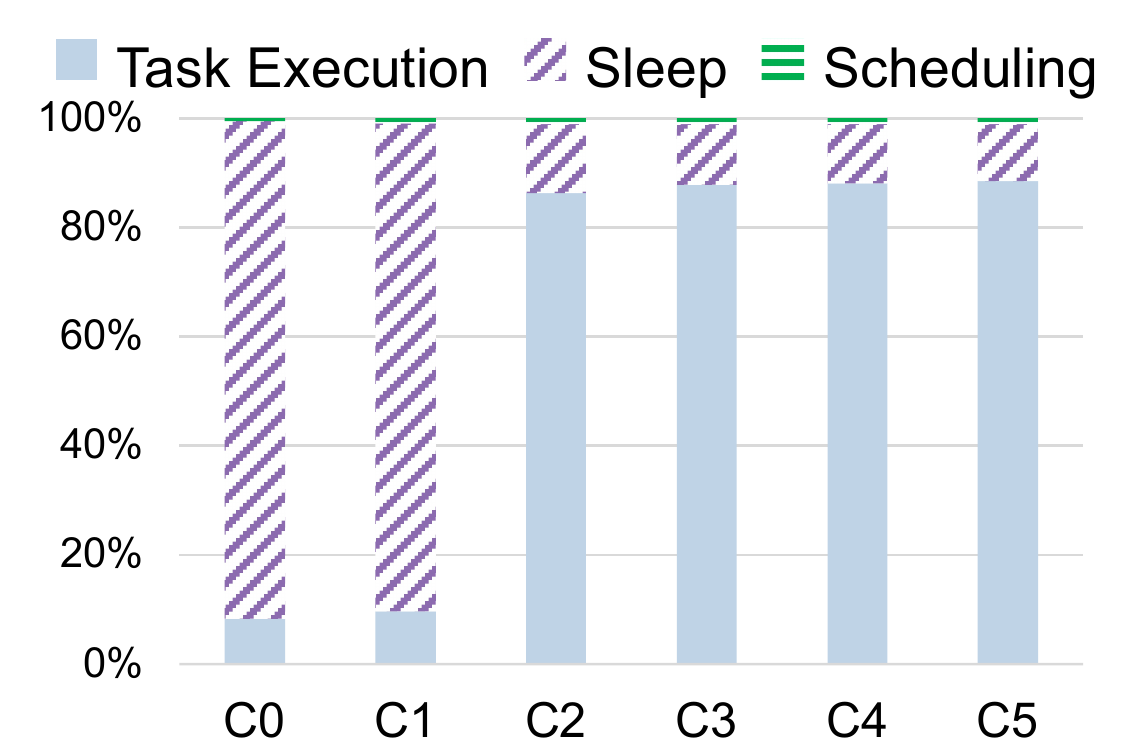}
\label{fig_rws_timed}}  
\subfloat[\scriptsize{CATS}]{\includegraphics[width=0.25\textwidth]{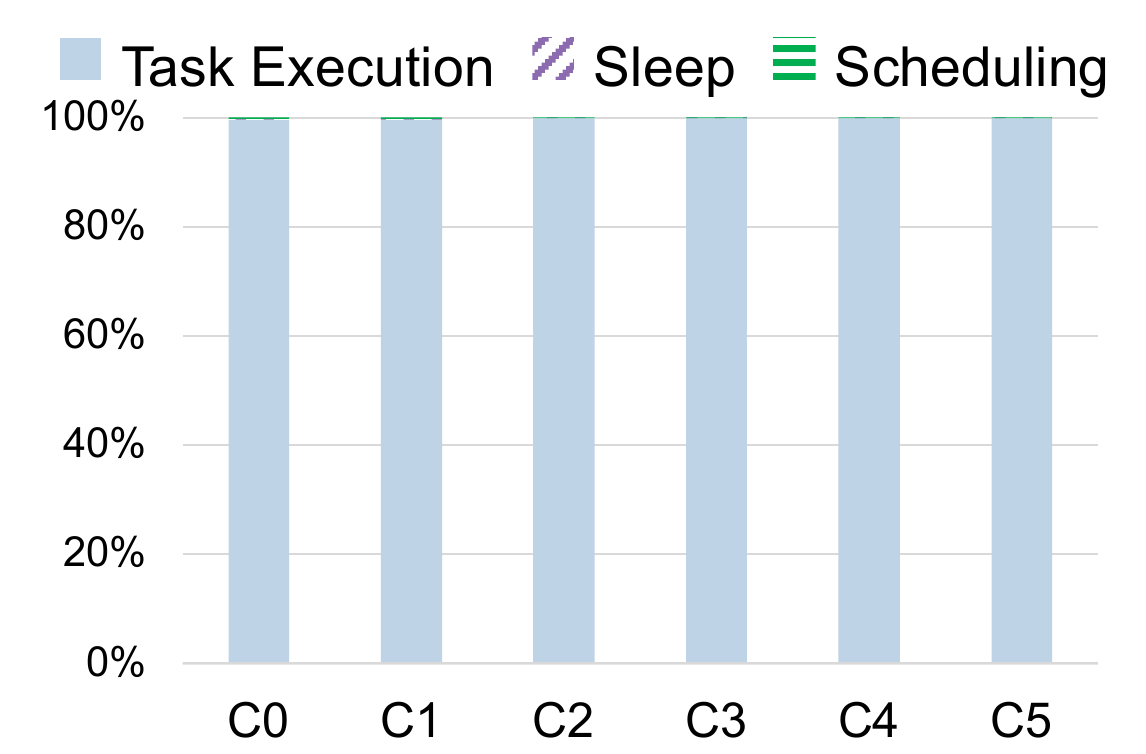}
\label{fig_cats_timed}}  
\subfloat[\scriptsize{CALC}]{\includegraphics[width=0.25\textwidth]{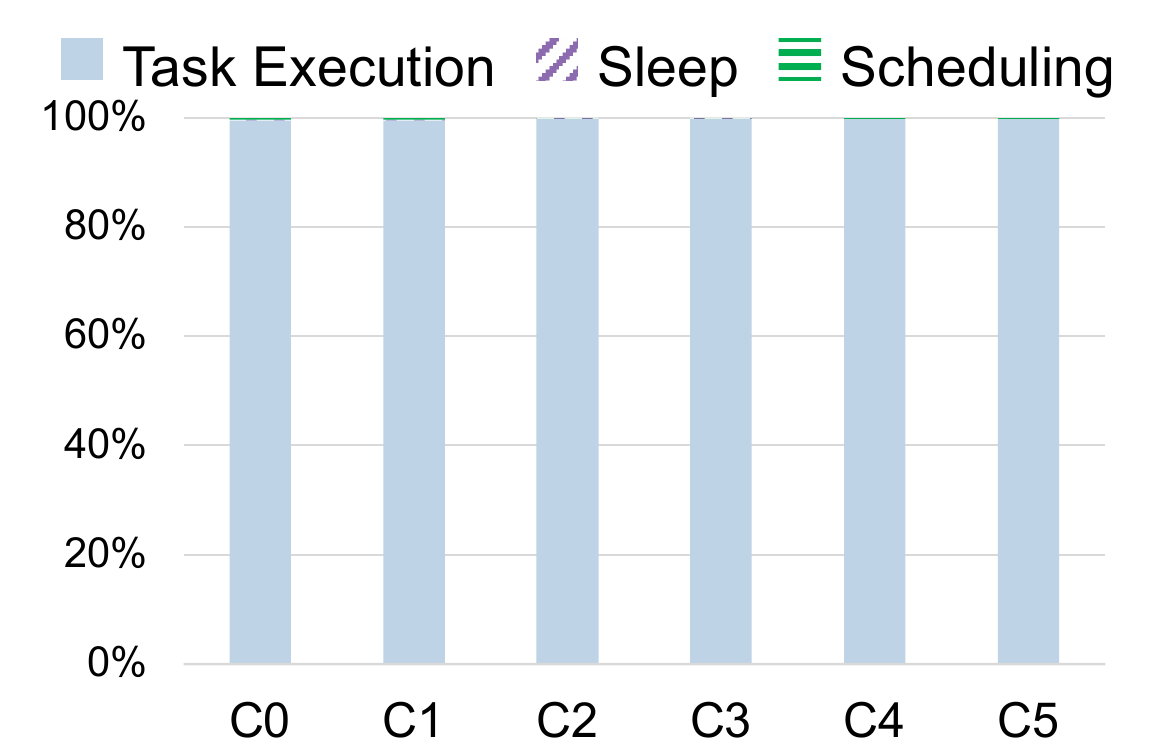}
\label{fig_calc_timed}}  
\subfloat[\scriptsize{ERASE}]{\includegraphics[width=0.25\textwidth]{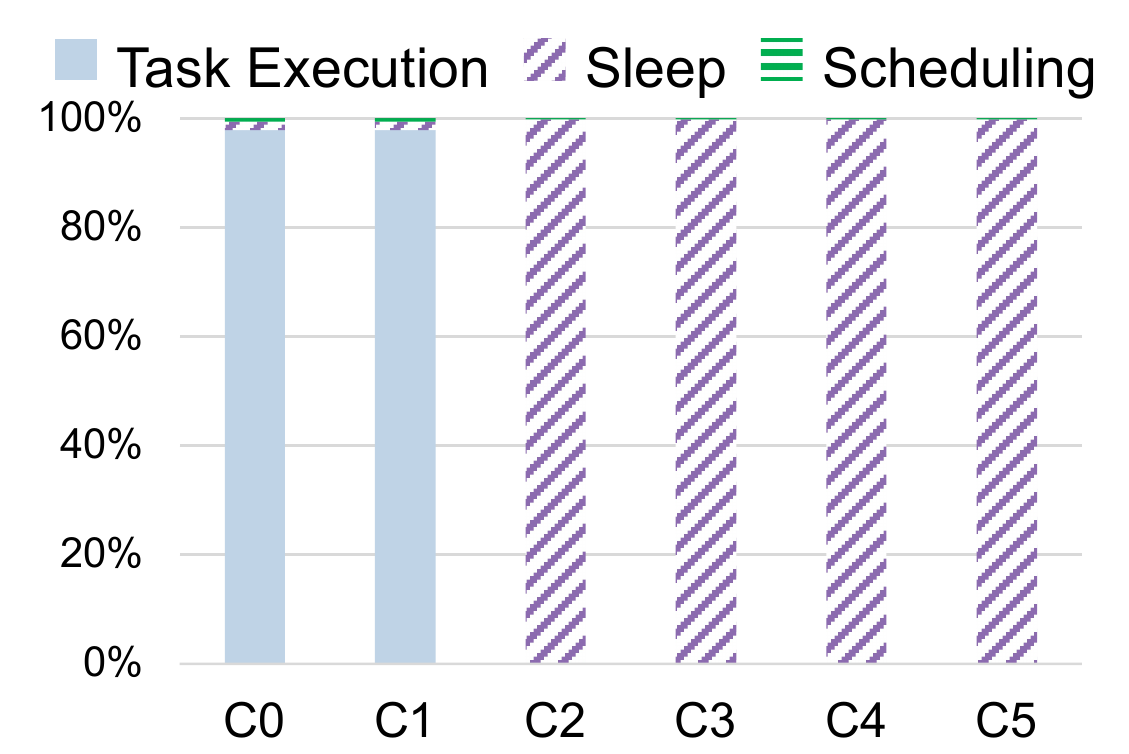}
\label{fig_erase_timed}}
\vspace{-3mm}
\caption{Time distribution of RWS, CATS, CALC and ERASE with Synthetic MatMul on Jetson TX2.} 
\vspace{-5mm}
\label{fig_mm_timedistribution}
\end{figure*}

\textbf{Analysis:} We choose MatMul with $dop$=4 in MAX\&MAX as an example to illustrate why ERASE ends up consuming less energy in comparison to the other schedulers we evaluate. 
Figure~\ref{fig_mm_taskdistribution} presents the task distribution on each execution place chosen by the four schedulers, where C0 and C1 denote two Denver cores with a maximum possible resource width of two and C2 to C5 denote the four A57 cores with a maximum possible resource width of four. 
Figure~\ref{fig_mm_timedistribution} shows the time distributed between scheduling activity, sleep and actual task execution when using these schedulers.
It can be observed that with RWS the majority of tasks are executed on the four A57 cores.
Consequently, the two Denver cores are idle and are put to sleep for energy savings.
CATS permits work stealing if the Denver cores attempt to steal tasks from the A57 cores mainly to permit migration of critical tasks to high performance cores. 
Consequently, more than 65\% of tasks execute on the high-performance Denver cores, leading to 60\% performance improvement compared to RWS.
The result of CALC shows that 27\% ((C0,2) and (C2,4)) of tasks exploit the task moldability. 
As discussed earlier in Section~\ref{Type_Moldability}, executing a single MatMul task with two Denver cores is the most energy efficient configuration on this platform and this results in an additional energy reduction of 7\% in comparison to CATS. 
For CATS and CALC, almost 100\% of the time is spent in executing tasks as shown in Figures~\ref{fig_cats_td} and~\ref{fig_calc_td}. 
ERASE correctly predicts and executes most of the tasks on Denver cores with a resource width of two since it is the most energy efficient option thereby consuming the least energy when compared to the other schedulers.
In MIN\&MAX, executing MatMul tasks with two Denver cores also consumes the least energy. 
However, the performance is worse than other schedulers with high $dop$ since all tasks are executed with the lowest frequency on Denver cores while the A57 cores are idle.

\subsubsection{Applications.}
\begin{figure}[t]
\centering
\vspace{-6mm}
\subfloat[\scriptsize{VGG-16 - Energy}]{\includegraphics[width=0.25\textwidth]{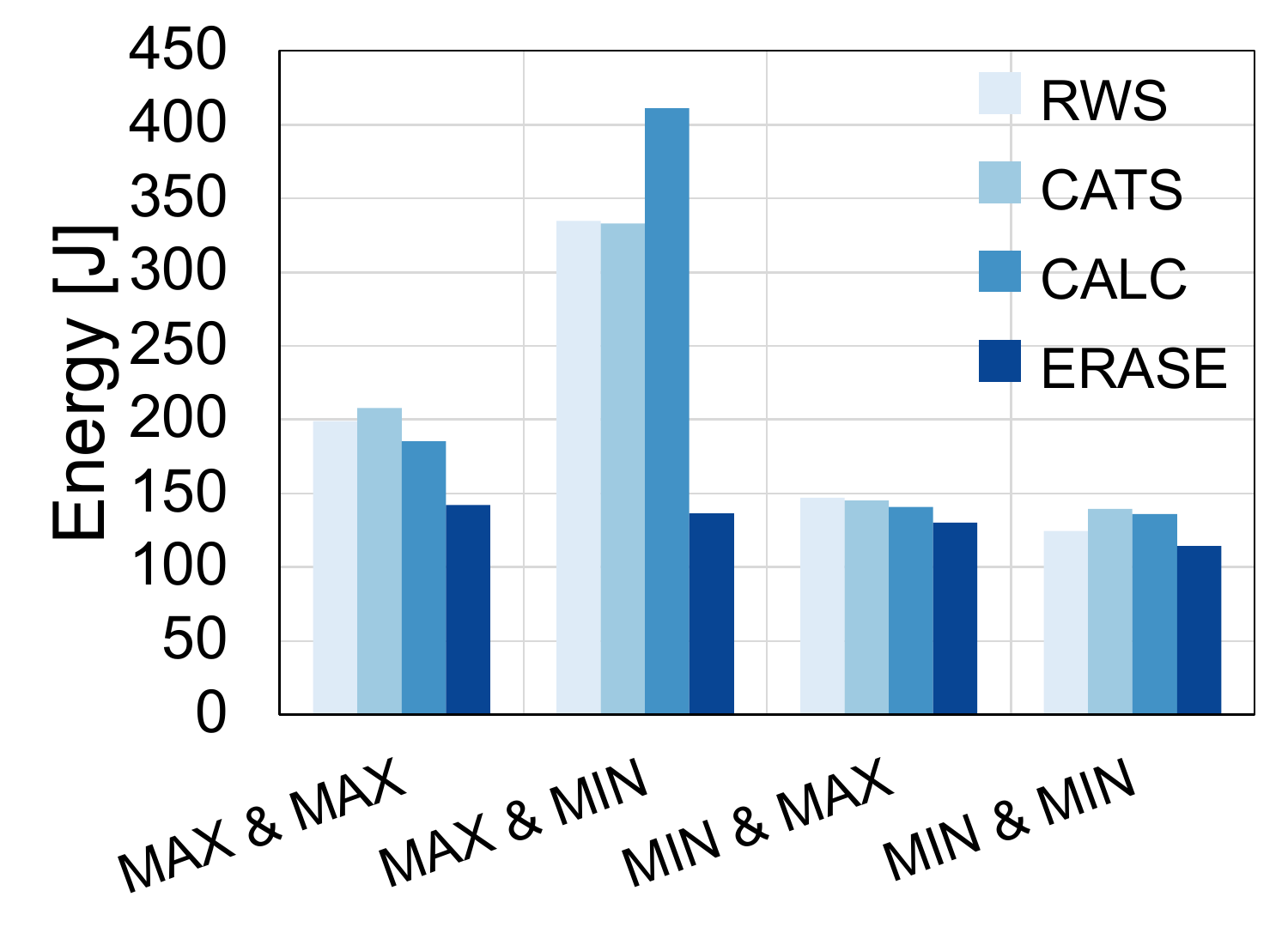}
\label{fig_vgg16}}
\subfloat[\scriptsize{VGG-16 - Execution Time}]{\includegraphics[width=0.25\textwidth]{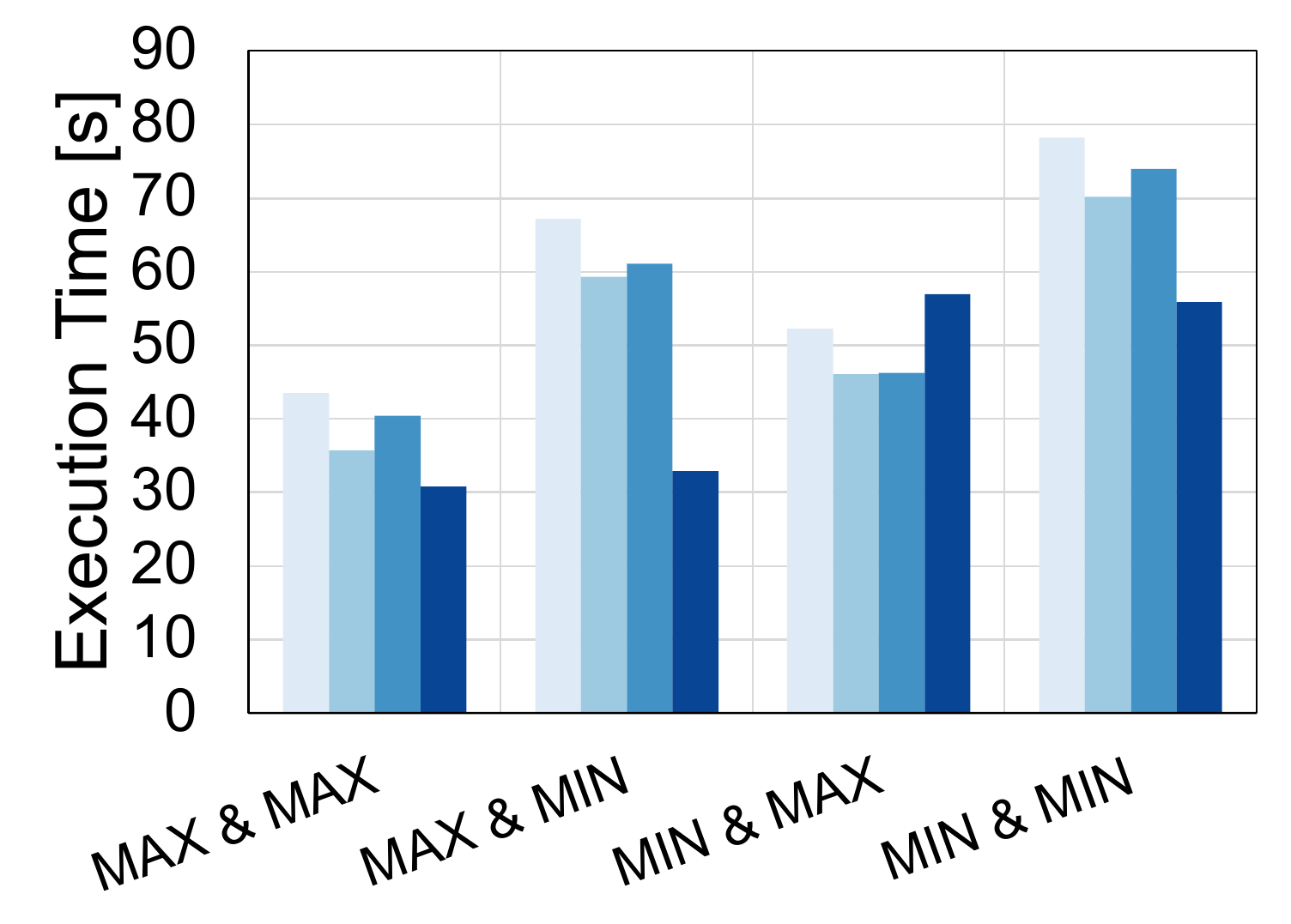}
\label{fig_vgg16_perf}}
\subfloat[\scriptsize{Alya 50K - Energy}]{\includegraphics[width=0.25\textwidth]{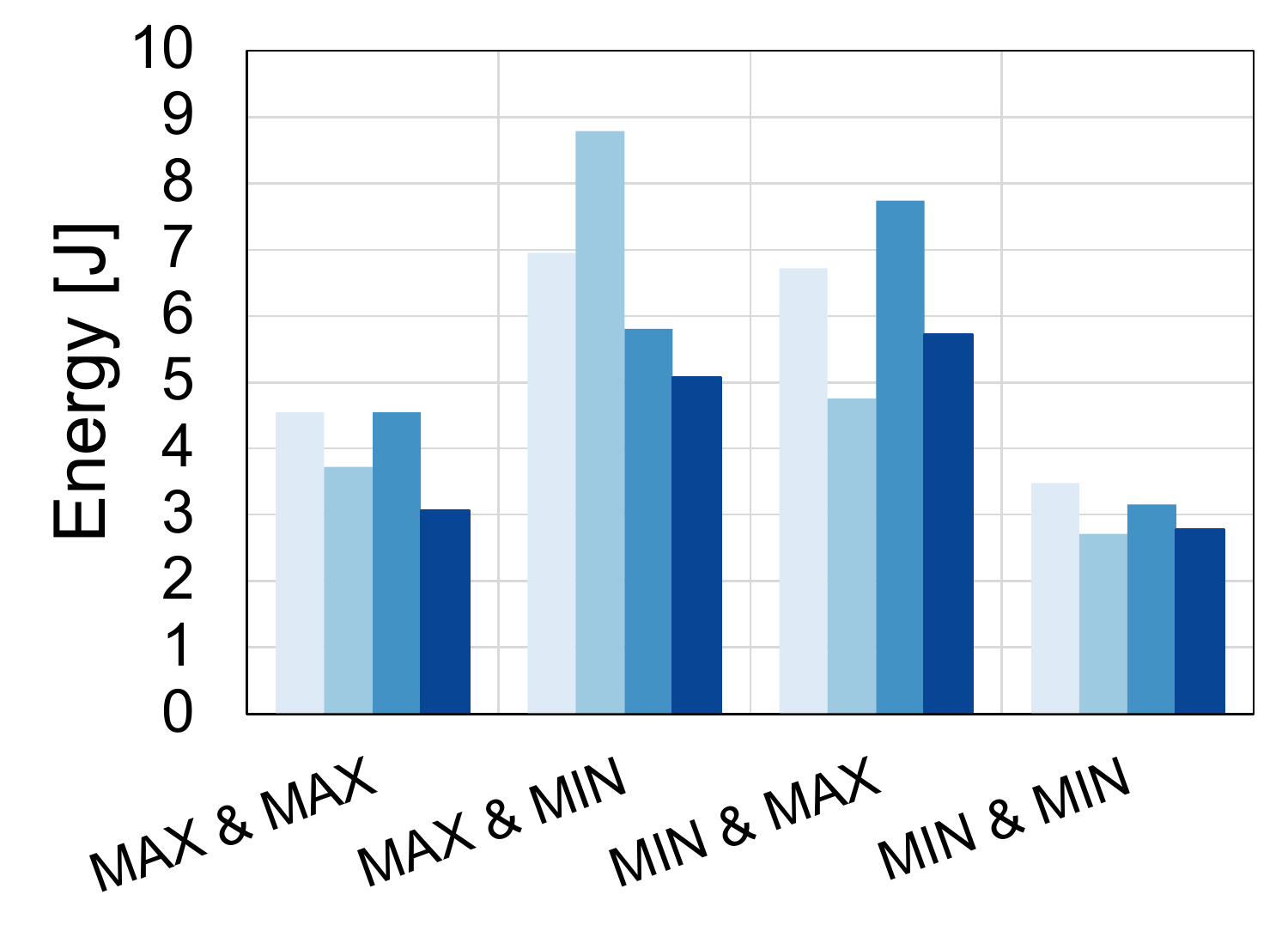}
\label{fig_alya_50k_E}}
\subfloat[\scriptsize{Alya 50K - Execution Time}]{\includegraphics[width=0.25\textwidth]{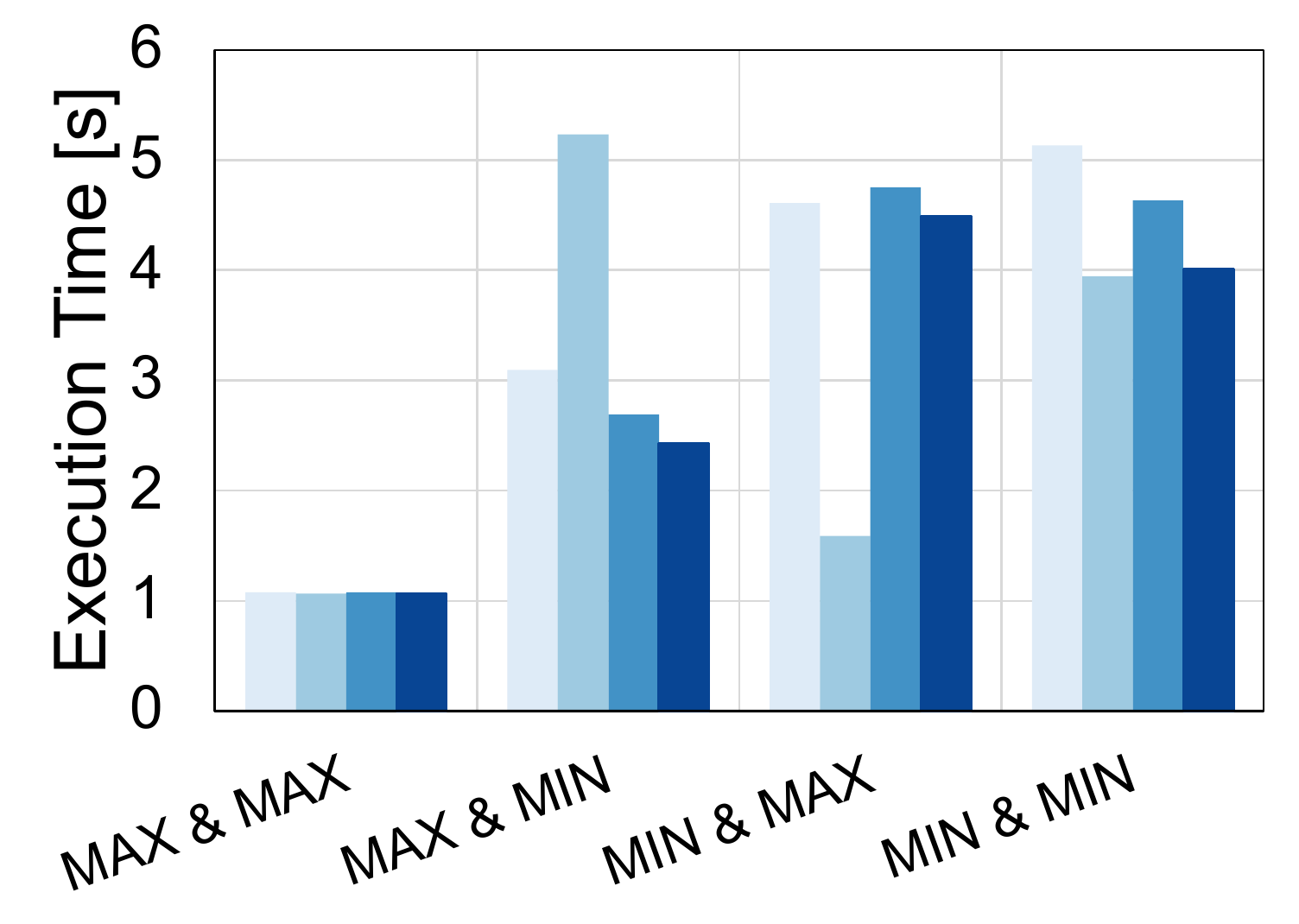}
\label{fig_alya_50k_P}}
\vspace{-3mm}
\subfloat[\scriptsize{Alya 100K - Energy}]{\includegraphics[width=0.25\textwidth]{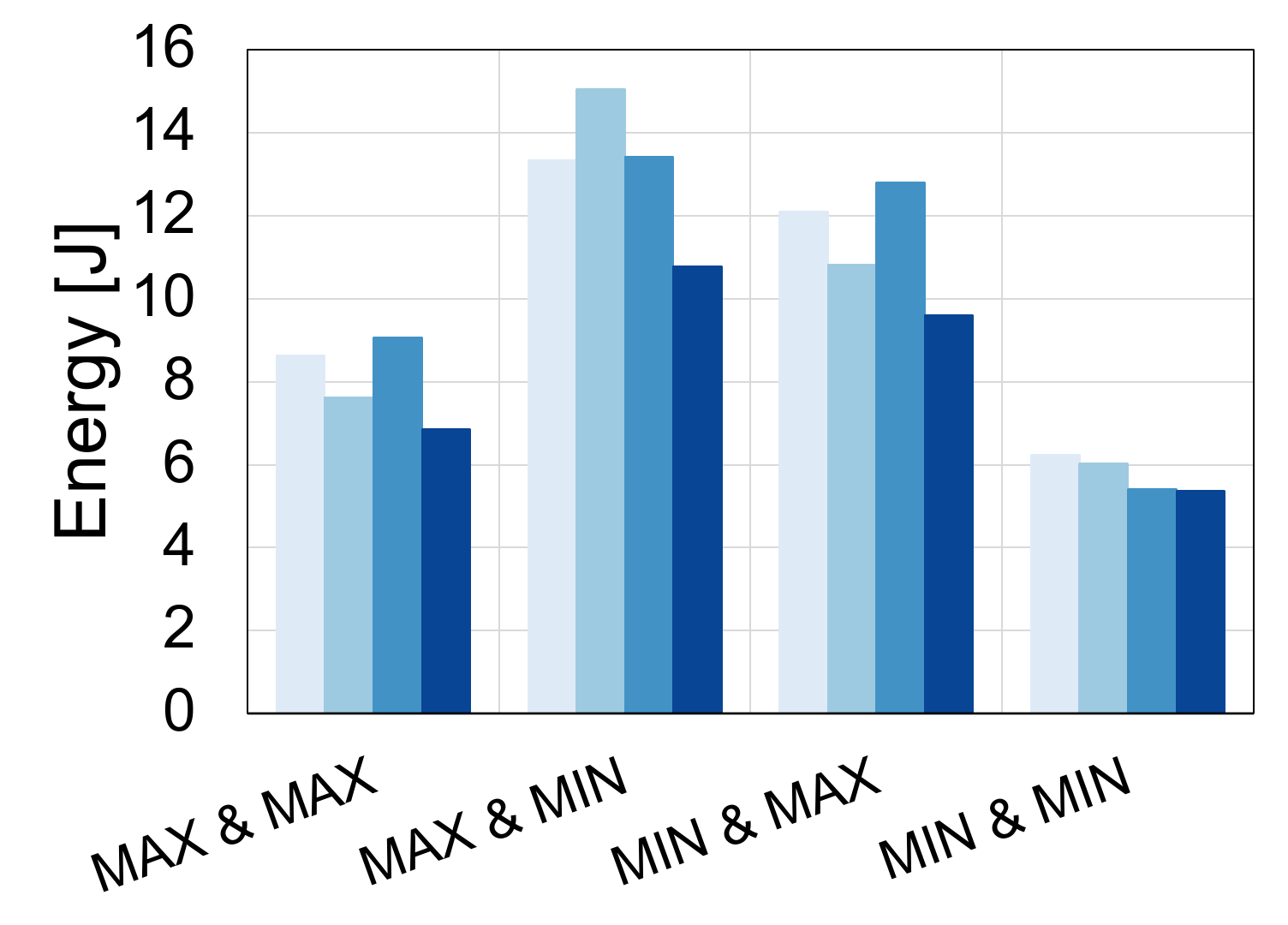}
\label{fig_alya_100k_E}}
\subfloat[\scriptsize{Alya 100K - Execution Time}]{\includegraphics[width=0.25\textwidth]{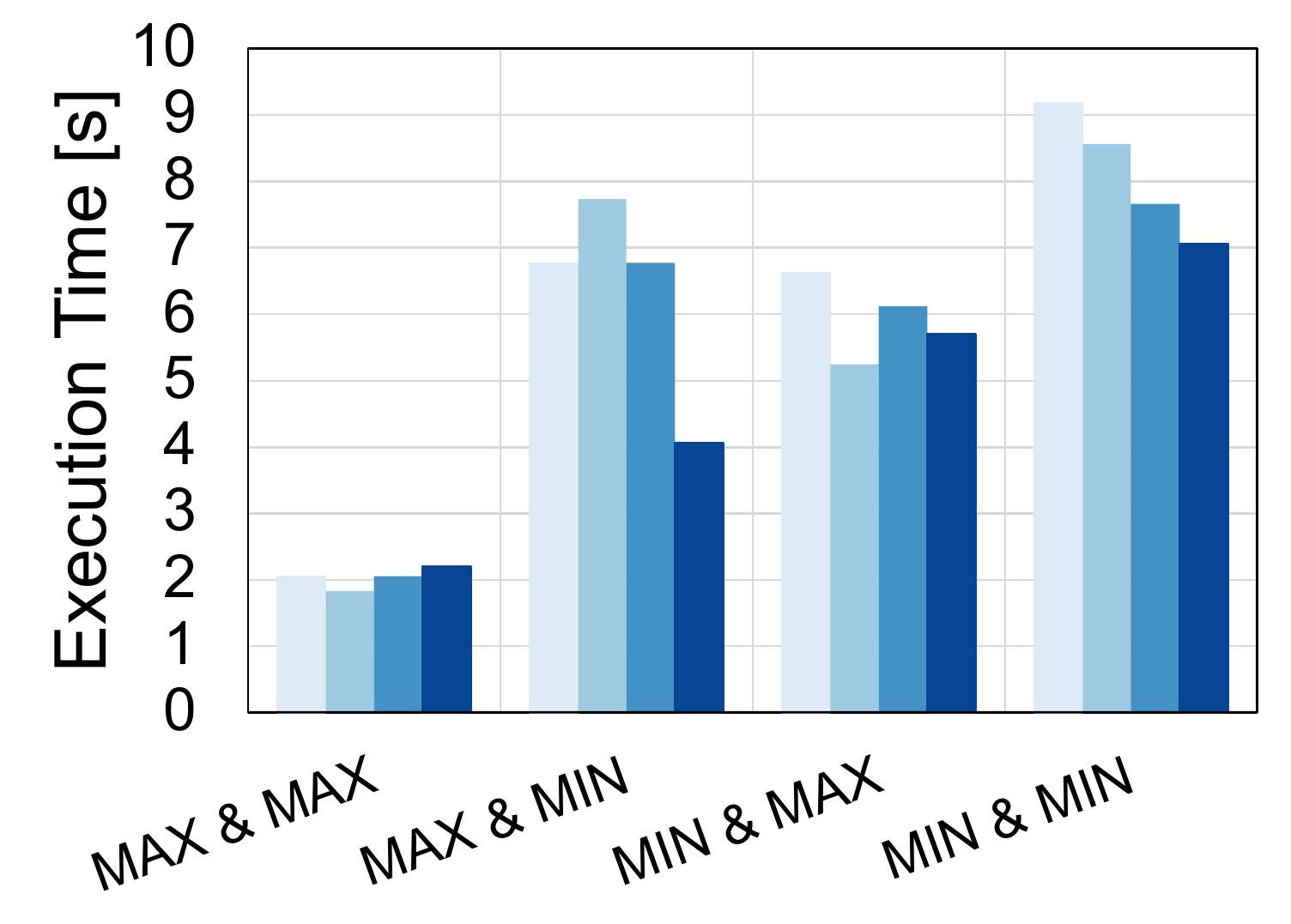}
\label{fig_alya_100k_P}}
\subfloat[\scriptsize{Alya 200K - Energy}]{\includegraphics[width=0.25\textwidth]{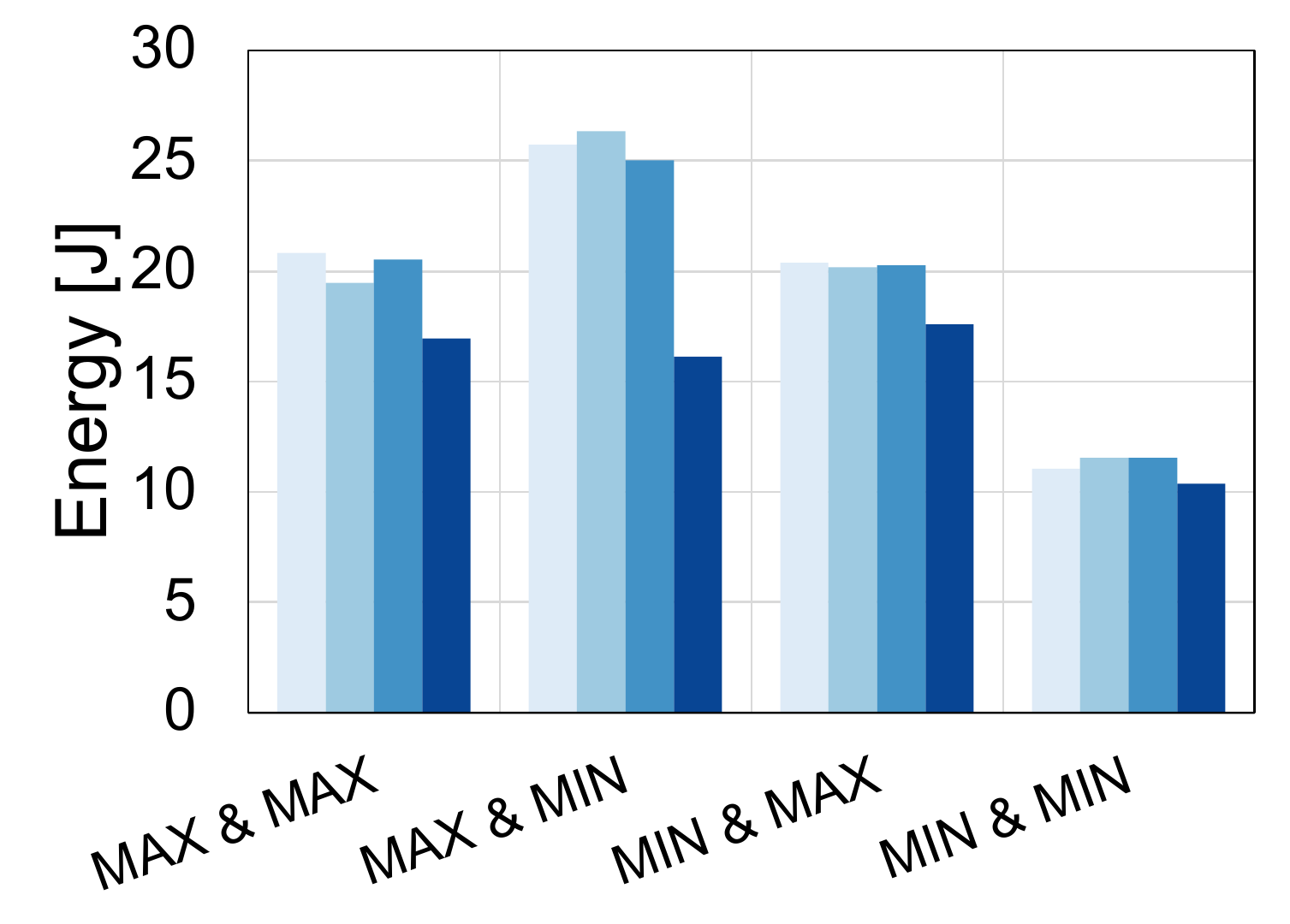}
\label{fig_alya_200k_E}}
\subfloat[\scriptsize{Alya 200K - Execution Time}]{\includegraphics[width=0.25\textwidth]{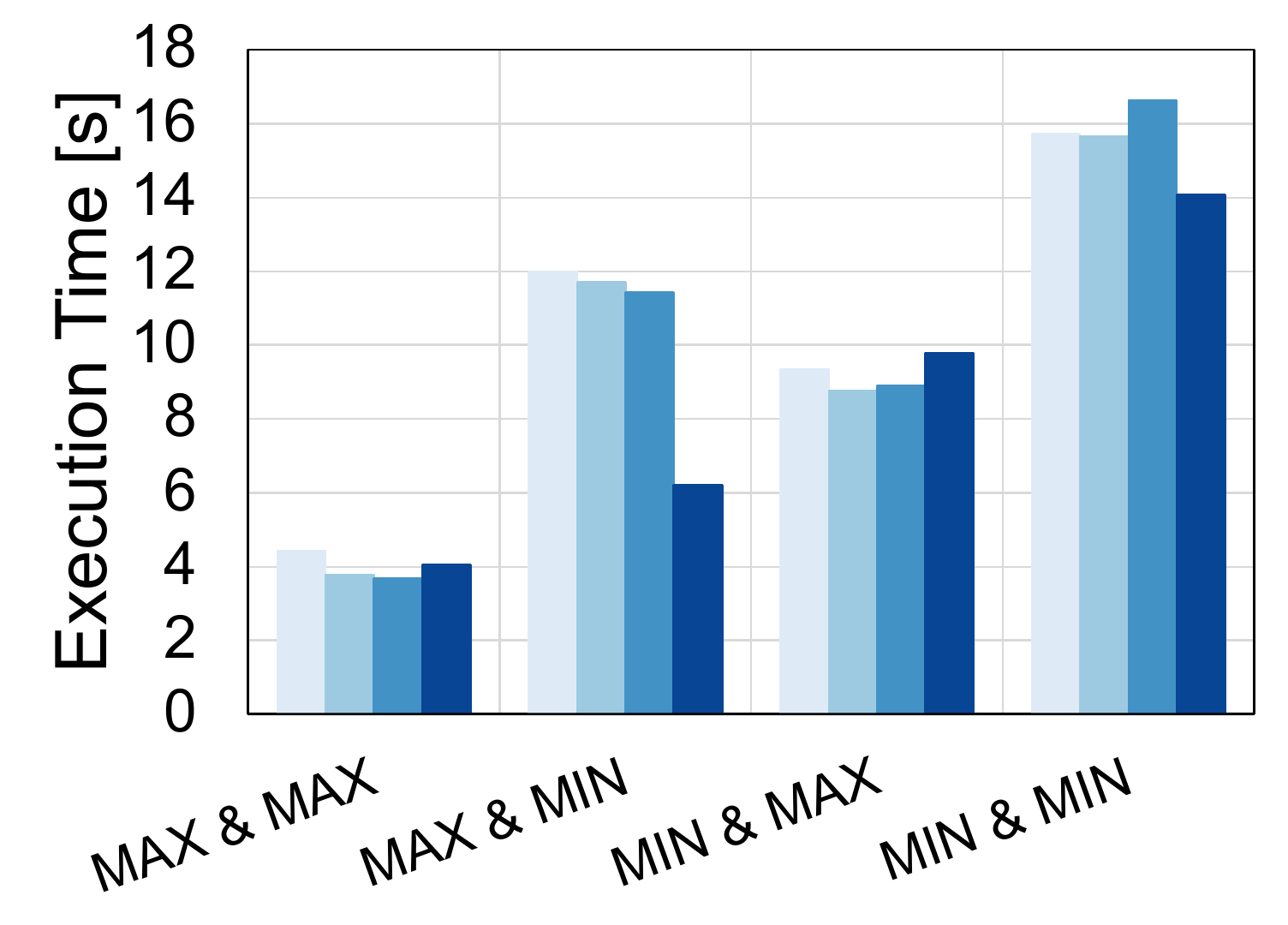}
\label{fig_alya_200k_P}}
\vspace{-3mm}
\subfloat[\scriptsize{DotProduct - Energy}]{\includegraphics[width=0.25\textwidth]{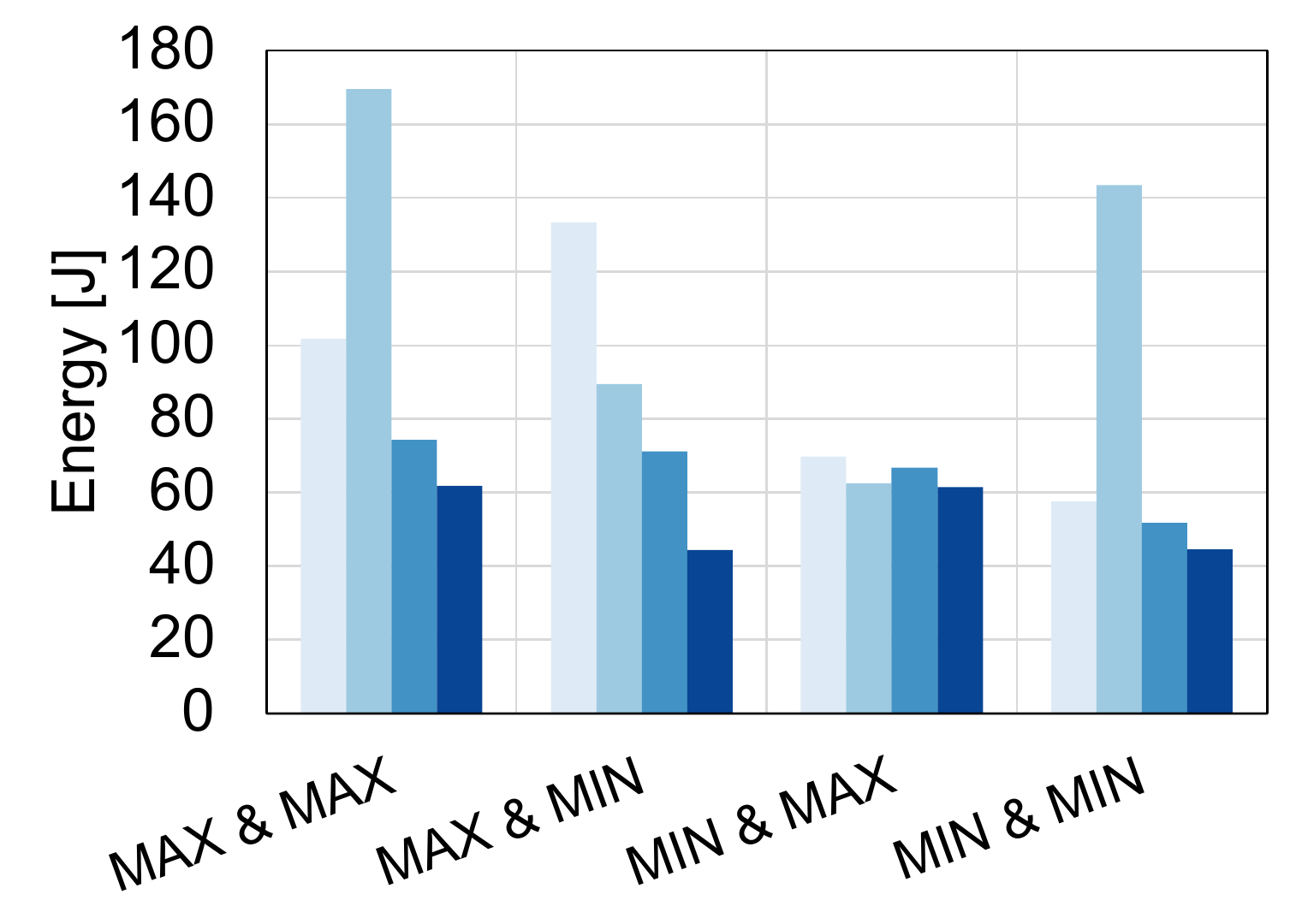}
\label{fig_dotP_E}}
\subfloat[\scriptsize{DotProduct - Execution Time}]{\includegraphics[width=0.25\textwidth]{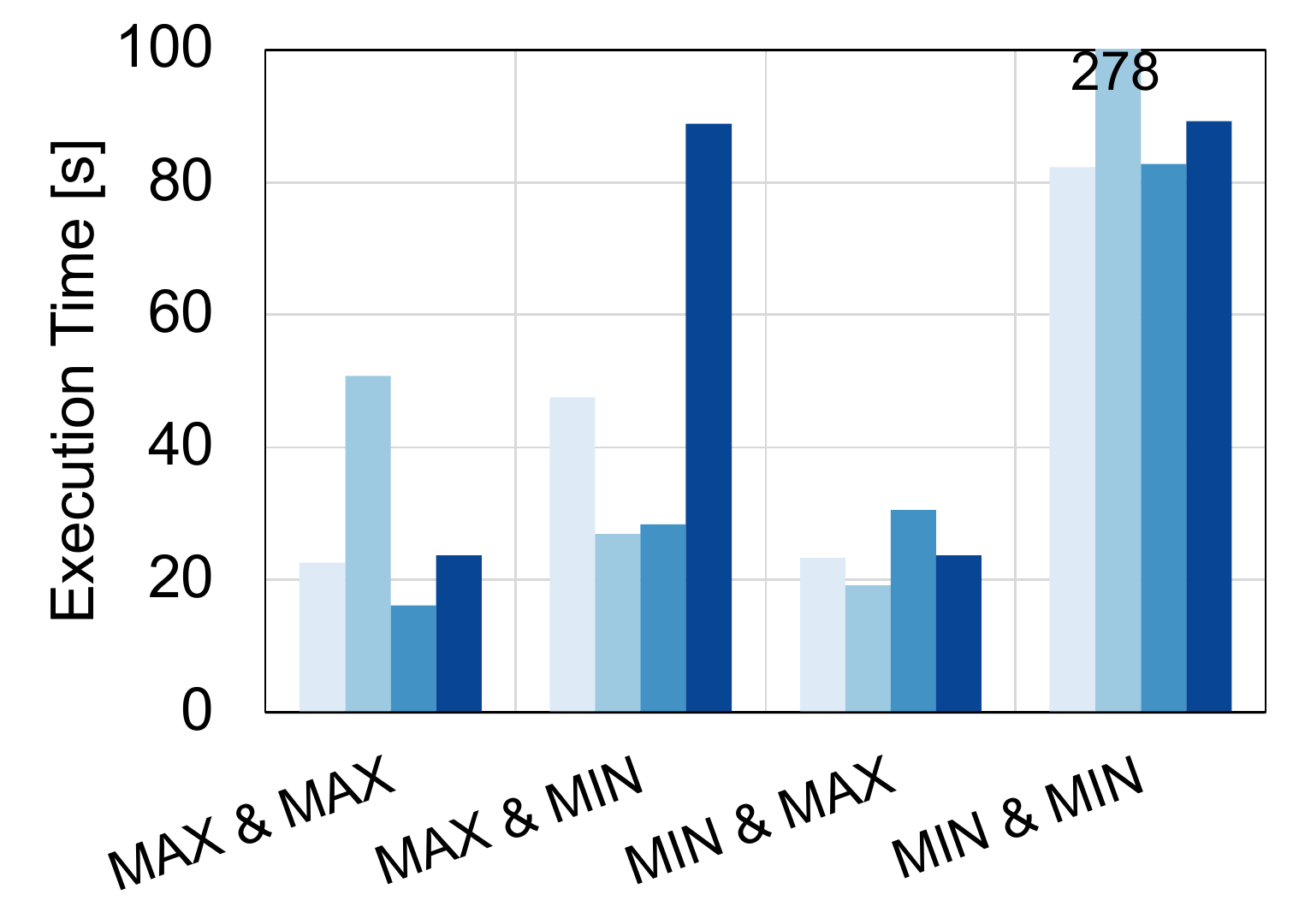}
\label{fig_dotP_P}}
\subfloat[\scriptsize{BioInfection - Energy}]{\includegraphics[width=0.25\textwidth]{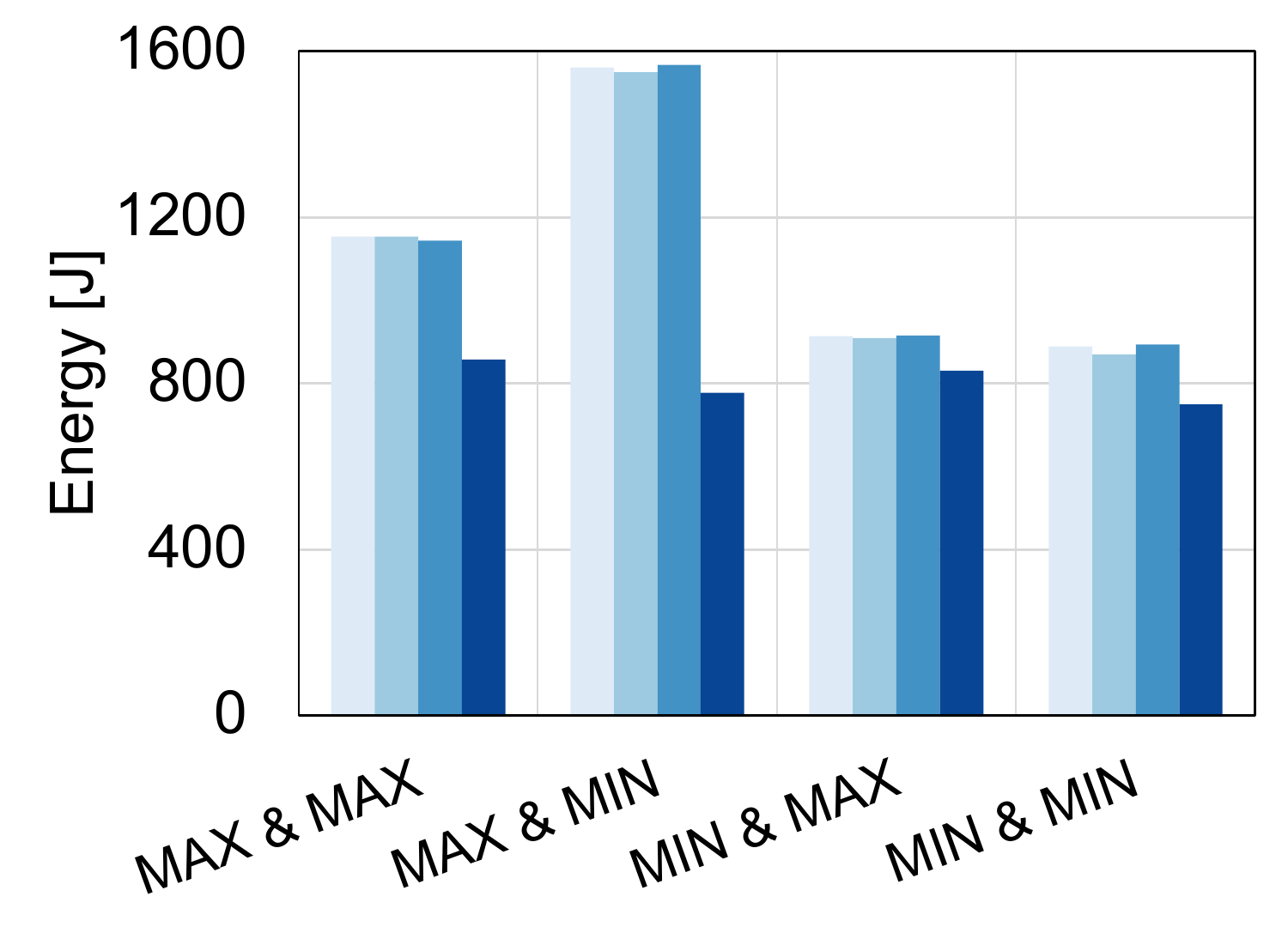}
\label{fig_BioAPP_E}}
\subfloat[\scriptsize{BioInfection - Execution Time}]{\includegraphics[width=0.25\textwidth]{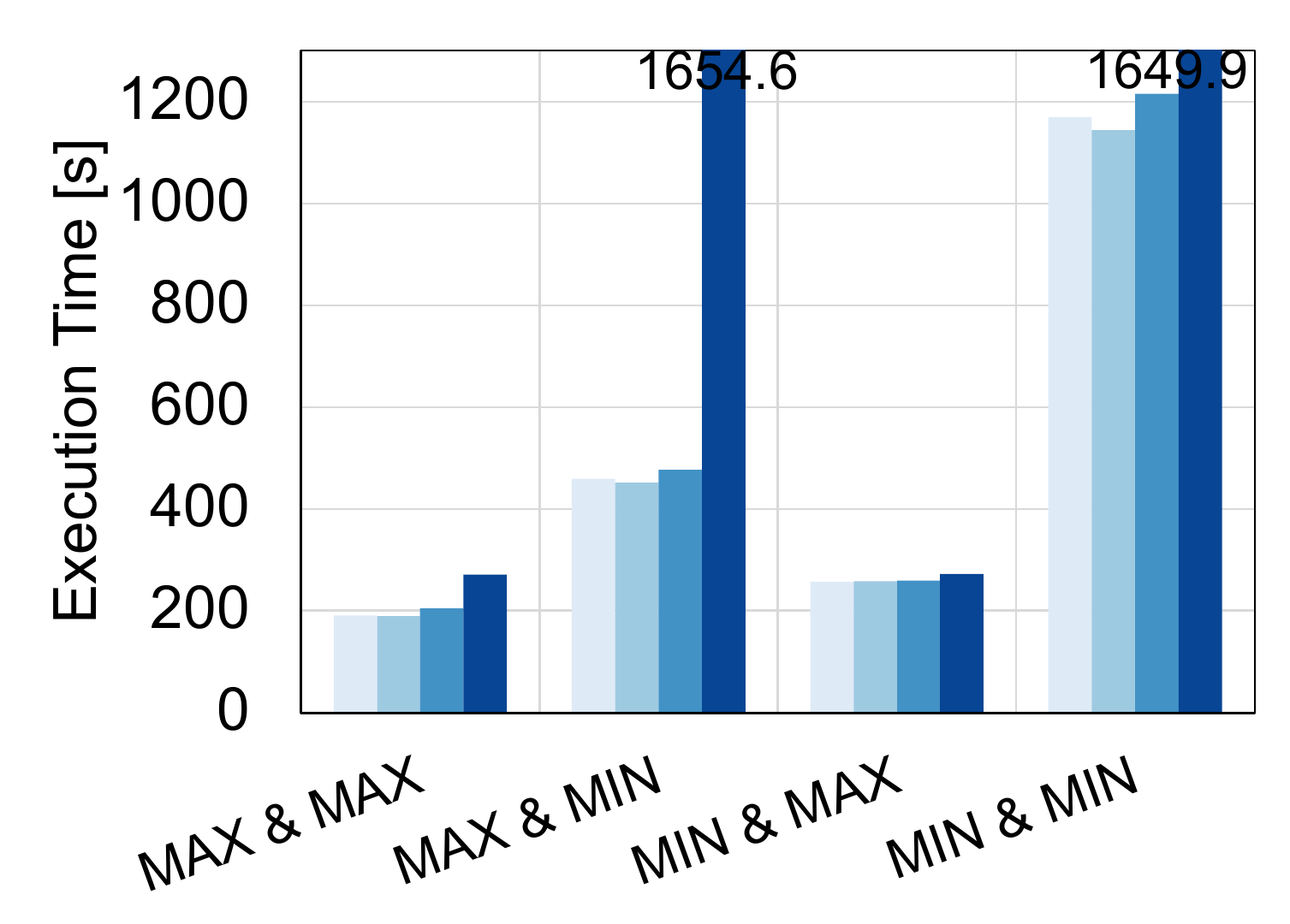}
\label{fig_BioAPP_P}}
\vspace{-3mm}
\caption{Energy consumption and execution time comparison of VGG-16, Alya, Dot Product and Biomarker Infection Research between RWS, CATS, CALC and ERASE on Jetson TX2.}
\vspace{-5mm}
\label{real_applications}
\end{figure}

Figure~\ref{real_applications} shows 
the four schedulers when running VGG-16, Alya, Dot product and Biomarker Infection on TX2. 
The results indicate that ERASE is the most energy efficient scheduler across the majority of environment settings.
In comparison to RWS, CATS and CALC, ERASE achieves between 8\% $\sim$ 59\% energy reduction and outperforms them by 19\% on average across four different environment settings on VGG-16.
For Alya with three different input sizes, ERASE reduces energy consumption by 18\% and improves performance by 17\% on average, compared to the other three schedulers.
Additionally, Dot Product and Biomarker Infection respectively achieve 34\% and 25\% energy savings on average compared to the other three schedulers.
We observe that the energy reduction is significant when the performance gap of two clusters is large (e.g.~52\% on average with Dot Product and configuration MAX(D)\&MIN(A)). 
In contrast, less energy savings are observed when the performance gap is small (e.g.~7\% on average with Dot Product and configuration MIN(D)\&MAX(A)). 
Similar trends can also be observed in other tested cases.

\subsection{Evaluation on Symmetric Platform} \label{haswell_eva}

On a symmetric platform,
the scope of energy efficient task scheduling is dominated by task moldability, i.e.~the selection of an appropriate resource width for executing a task. 
Note that CATS is only designed for asymmetric platforms, thus it is not included in the evaluation on Tetralith.


\subsubsection{Synthetic Benchmarks.}
\begin{figure*}[t]
\centering
\vspace{-6mm}
\subfloat[\scriptsize{MatMul - Energy}]{\includegraphics[width=0.25\textwidth]{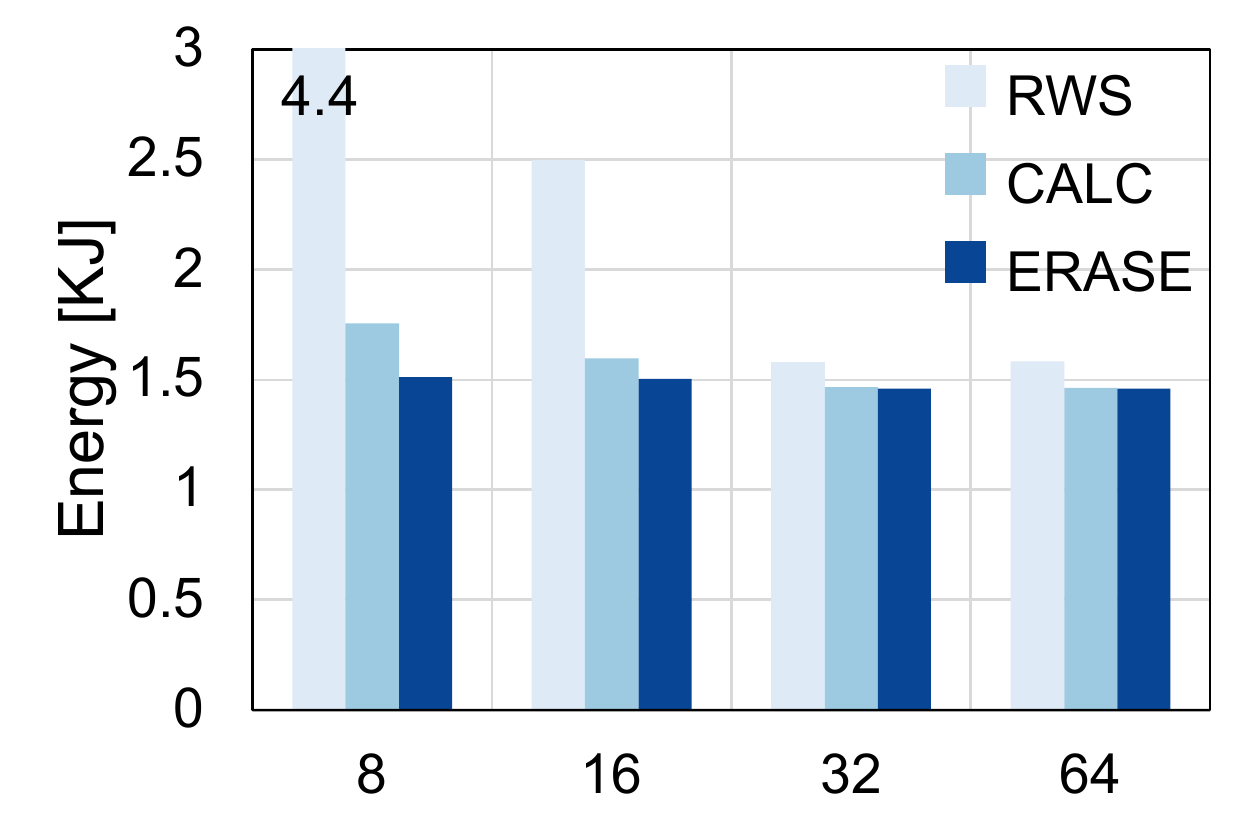}
\label{fig_mm_tetralith_energy}}
\subfloat[\scriptsize{MatMul - Execution Time}]{\includegraphics[width=0.25\textwidth]{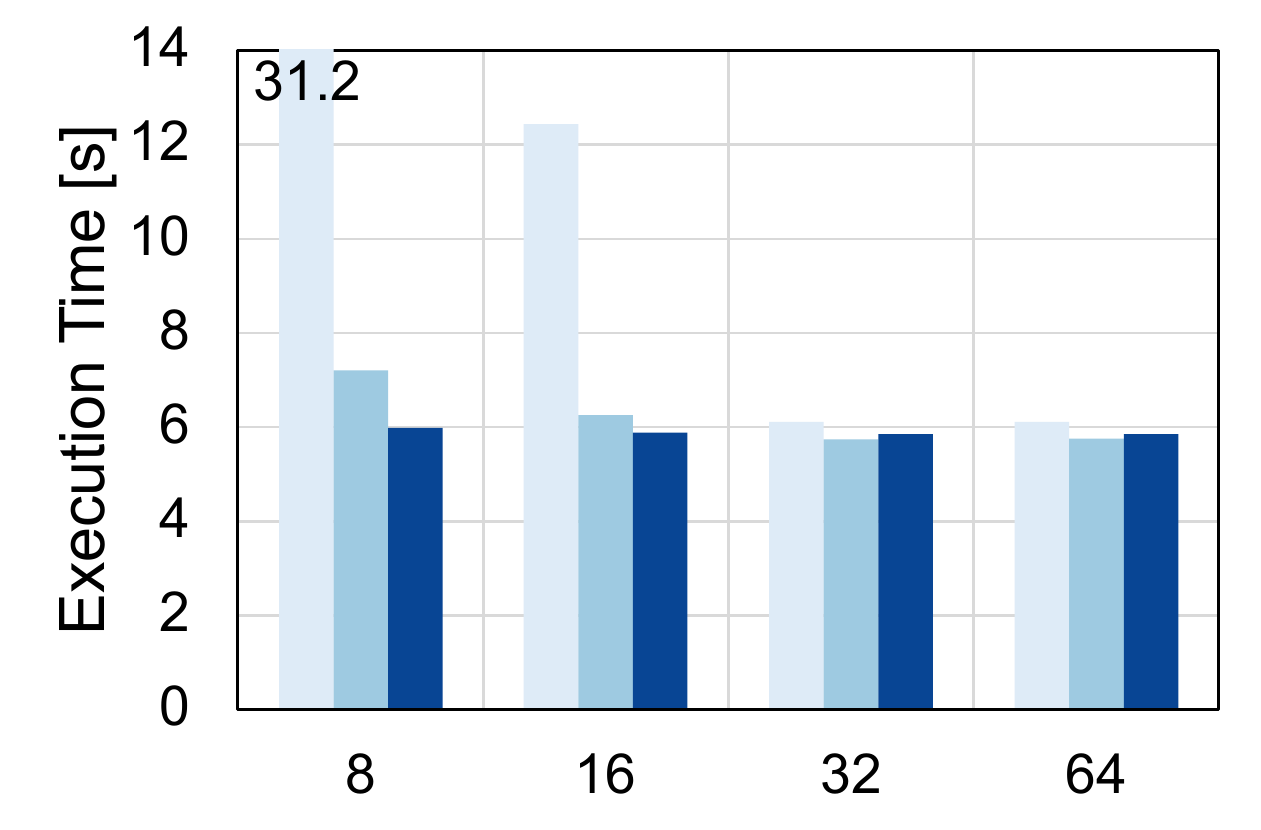}
\label{fig_mm_tetralith_time}}
\subfloat[\scriptsize{Copy - Energy}]{\includegraphics[width=0.25\textwidth]{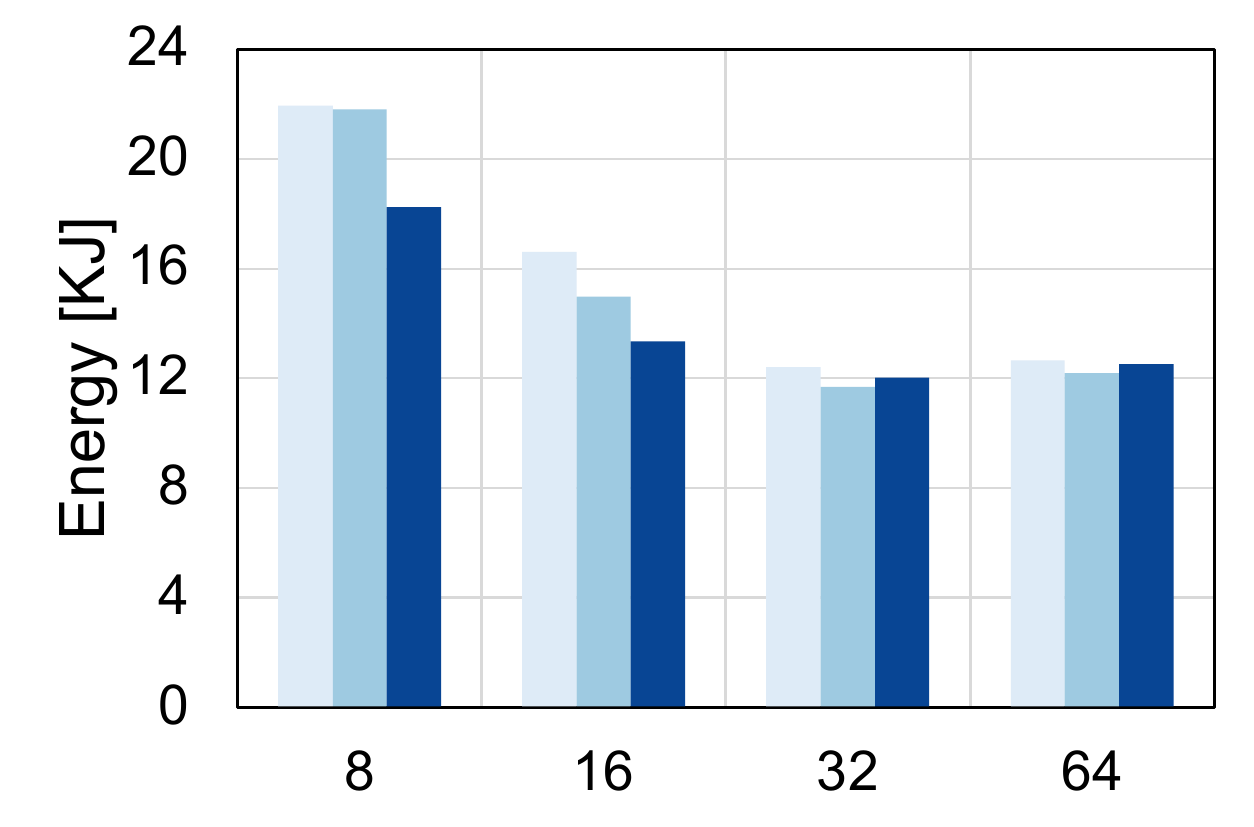}
\label{fig_cp_tetralith_energy}}
\subfloat[\scriptsize{Copy - Execution Time}]{\includegraphics[width=0.25\textwidth]{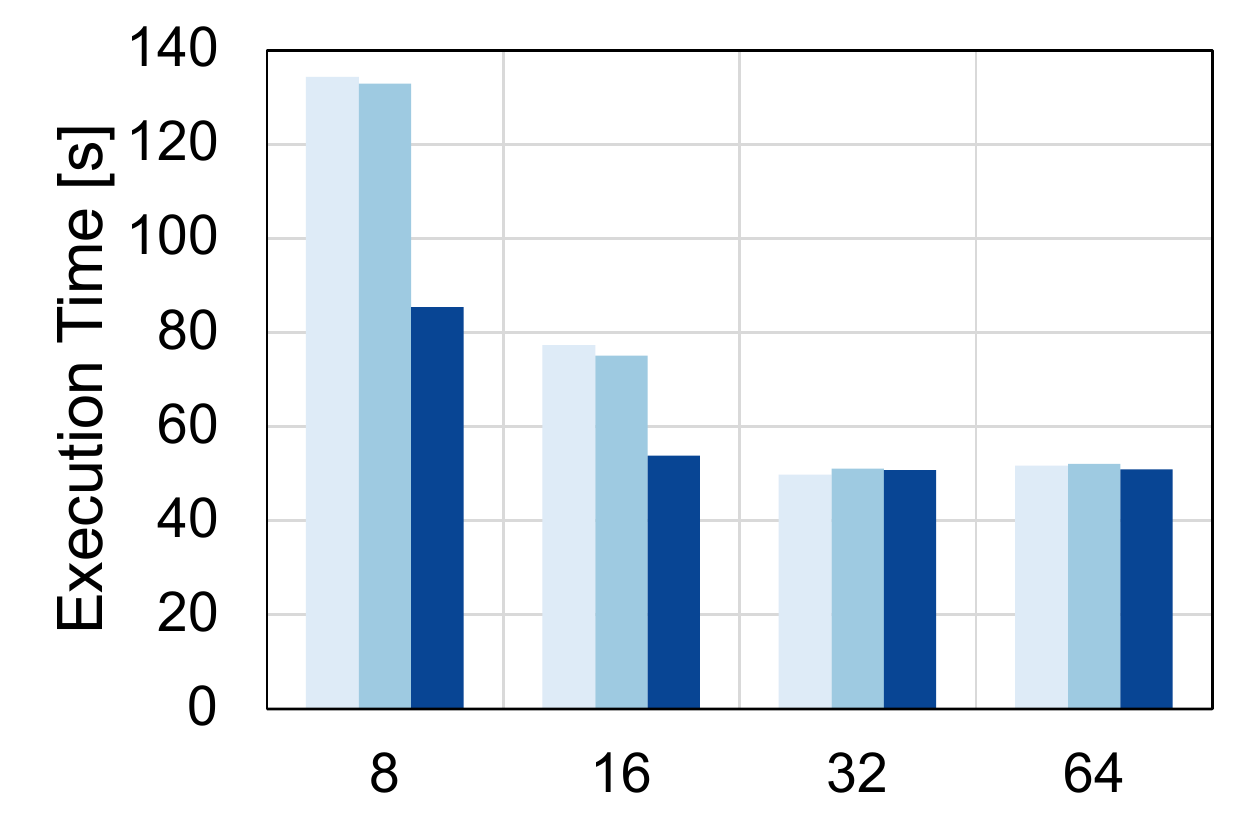}
\label{fig_cp_tetralith_time}}
\\
\vspace{-3mm}
\subfloat[\scriptsize{Stencil - Energy}]{\includegraphics[width=0.25\textwidth]{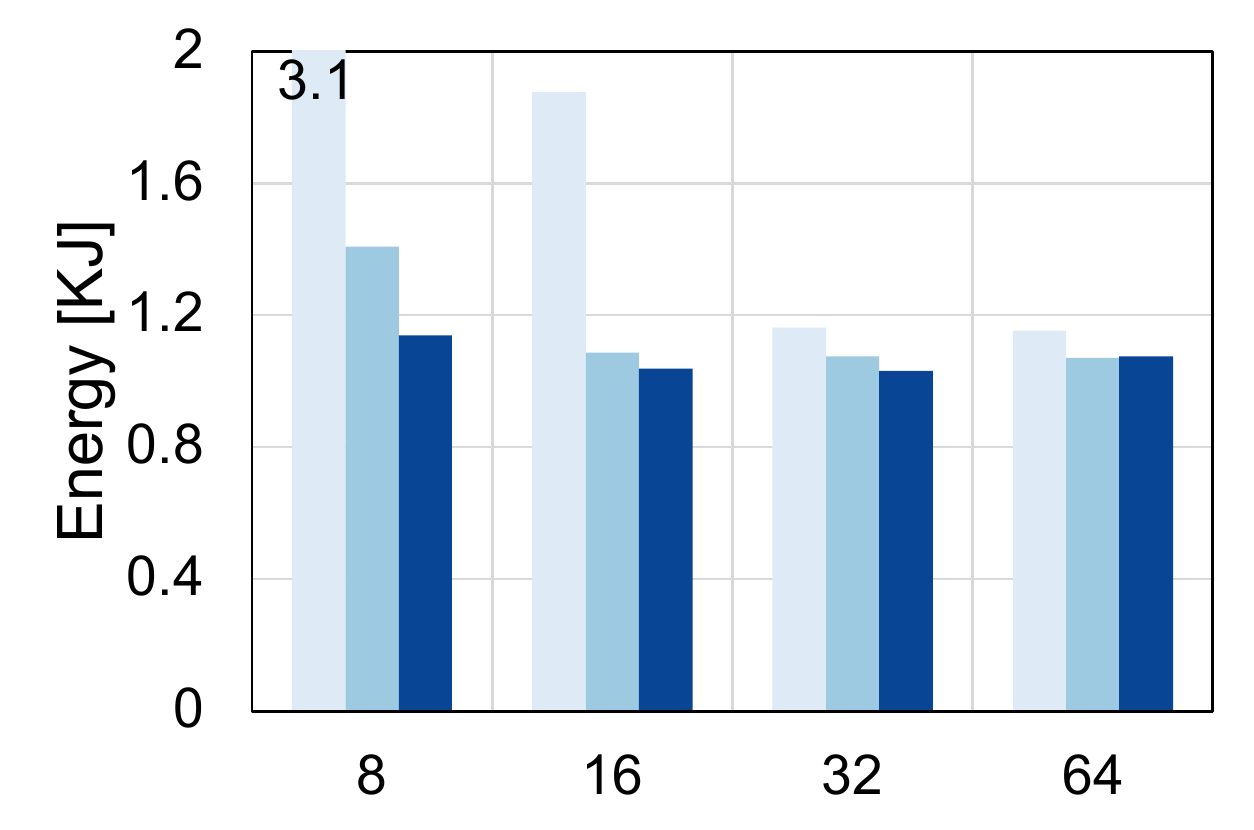}
\label{fig_st_tetralith_energy}}
\subfloat[\scriptsize{Stencil - Execution Time}]{\includegraphics[width=0.25\textwidth]{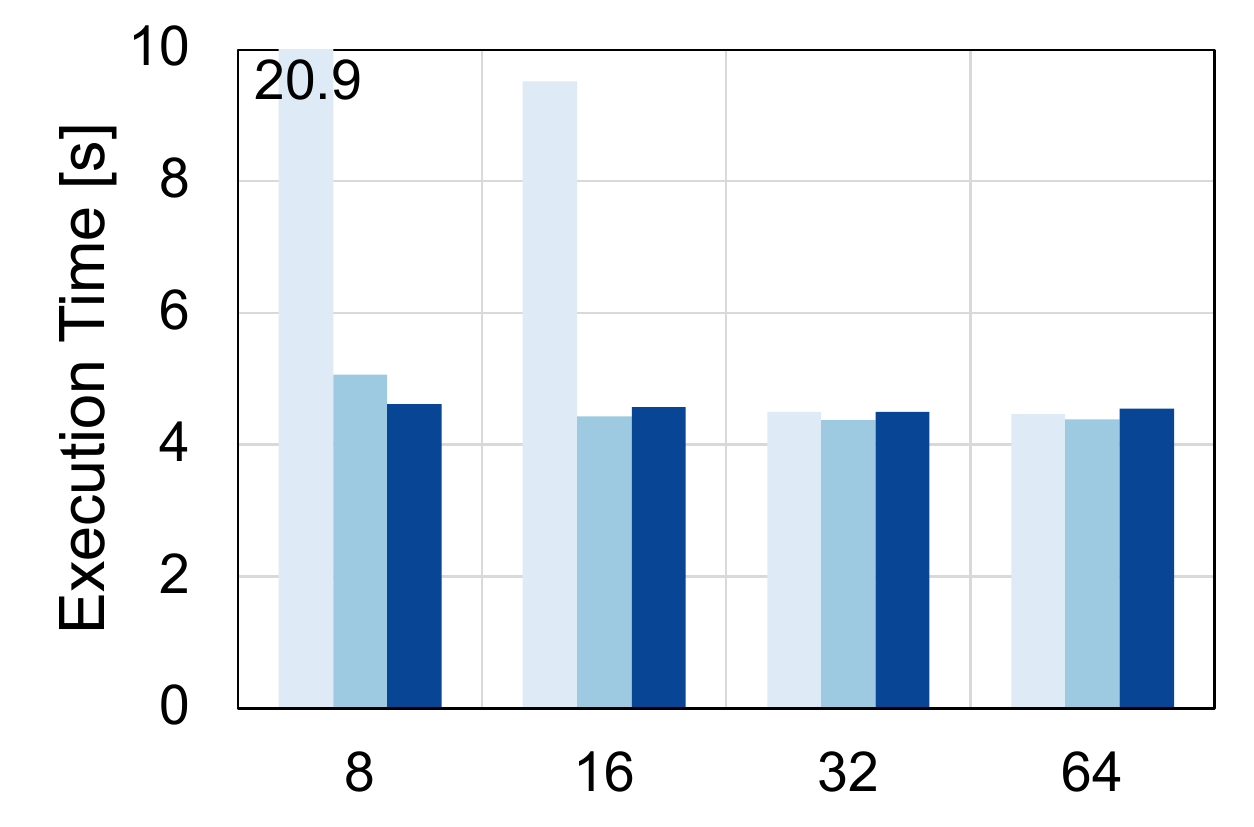}
\label{fig_st_tetralith_time}}
\subfloat[\scriptsize{MatMul(dop=8)}]{\includegraphics[width=0.25\textwidth]{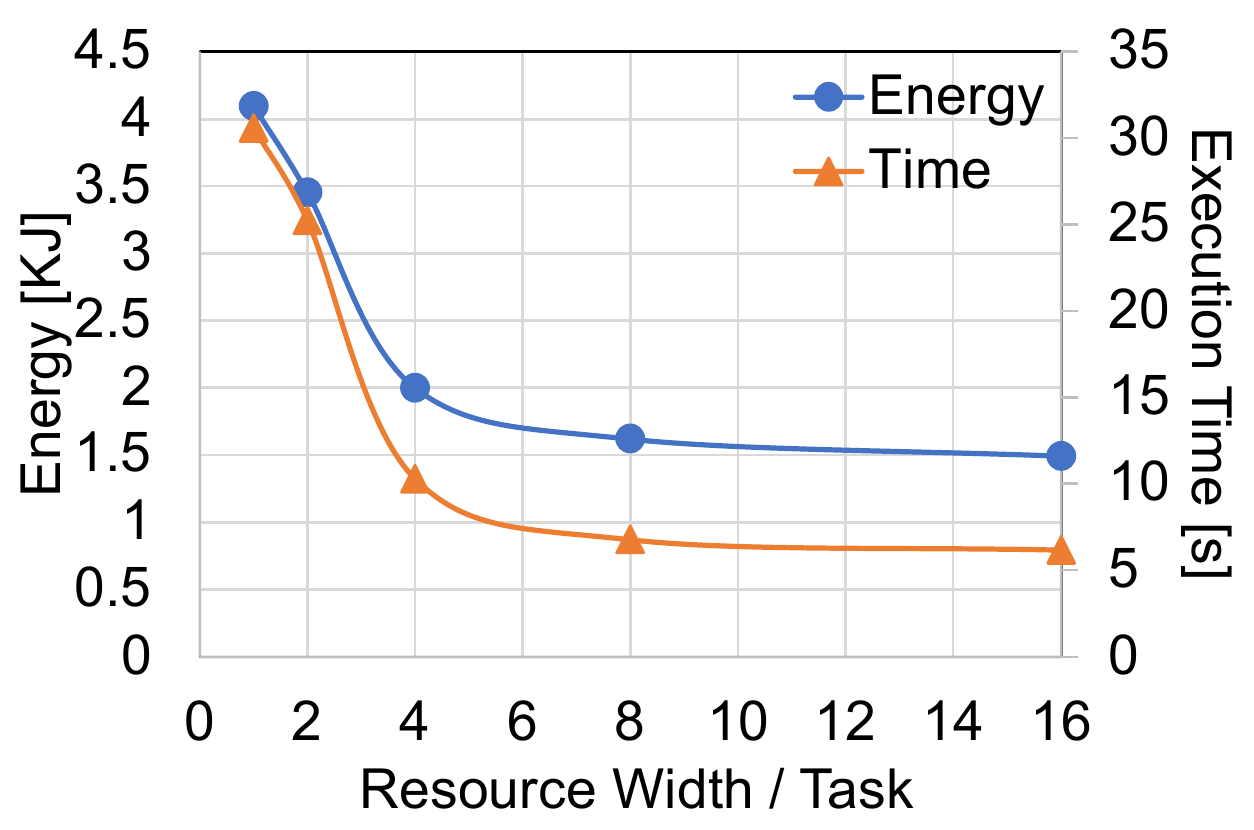}
\label{fig_mm_width}}
\subfloat[\scriptsize{Copy(dop=8)}]{\includegraphics[width=0.25\textwidth]{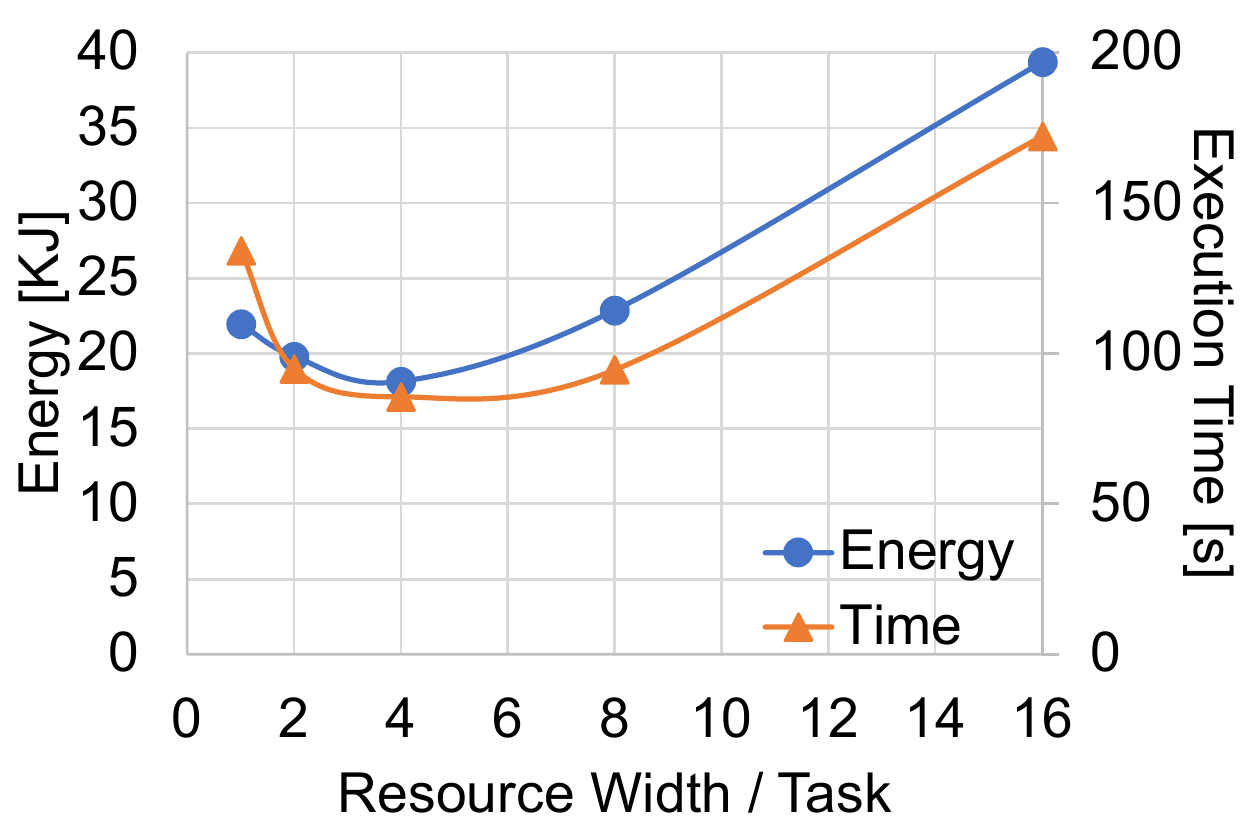}
\label{fig_cp_width}}
\vspace{-3mm}
\caption{Energy and execution Time comparison of RWS, CALC and ERASE on Tetralith.}
\vspace{-4mm}
\label{tetralith_results}
\end{figure*}

Figures~\ref{fig_mm_tetralith_energy} to~\ref{fig_st_tetralith_time} present the energy and execution time comparison of RWS, CALC and ERASE
with $dop$ of 8 to 64.
The results show that ERASE consumes 15\% less energy and achieves 18\% performance improvement on average when compared to the other two schedulers across different $dop$. 
The energy reduction is significant specially when $dop$ is low.

\textbf{Analysis:} The baseline scheduler RWS only utilizes a single core for each task, thus with low task parallelism, idle cores consistently attempt work stealing. 
With our exponential back-off sleep strategy, RWS reduces
energy consumption by 14\% (11\%) on average than baseline with $dop$=8 (16).
Even so, RWS still performs worse than the other two schedulers with low parallelism due to resource under-utilization.
In contrast, CALC permits task moldability and the wider width selection effectively reduces the under-utilization periods. 
Therefore, fewer cores are idle in CALC and the back-off sleep has limited effects on energy consumption.

Figure~\ref{fig_mm_width} and ~\ref{fig_cp_width} presents the energy consumption and execution time comparison, where each MatMul and Copy task is executed with preset specific width (i.e.~1, 2, 4, 8, 16) at $dop$ of 8.
For MatMul, the majority of tasks with CALC execute with resource widths of 8. This minimizes the parallel execution cost, improving  performance by efficiently using available resources and reduces the energy waste from under-utilization compared to RWS. 
In comparison, ERASE executes 99\% of MatMul tasks with a resource width of 16, since execution with width 16 gains more performance benefits with less power costs compared to the execution with width 8. 
This results in the resource width selection of 16 thereby achieving energy minimization and also faster execution (as shown in Figure~\ref{fig_mm_width}). 
In the case of Copy, more than 97\% of tasks in CALC execute with width 1 for parallel execution cost minimization, since moldable execution does not bring linear performance improvement. 
Therefore, the energy and execution time are almost the same as RWS.
We observe that ERASE 
is able to figure out the best resource width for different DAG parallelism, specifically width 4 (2) for $dop$=8 (16).
As shown in Figure~\ref{fig_cp_width}, execution with lower widths (i.e.~1 and 2) does not effectively utilize all resources and sleeping does not bring enough energy benefits to compensate for the performance loss. 
In contrast, execution with higher width (i.e.~8 and 16) leads to high power costs with limited performance improvement. 

\subsubsection{Applications.}
The evaluation of Heat and Sparse LU showed in Table~\ref{tab:slu_heat} indicates again that ERASE consumes less energy than other scheduling policies.

\textbf{Analysis:} We compare the distribution of resource widths for tasks with CALC and ERASE when running Heat to understand why ERASE performs better in comparison to the other schedulers.
In the case of CALC, 89\% of tasks execute with a resource width of 1 due to reduced parallel execution cost associated with the execution. 
In the case of ERASE, the majority of tasks execute with wider resource widths and the numbers of tasks that execute with a resource width of 4 and 16 are almost equal.
This happens because in Heat there are two different types of kernels with different scalability characteristics: \textit{jacobi} and \textit{copy}. 
The \textit{jacobi} kernel is compute-bound and while \textit{copy} is memory-bound, 
ERASE is able to determine the best width for each kernel.
During the run, 99\% of \textit{jacobi} tasks execute with width 16,
while 98\% of the \textit{copy} tasks execute with width 4.

\begin{table*}[h]
\scriptsize
\vspace{-3mm}
\caption{Energy and execution time of RWS, CALC and ERASE with Sparse LU and Heat on Tetralith}
\vspace{-3mm}
\begin{center}
\begin{tabular}{p{2.3cm}p{.8cm}p{.8cm}p{.8cm}p{.5cm}p{.8cm}p{.8cm}p{.8cm}} \toprule
\textbf{Application} & \multicolumn{3}{c}{\textbf{Sparse LU}} & & \multicolumn{3}{c}{\textbf{Heat}} \\ \midrule
& RWS & CALC & ERASE & & RWS & CALC & ERASE \\ \midrule
\textbf{Energy[J]} & 15446 & 11597 & 10489 & & 10853 & 10729 & 8972 \\ \midrule
\textbf{Execution Time[s]} & 63.2 & 55.1 & 41.9 & & 54.2 & 55.7 & 62.6  \\ \bottomrule
\end{tabular}
\label{tab:slu_heat}
\vspace{-4mm}
\end{center}
\end{table*}

\subsection{DVFS Awareness of ERASE} \label{dyna_reaction}
\begin{figure}[!t]
\centering
\vspace{-5mm}
\subfloat[\scriptsize{ERASE without reacting to dynamic frequency changes}]{\includegraphics[width=\textwidth, left]{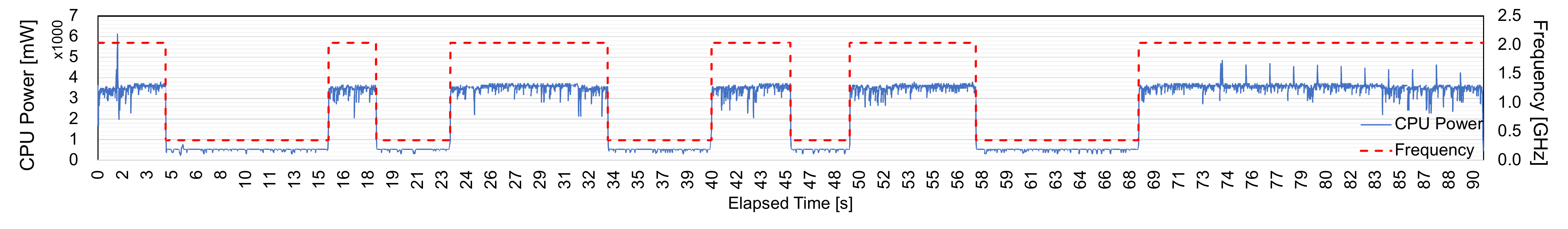}
\label{fig_noreaction}}
\vspace{-3mm}
\\
\subfloat[\scriptsize{ERASE reacting to dynamic frequency changes}]{\includegraphics[width=\textwidth, left]{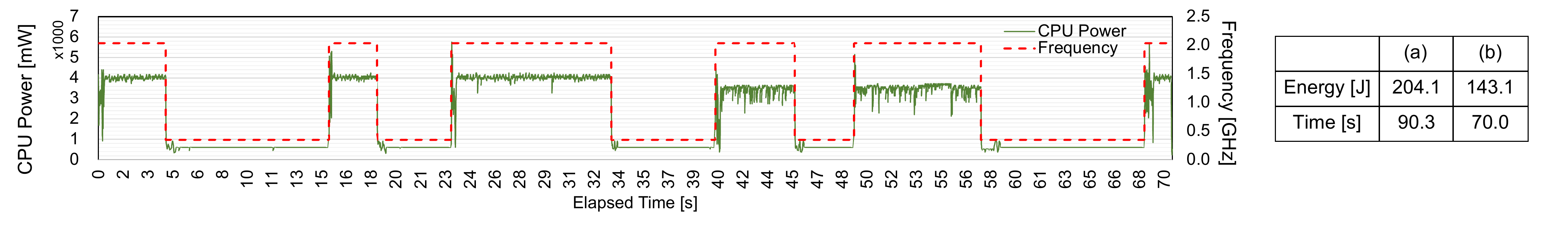}
\label{fig_reaction}}
\vspace{-3mm}
\caption{The comparison of ERASE execution with MatMul between DVFS awareness and non-awareness.}
\vspace{-4mm}
\label{fig_erase_dynaDVFS}
\end{figure}

To analyze the adaptability of ERASE to externally controlled DVFS changes, we execute MatMul on Jetson TX2 with randomly inducing frequency changes (shown as red dotted curves in Figure~\ref{fig_erase_dynaDVFS}).
We generate the DVFS changes externally in a different running process and alternate the frequencies for both Denver and A57 clusters between the highest and the lowest (2.04, 0.35 GHz) by writing the specific frequency into system files and then sleeping a random time between 3s to 12s.

Figure~\ref{fig_noreaction} shows the case where ERASE models are unaware of the frequency change and keep using the same model settings detected at the beginning of execution.
While Figure~\ref{fig_reaction} shows the case where ERASE models detect the dynamical frequency changes and automatically adapt to the new frequency instantly.
The results show that without DVFS awareness, the majority of tasks (>98\%) are executed with sub-optimal execution place settings. 
In comparison to the DVFS unaware case, ERASE achieves 30\% of energy savings and 23\% performance improvement when the models are capable of reacting and adapting to DVFS and recalculate the best execution places accordingly. 
As shown in Figure~\ref{fig_reaction}, at the beginning of the execution, CPU power consumption shows a short period of variation ranging from 1W to 5W, indicating that ERASE performance model is in the training phase.
After that, when the frequency changes (either from lowest to highest or from highest to lowest), performance models are retrained instantly, resulting in CPU power variations that can only be observed over a short interval immediately following the change before stabling.
After training the performance model, the corresponding power model for the newly detected frequency is used for new execution place predictions.

To evaluate the overhead associated with adapting to externally controlled DVFS
and to determine how quickly ERASE can react to DVFS, we compare energy and execution time between ERASE and the ideal scenario where the best execution place for each task is determined by offline analysis thereby precluding the use of power and performance models.
Our evaluation shows that the differences in energy and execution time between ERASE and the ideal case are 0.6\%.

\subsection{Model Accuracy} \label{acc_eval}
We evaluate the accuracy of the performance model and the power model used in ERASE using benchmarks listed in Section~\ref{syntheticdag}.
Moreover, in order to explore the accuracy of combining the two models, we evaluate the accuracy of predicted energy consumption.
We adopt the statistical metric $\mathbf{MAPE}$ (Mean Absolute Percentage Error) as a measure of accuracy, which is computed as follows:
\vspace{-2mm}
\begin{equation}  \footnotesize
    \vspace{-1mm}
    MAPE = \frac{\sum\limits_{i=0}^{\tau-1} | \frac{real_{i} - predicted_{i}}{real_{i}}|}{\tau} \times 100, i = task\ 0, ..., \tau-1.
\end{equation}

\textbf{Performance Model Accuracy:}
We evaluate the prediction accuracy by comparing the predicted execution time obtained from the performance model to the actual execution time of each task.
Our evaluation on all seven benchmarks shows that the average MAPE for execution time prediction is 2.2\%. In other words, the model achieves 97.8\% prediction accuracy on average.

\begin{figure*}[!t]
\centering
\vspace{-4mm}
\subfloat[\scriptsize{on Denver cluster}]{\includegraphics[width=0.5\textwidth]{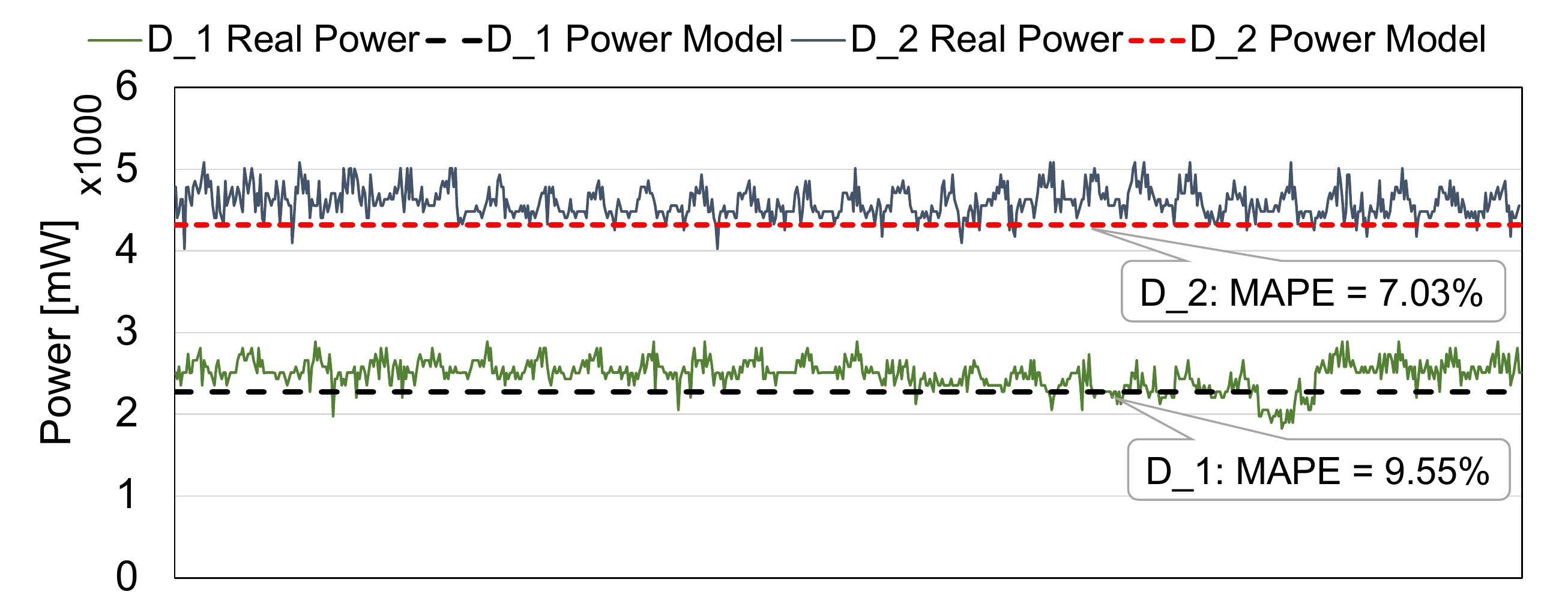}
\label{fig_sparselu_denver}}
\subfloat[\scriptsize{on A57 cluster}]{\includegraphics[width=0.5\textwidth]{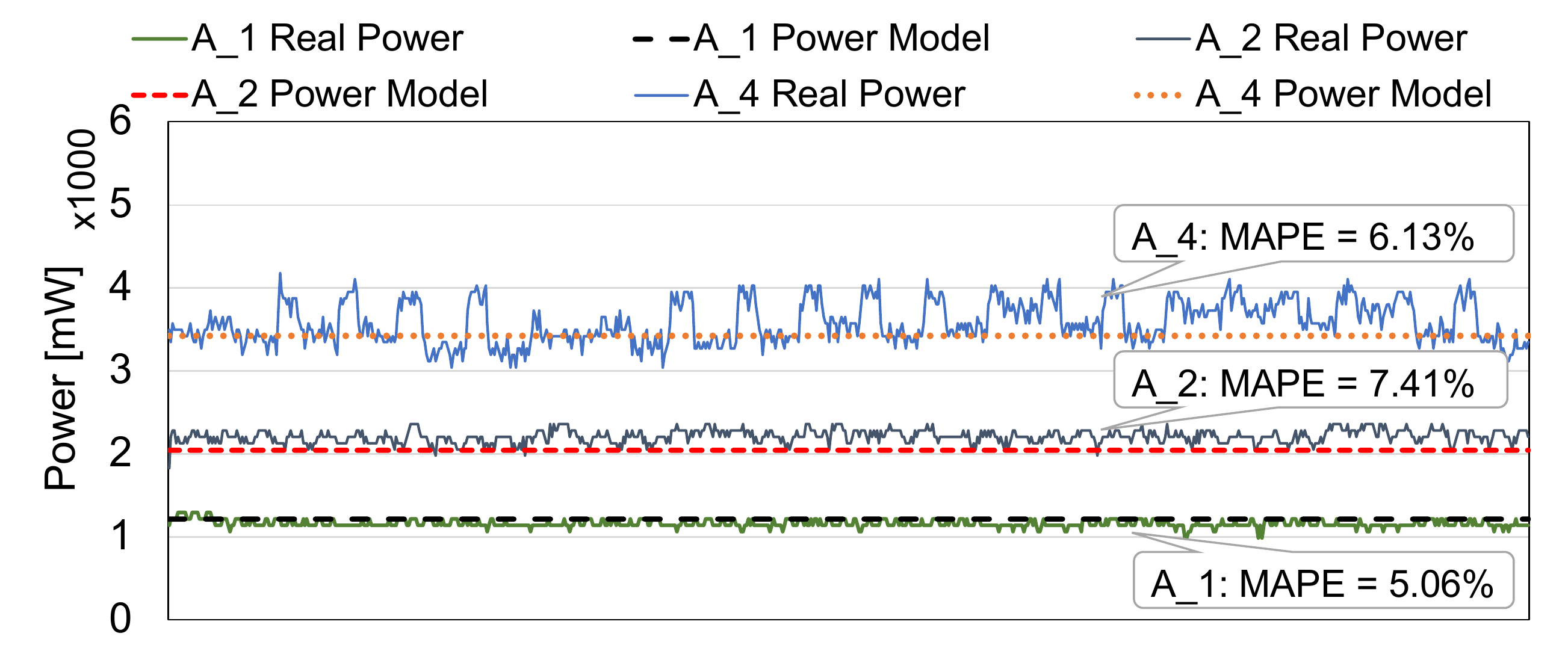}
\label{fig_sparselu_a57}}
\vspace{-3mm}
\caption{Power model accuracy evaluation by comparing real power and modeled power when running Sparse LU on different execution places of Jetson TX2 (frequency=2.04GHz).}
\vspace{-3mm}
\label{fig_powermodel}
\end{figure*}

\textbf{Power Model Accuracy:}
We evaluate the prediction accuracy 
by comparing the power consumption estimates obtained from the power profile to each power sample collected from the embedded INA3221 power sensor during the benchmark execution.
Table~\ref{tab:poweraccuracy} shows the prediction deviations of four evaluated benchmarks from three categories, including the results measured in maximum frequency, minimum frequency and by averaging the deviations measured with all 12 possible frequency degrees on the Jetson TX2 platform.
For task type categorization, we apply the methodology described in Section~\ref{charac} to empirically determine the following thresholds for Jetson TX2: \textit{AI}<6.25 for memory-boundness, 6.25<\textit{AI}<18.75 for cache-intensive and \textit{AI}>18.75 for compute-boundness.
We present the AI values of each benchmark on the two frequency levels in the table.
The results indicate that, despite its simplicity, this low overhead power modeling approach achieves reasonable and reliable accuracy for estimating power consumption of tasks.
Specifically, we take a real application - Sparse LU as 
an example to show the comparison between the modeled power and the real power trace when running with different execution configurations.
The corresponding results are shown in Figure~\ref{fig_powermodel}.
All tasks from the four different kernels in Sparse LU (see Section~\ref{syntheticdag}) with 2.04GHz are categorized as compute-bound, since the average AI of all Sparse LU tasks is 31.5. 
Thus, at runtime, the corresponding compute-bound power profile values are used as power estimates.
The power comparison results on Sparse LU show that the power prediction 
deviations are within 10\% (7\% on average) across different execution places.

\begin{table*}[t]
\scriptsize
\vspace{-3mm}
\caption{Power prediction deviations evaluated with four benchmarks on Jetson TX2}
\vspace{-3mm}
\begin{center}
\begin{tabular}{p{.35cm}p{.75cm}p{.75cm}p{.75cm}p{.75cm}p{.75cm}p{.75cm}p{.75cm}p{.75cm}p{.75cm}p{.75cm}p{.75cm}p{.75cm}} \toprule
& \multicolumn{3}{c}{\textbf{MatMul}} & \multicolumn{3}{c}{\textbf{Copy}} & \multicolumn{3}{c}{\textbf{Stencil}} & \multicolumn{3}{c}{\textbf{Sparse LU}} \\ \midrule
& 2.04GHz & 0.35GHz & Average & 2.04GHz & 0.35GHz & Average & 2.04GHz & 0.35GHz & Average & 2.04GHz & 0.35GHz & Average \\ \midrule
\textbf{D\_1} &	1.88\%  &	4.33\%  &   1.25\%	&   6.64\%  &   6.24\%  &   3.10\% & 1.14\% &   5.07\% &  1.47\%  &   9.55\%  &   3.87\%  &   2.82\%	 \\ \midrule
\textbf{D\_2} &  5.62\%  &   0.13\%  &   3.54\%	&   22.87\% &   17.90\% &   17.18\% & 6.19\%  & 3.74\%   &  3.81\% &   7.03\%  &   0.92\%  &   2.61\%   \\ \midrule
\textbf{A\_1} &  6.46\%  &   1.13\%  &   3.88\%	&   7.34\%	&   2.62\%  &   6.03\%  & 4.16\%  &  2.05\% &  4.22\% &   5.06\%  &   0.13\%  &   2.58\% \\ \midrule
\textbf{A\_2} &  3.96\%  &   3.16\%  &   2.94\%	&   7.71\%  &   8.56\%  &   7.16\%  & 7.92\%  &  6.11\% &  5.28\%  &   7.41\%  &   9.30\%  &   2.86\%  \\ \midrule
\textbf{A\_4} &  0.46\%  &   1.50\%  &   4.54\%  &   5.02\%  &   2.28\%  &   5.31\%   & 7.32\% & 4.55\% &  5.10\% &  6.13\%  &   9.68\%  &   5.92\% \\ \midrule
\multirow{2}{*}{\textbf{AI}}  &   423.8   &   400.6   &   ---  &  2.07   &  1.03   &   ---  &  17.23    &   16.88   & --- &   31.5   & 23.8    &  ---  \\
& \multicolumn{3}{c}{compute-bound} & \multicolumn{3}{c}{memory-bound} & \multicolumn{3}{c}{cache-intensive} & \multicolumn{3}{c}{compute-bound}\\ \bottomrule
\end{tabular}
\label{tab:poweraccuracy}
\end{center}
\vspace{-3mm}
\end{table*}


\begin{figure*}[t!]
\centering
\includegraphics[width=0.7\textwidth]{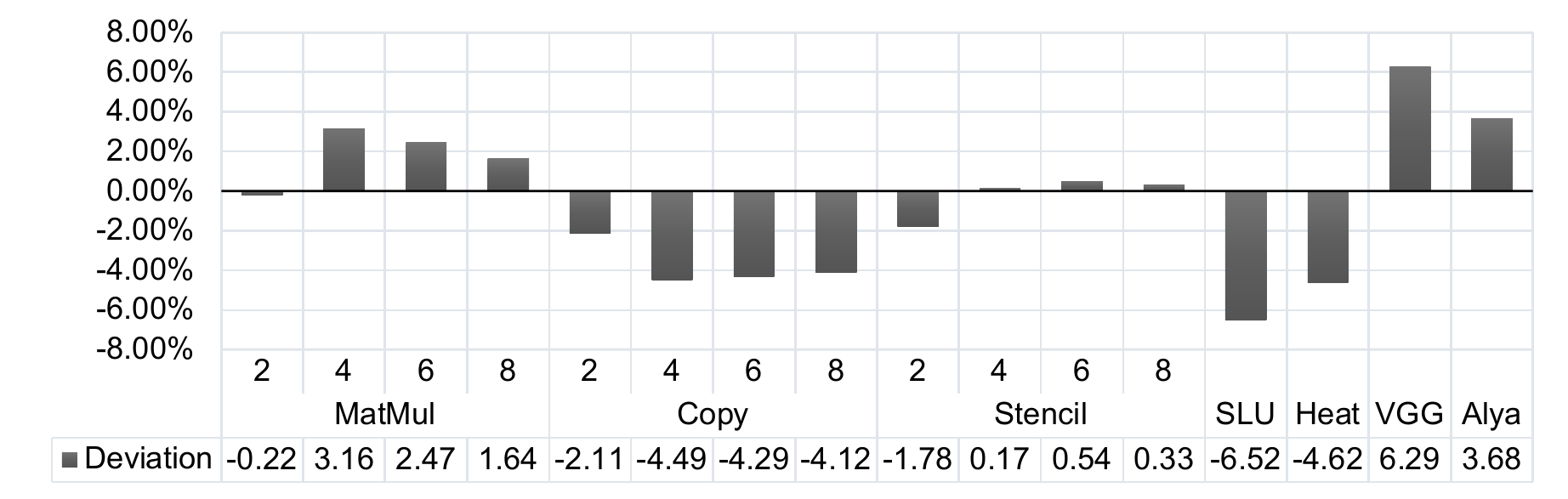}
\vspace{-3mm}
\caption{ERASE energy prediction deviations with all evaluated benchmarks on Jetson TX2.} 
\vspace{-4mm}
\label{fig_erase_accuracy}
\end{figure*}

\textbf{Energy Prediction Accuracy:}
Figure~\ref{fig_erase_accuracy} presents the predicted deviations of energy consumption measured in MAPE across all evaluated benchmarks. 
The MAPE value of each benchmark is computed by comparing the predicted energy consumption 
of each task to the real energy consumption.
The evaluation shows that the differences between predicted energy predictions and the real energy consumption are within 7\% (3.5\% on average).

\section{Related Work}\label{RelatedWork}
Energy efficient task scheduling has received some attention in recent years.
Existing proposals rely on DVFS reconfiguration, scheduling strategies or combinations thereof.
There are some prior works 
applying to the environments with a centralized scheduler for all cores.
Costero et al.~\cite{LuisCostero-HPCS2017-DVFS/Energy/RatioDecision} developed policies for frequency scaling and task scheduling for asymmetric platforms.
First, they tune DVFS as a function of the ratio between running critical tasks and non-critical tasks.
Next, they schedule tasks either to the big or LITTLE cluster based on the current number of ready tasks (parallel slackness).    
Sanjeev et al.~\cite{baskiyar2010energy} proposed EADAGS, a two-phase task scheduler to minimize the total execution time and energy consumption using dynamic voltage scaling and decisive path scheduling. 
In the first phase, EADAGS creates a single queue and puts critical tasks into the queue and then schedules critical tasks one by one to a processor that minimizes the execution time of the task. 
Then the energy consumption is calculated and it tunes down the voltage of processors if the decision does not bring the performance slowdown.
These methods using a centralized queue are more suitable for small systems and they present the issues of poor scalability and increasing communication cost between cores and the centralized queue on larger shared memory systems. 

In contrast, 
work stealing allows each core to have its own work queue and idle cores greedily steal tasks from others to keep workload balancing, which 
achieves better scalability and reduces the contention~\cite{blumofe-jacm99, ChenQuan-IPDPS2012-WorkStealing/Asymmetry/Performance}.
Haris et al.~\cite{HarisRibic-WorkStealing/PercoreDVFS/Tempo} proposed HERMES,
which exploited DVFS per core by checking the task stealing sequences between cores and workloads on two AMD machines. 
However, their works~\cite{HarisRibic-WorkStealing/PercoreDVFS/Tempo, HarisRibic-AEQUITAS} target only symmetric platforms and do not consider the opportunities on asymmetric architectures.
Torng et al.~\cite{Christopher-ISCA2016-Asymmetric/PercoreDVFS/Parallelism} proposed AAWS that targets for improving the performance of CPU-bound applications in work stealing runtime on asymmetric platforms. It used work-pacing, work-sprinting and work-mugging strategies along with the per-core DVFS capability to map the critical tasks to the most appropriate resources and balance the workloads among cores.
For moldable streaming tasks, Melot et al.~\cite{melot_taco_2015} proposed a heuristic for ``crown'' frequency scaling using a simulated annealing technique to achieve energy efficient mapping on symmetric manycore systems. 
In contrast, ERASE achieves energy reduction by task type awareness, exploiting moldability and task parallelism detection for both static and dynamic frequency settings.
\section{Conclusion} \label{conclusion}
We propose ERASE - an intra-application task scheduler for reducing energy consumption of executing fine-grained task-based parallel applications and discuss the design and implementation on top of a work stealing runtime. 
The design of the scheduler is motivated by the following observations:
(1) leveraging DVFS in per task basis is ineffective when using fine-grained tasking and in environments with cluster/chip-level DVFS;
(2) task moldability, wherein a single task can execute on multiple threads/cores via work-sharing, can help to reduce the energy consumption;
(3) combining task type-aware execution and task moldability is crucial for energy savings, particularly on asymmetric platforms.
The scheduler works by estimating the energy consumption of different execution places at a per-task level and performs task mapping decisions so as to minimize the energy consumption of each task. 
It also adaptively puts cores that repeatedly execute work stealing loop without success to sleep state by applying an exponential back-off sleeping strategy. 
In addition, ERASE is capable of adapting to both given static frequency settings and externally controlled DVFS.
We evaluate ERASE by comparing it to several state-of-the-art scheduling techniques on two different platforms. 
The results indicate that ERASE can provide significant energy savings in comparison, across a range of benchmarks and system environment settings.

\begin{acks}
This work has received funding from the \grantsponsor{1}{European Union Horizon 2020 research and innovation programme}{https://legato-project.eu/} under grant agreement No.~\grantnum[https://legato-project.eu/]{1}{780681}.
This work has also received funding from the \grantsponsor{2}{European High-Performance Computing Joint Undertaking (JU)}{https://eprocessor.eu} under grant agreement No.~\grantnum[https://eprocessor.eu]{2}{956702}. 
The JU receives support from the European Union’s Horizon 2020 research and innovation programme and Spain, Sweden, Greece, Italy, France, Germany.
The computations were enabled by resources provided by the Swedish National Infrastructure for Computing (SNIC), partially funded by the \grantsponsor{3}{Swedish Research Council}{https://www.vr.se/} through grant agreement No.~\grantnum[https://www.vr.se/]{3}{2018-05973}.
\end{acks}

\bibliographystyle{ACM-Reference-Format}
\bibliography{jrlib-miquel,Reference}

\end{document}